\documentclass{article}
\usepackage{epsf}
\usepackage{epsfig}
\usepackage{amssymb}
\usepackage{amsthm}
\usepackage{amsmath}
\usepackage{epstopdf}
\usepackage{tikz}
\usetikzlibrary{patterns}
\usepackage[mathscr]{eucal}
\usepackage{amsthm}
\usepackage{amsmath}
\usepackage{amssymb}
\usepackage{mathrsfs}
\usepackage{stmaryrd} 
\usepackage{cite} 

\newtheorem{remark}{Remark}
\newtheorem{example}{Example}

\usepackage{color}
\newcommand{\aSt}{\mathcal{W}}

\newcommand{\jesper}[1]{$\framebox{\tiny j}$\ \textbf{\texttt{{\color{blue}\footnotesize#1}}}}
\newcommand{\IrrJTL}[2]{\mathcal{X}_{#1,#2}}
\newcommand{\AStTL}[2]{\aSt_{#1,#2}}
\newcommand{\StJTL}[2]{\mathcal{W}_{#1,#2}}
\newcommand{\bAStTL}[2]{\overline{\mathcal{W}}_{#1,#2}}
\newcommand{\ATL}[1]{\mathsf{T}^a_{#1}}
\newcommand{\q}{\mathfrak{q}}
\newcommand{\rJTL}[1]{\mathsf{JTL}_{#1}}

\newcommand{\oZ}{\mathbb{Z}}

\newcommand{\qq}{\mathrm{q}}

\newcommand{\bq}{\bar{q}}

\newcommand{\IrrV}[1]{\mathsf{X}_{#1}}
\newcommand{\IrrVb}[1]{\overline{\mathsf{X}}_{#1}}
\newcommand{\G}{\mathrm{G}}
\begin{document}

\title{Bootstrap approach to geometrical four-point functions \\ in the two-dimensional critical $Q$-state Potts model:\\ A study of  the $s$-channel spectra}

\author{Jesper Lykke Jacobsen$^{1,2,3}$ and Hubert Saleur$^{3,4}$ \\
[2.0mm]
  ${}^1$Laboratoire de Physique Th\'eorique, D\'epartement de Physique de l'ENS, \\
  \'Ecole Normale Sup\'erieure, Sorbonne Universit\'e, CNRS, \\ PSL Research University, 75005 Paris, France \\
  ${}^2$Sorbonne Universit\'e, \'Ecole Normale Sup\'erieure, CNRS, \\ Laboratoire de Physique Th\'eorique (LPT ENS), 75005 Paris, France \\
  ${}^3$Institut de Physique Th\'eorique, Universit\'e Paris Saclay, CEA, CNRS,
  91191 Gif-sur-Yvette, France \\
  ${}^4$Department of Physics,
  University of Southern California, Los Angeles, CA 90089-0484,
   USA }


\maketitle

\begin{abstract}

We revisit in this paper the problem of connectivity correlations in the Fortuin-Kasteleyn cluster representation of the two-dimensional $Q$-state Potts model
conformal field theory. In a recent work \cite{Ribault}, results for the four-point functions were obtained, based on the bootstrap approach, combined with simple conjectures for the spectra in the different fusion channels. In this paper, we test these conjectures  using  lattice algebraic considerations combined with extensive numerical studies of correlations on infinite cylinders. We find that the spectra  in the scaling limit are much richer than those proposed in \cite{Ribault}: they involve in particular  fields with conformal weight $h_{r,s}$ where $r$ is dense on the real axis. 
\end{abstract}

\tableofcontents

\section{Introduction}

In a very interesting recent paper \cite{Ribault}, a proposal was put forward for some of the four-point correlation functions of the percolation problem in two dimensions. This proposal was part of a more general conjecture addressing various geometrical objects involving four points in the diagrammatic  formulation \cite{FK72} of the $Q$-state Potts model \cite{Potts52}. The case $Q=1$ corresponds to percolation, and the proposal in \cite{Ribault} covers such objects as the probability that two of these points belong to one cluster, and the two others to another cluster. Obtaining closed-form expressions for such objects is one of the holy grails in the field. It is a far from obvious endeavour because the conformal field theory (CFT) describing percolation (and more generally geometrical features of the $Q$-state Potts model) is not well understood: it is non-unitary, probably involves logarithms (even for $Q$ generic), and involves operators which are not degenerate, precluding the use of the differential equations approach \`a la BPZ \cite{yellowpages}.

The construction in \cite{Ribault} is elegant and powerful. It starts with a seemingly reasonable hypothesis for the spectrum of operators appearing in  the  fusion channels for the fusion of two order operators, and determines, using  a clever code,  the whole set of structure constants based on our knowledge of conformal blocks \cite{Zamoblocks1,Zamoblocks2} together with the imposition of crossing symmetry. The results are then checked against  Monte Carlo simulations, with, it is claimed,  reasonably good agreement.

Although the results in \cite{Ribault} are appealing, they are not really consistent with what is known about the Potts model CFT and, in particular, percolation. Early work \cite{DFSZ} has revealed indeed a much richer spectrum than the one postulated in   \cite{Ribault}, which covers only a very tiny set of the known  full operator content of the theory. Of course, it could be that  by some accident, the order operator in the $Q$-state Potts model does not couple to as many fields as one would expect,  at least in the scaling limit. But it could also be that something is simply missing in the work of \cite{Ribault}, despite the apparent numerical effectiveness of their proposal. 

To investigate this question  requires a long and detailed analysis, of which we present the results here. In a nutshell, we have gathered direct, in our opinion unquestionable evidence that the spectrum of the $Q$-state Potts model is as complex as could have been feared, that many more fields  appear in the OPE of order operators in the Potts model than was conjectured in \cite{Ribault}, and that the proposal in that paper, appealing as it may be, simply cannot be correct. It  is, at best, a good numerical approximation to the true expressions for the four-point functions. 

Our paper is organised as follows. In section \ref{sec:Potts-corr} we remind the reader of basic facts and results about the $Q$-state Potts model and its geometrical formulation. Algebraic aspects---which constitute a crucial part of our approach---are discussed in section \ref{sec:latt-alg}. 
Section \ref{sec:four-point} summarises our method of analysis, and how we extract exponents as well as amplitudes, from lattice data. Section \ref{sec:results} discusses our results for the spectra in the intermediate channels of four-point functions. A comparison with results in \cite{Ribault} is provided in section \ref{sec:comp-ribault}. In section \ref{sec:conclusion}, we return to the issue of divergences in the amplitudes, and re-analyse briefly our results as well as those of \cite{Ribault} from the point of view of degeneracies, and, potentially, logarithmic CFTs.

Since a good part of our analysis is based on extracting amplitudes from lattice data, a lot of technical aspects have to be considered both to make the program possible, and to check its validity and its limits. We have thus gathered quite a bit of material in a series of appendices. Appendix \ref{sec:appA} discusses in detail many aspects of the numerical algorithms and other techniques used to obtain our results, while appendix \ref{sec:appB} goes over a series of checks, including detailed comparisons, in particular, with known results for $Q=0,2,4$. 

\section{Potts model and its correlation functions}
\label{sec:Potts-corr}

We consider the $Q$-state Potts model \cite{Potts52} defined on a graph $G=(V,E)$ with vertices $V$ and edges $E$.
There is a spin $\sigma_i = 1,2,\ldots,Q$ attached to each vertex $i \in V$ and an interaction energy $-K \delta_{\sigma_i,\sigma_j}$
attached to each edge $(ij) \in E$. The partition function (in units where the inverse temperature is absorbed into the coupling constant $K$) reads
\begin{equation}
 Z = \sum_{\{ \sigma \}} \prod_{(ij) \in E} {\rm e}^{K \delta_{\sigma_i,\sigma_j}} \,.
\end{equation}

Note that this initial formulation supposes $Q$ to be a positive integer, $Q \in \mathbb{N}$.
This constraint can be lifted in a rewriting of $Z$ due to Fortuin and Kasteleyn (FK) \cite{FK72}. Indeed, write
${\rm e}^{K \delta_{\sigma_i,\sigma_j}} = 1 + v \delta_{\sigma_i,\sigma_j}$ with temperature parameter $v = {\rm e}^K - 1$,
expand the product $\prod_{(ij) \in E}$, and perform the sum over all spins $\{ \sigma \}$ to obtain
\begin{equation}
 Z = \sum_{A \subseteq E} v^{|A|} Q^{k(A)} \,, \label{Z_FK}
\end{equation}
where the sum is over all $2^{|E|}$ subsets of $E$, and $|A|$ denotes the number of edges in the subset.
Moreover, $k(A)$ denotes the number of connected components (also called FK clusters) in the subgraph $G_A = (V,A)$.

In the remainder of the paper we take $G$ to be the two-dimensional square lattice. The temperature parameter will be taken at its critical value,
$v_{\rm c} = \sqrt{Q}$ \cite{Potts52,Baxter73}, so that the model is conformally invariant in the continuum limit. In this latter limit, we are interested in the geometry of the
infinite plane, so that boundary effects are immaterial. We shall often use the trick of transforming this into the geometry of a cylinder, via an appropriate
conformal mapping (details will be given below). This cylinder geometry is convenient for imposing the lattice discretisation, which is our main tool of
algebraic and numerical investigations. In that case we always take the square lattice $G$ to be axially oriented with respect to the cylinder axis,
so that the row-to-row transfer matrix describes the (imaginary) time evolution of $L$ Potts spins.

\subsection{Loop model}

An equivalent formulation of $Z$ is given \cite{BKW76} by the loop model on the medial lattice ${\cal M}(G) = (V_{\cal M},E_{\cal M})$. The vertices $V_{\cal M}$
of ${\cal M}(G)$ are situated at the mid-points of the original edges $E$, and two vertices in $V_{\cal M}$ are connected by an edge in $E_{\cal M}$
whenever the former stand on edges in $E$ that are incident on a common vertex from $V$. In particular, when $G$ is a square lattice, ${\cal M}(G)$
is just another square lattice, tilted through an angle $\frac{\pi}{4}$ and scaled down by a factor of $\sqrt{2}$.
There is a bijection between edge subsets $A \subseteq E$ and completely-packed loops on ${\cal M}(G)$. The loops are defined so that they turn
around the FK clusters and their internal cycles (alternatively they separate the FK clusters from their duals). One has then
\cite{BKW76}
\begin{equation}
 Z = Q^{|V|/2} \sum_{A \subseteq E} \left( \frac{v}{n} \right)^{|A|} n^{\ell(A)} \,, \label{Z_loop}
\end{equation}
where $\ell(A)$ denotes the number of loops. The loop fugacity is
\begin{equation}
 n = Q^{1/2} = \mathfrak{q} + \mathfrak{q}^{-1} \,, \label{loop-fugacity}
\end{equation}
where the (quantum group related) parameter $\mathfrak{q}$ will be used intensively below. The loop model will be convenient to make contact with
the Temperley-Lieb (TL) algebra \cite{TL71}, which will be discussed below.
Note also that on a square lattice, we have simply $\frac{v_{\rm c}}{n} = 1$, so at the critical point $Z$ depends only on $\ell(A)$.

\subsection{Correlation functions}

The Potts model allows for the definition of various correlation functions, depending on whether one uses its formulation in terms of Potts spins,
FK clusters or TL loops.%
\footnote{One can even use the spin and FK cluster formulations simultaneously to define new correlation functions \cite{VJ_spin_FK}.}
The spin correlators are naturally defined in terms of the order parameter (or spin) operator
\begin{equation}
 {\cal O}_a(\sigma_i) \equiv Q \delta_{\sigma_i,a} - 1 \,. \label{spin-op}
\end{equation}
More interesting and general results can however be obtained by moving to the cluster or loop formulations, in which the correlation functions acquire
a geometrical content. In the same vein, the spin correlators can be analytically continued from $Q \in \mathbb{N}$ to arbitrary real values, in which
case they also acquire a geometrical interpretation \cite{DJS_spin1,DJS_spin2,DelfinoViti} to which we shall return in a short while. Such generalisations to $Q \in \mathbb{R}$
are not only useful, but actually indispensable in our case, since our main objective is to study correlation functions in the generic case where
$\mathfrak{q}$ is not a root of unity.

Correlation functions in the loop formulation are of either electric or magnetic type, where the terminology refers to the Coulomb gas approach to CFT
\cite{Nienhuis_CG,DFSZ}.
Let $i_1,i_2,\ldots,i_N$ be a number of distinct marked vertices. Electric correlators are defined for $i_k \in V$ by appropriately modifying the weight of loops that
contain a subset of marked vertices on their inside, and the remainder on their outside. Magnetic correlators are defined for $i_k \in V_{\cal M}$ by
specifying whether given vertices belong to the same or different loops; one can also increase the set of possibilities by allowing for topological defects
that insert a number of open loop strands at each marked vertex \cite{DS_watermelon}. While these electromagnetic correlation functions have been intensively studied for
$N=1,2$ in a variety of contexts, we wish here to recall only one recent result. Namely, the electric $N=3$ correlation functions have been shown to be
related, for generic values of $n \in [0,2]$, to the so-called DOZZ formula for the structure constants within Liouville field theory \cite{DelfinoViti,PiccoSantachiaraVitiDelfino,IkhlefJacobsenSaleur}.

Our main interest here is however correlation functions defined in terms of the FK clusters.
Let again $i_1, i_2,\ldots, i_N \in V$ be a number of distinct marked vertices, and let ${\cal P}$ be a partition of a set of $N$ elements. We then
define
\begin{equation}
 P_{\cal P} = \frac{1}{Z} \sum_{A \subseteq E} v^{|A|} Q^{k(A)} {\cal I}_{\cal P}(i_1,i_2,\ldots,i_N | A) \,, \label{P_corr_def}
\end{equation}
where $Z$ is given by (\ref{Z_FK}), and ${\cal I}_{\cal P}(i_1,i_2,\ldots,i_N | A)$ is the indicator function that, $\forall k,l \in \{1,\ldots,N\}$,
vertices $i_k$ and $i_l$ belong to
the same connected component in $A$ if and only if $k$ and $l$ belong to the same block of the partition ${\cal P}$. It is convenient to denote
${\cal P}$ by an ordered list of $N$ symbols ($a,b,c,\ldots$) so that identical symbols refer to the same block. For instance, with $N=2$,
$P_{aa}$ is the probability that vertices $i_1,i_2$ belong to the same FK cluster, whereas $P_{ab} = 1 - P_{aa}$ is the probability that $i_1,i_2$
belong to two distinct FK clusters. In the context of four-point functions, we are therefore interested in the 15 probabilities
$P_{aaaa}, P_{aabb},\ldots,P_{abcd}$. The combinatorial properties of FK correlation functions were further discussed in \cite{DelfinoViti}.

It is natural to relate $P_{\cal P}$ to correlation functions of the spin operator. Define
\begin{equation}
 G_{a_1,a_2,\ldots,a_N} = \left \langle {\cal O}_{a_1}(\sigma_{i_1}) {\cal O}_{a_2}(\sigma_{i_2}) \cdots {\cal O}_{a_N}(\sigma_{i_N}) \right \rangle \,,
 \label{order-param-corr}
\end{equation}
where $a_1,a_2,\ldots,a_N$ is a list of (identical or different) symbols defining a set partition ${\cal P}$, and the expectation value $\langle \cdots \rangle$
is defined with respect to the normalisation $Z$. It is straightforward to formally relate the $G_{\cal P}$ to $P_{\cal P}$. Indeed, to evaluate the expectation value
of a product of Kronecker deltas, we initially suppose that $Q$ is integer, and use that spins on the same FK cluster are equal, while spins on different
clusters are statistically dependent. This leads to $Q$-dependent relations, which can finally be extended to real values of $Q$ by analytical continuation.
For instance, with $N=2$, one readily finds that
\begin{equation}
 G_{a_1,a_2} =  \left(Q \delta_{a_1,a_2} -1\right) P_{aa} \,. \label{G_2-point}
\end{equation}
In other words, the two-point function of the spin operator is proportional to the probability that the two points belong to the same FK cluster.
Therefore ${\cal O}_a(\sigma_i)$ effectively ``inserts'' an FK cluster at position $i \in V$ and ensures its propagation until it is ``taken out'' by
another spin operator.

\begin{remark}
In a recent series of works \cite{tensor1,tensor2,tensor3} we have introduced a more general class of operators
${\cal O}_{a_1,a_2,\ldots,a_N}(\sigma_{i_1},\sigma_{i_2},\ldots,\sigma_{i_N})$ that act on $N$ spins according to given irreducible representations
of the symmetric groups $S_Q$ and $S_N$. These operators can enforce the propagation of more than one FK cluster, with the set of propagating
clusters having specific symmetry properties. Some of the four-point functions to be considered below (namely $P_{abab} \pm P_{abba}$),
with the points being considered as regrouped in two pairs, 
actually coincide with two-point functions of such operators, each acting on a pair of spins ($N=2$).
\end{remark}

In the remainder of this paper we shall focus on the same subset of four-point correlation functions as was studied in \cite{Ribault}.
They are the functions $P_{\cal P}$ where the partition ${\cal P}$ contains only one or two blocks, namely: $P_{aaaa}$, $P_{aabb}$, $P_{abba}$ and $P_{abab}$.
The relation with the corresponding $G_{\cal P}$ read \cite[eqs.~(19)--(22)]{DelfinoViti}
\begin{subequations} \label{sumrules}
\begin{eqnarray}
G_{aaaa} &=&(Q-1)(Q^2-3Q+3) P_{aaaa}+(Q-1)^2(P_{aabb}+P_{abba}+P_{abab}) \,, \\
G_{aabb} &=&(2Q-3) P_{aaaa}+(Q-1)^2P_{aabb}+P_{abba}+P_{abab} \,, \\
G_{abba} &=&(2Q-3)P_{aaaa}+P_{aabb}+(Q-1)^2P_{abba}+P_{abab} \,, \\
G_{abab} &=&(2Q-3)P_{aaaa}+P_{aabb}+P_{abba}+(Q-1)^2P_{abab} \,.
\end{eqnarray}
\end{subequations}
As already stated above, for $Q$ arbitrary, the left-hand sides of these equations are only formally defined: it is in fact the right-hand sides that give them a meaning.
Note that this linear system has determinant $Q^4 (Q-1)(Q-2)^3 (Q-3)$, so it cannot be fully inverted for $Q=0,1,2,3$. 

By analogy with (\ref{G_2-point}) one would expect that, in the scaling limit, the four $P_{\cal P}$ of interest would be described by combinations of conformal
blocks for the spin operator. In particular, the function $P_{aaaa}$ corresponding to the four points being in the same cluster should become, in the scaling limit, a crossing-invariant such combination. The other three would maybe not be crossing-invariant individually, but might be related with each other by crossing (or give rise,
after proper combinations, to other crossing-invariant objects).

Clearly, to implement the bootstrap programme, one needs an idea of the set of
conformal blocks that may appear in these geometrical correlation functions. 
The key question in this problem---the one that we shall pursue in the remainder of this paper---is therefore what happens in the $s$-channel of each of
these four correlation functions, when two order operators are brought close to each other. Note that, since the conformal field theories we are dealing with are not unitary, the behavior of the $G$ or $P$ functions might be more complicated than in the unitary cases, and involve, in particular, logarithmic terms. Examples of such behaviours are already known for two- and three-point functions \cite{tensor1,tensor2,tensor3}.

\begin{remark}
An important note: unless otherwise specified we will use the same notation (such as $P_{\cal P}$ and $G_{\cal P}$) for correlation functions
defined on the lattice and  for their scaling limits. 
\end{remark}

\section{Lattice algebras}
\label{sec:latt-alg}

As mentioned above, our main exploratory tool for unravelling the structure of four-point functions is to impose a lattice discretisation and study the
Potts model in the cylinder geometry. The algebraic object that propagates a row of $L$ Potts spins axially along the cylinder axis is a linear operator
called the row-to-row
transfer matrix $T$. In this section we discuss how $T$ can be used to build the partition function $Z$, and defer the more technical question about
how to build the correlation functions $P_{\cal P}$ to Appendix~\ref{sec:appA1}. Both the algebraic definition of $T$ and the space of states on which it acts depend
subtly on the degrees of freedom defining the model. We are here interested in two different representations, viz.\ in terms of FK clusters and TL spins,
which we now describe in turn. The key technical point is to impose a weight $Q$ per cluster in the former case, or a weight $n$ per loop in the latter.

\subsection{FK clusters and the join-detach algebra}
\label{sec:join-detach}

To build a row of an axially oriented square lattice, the transfer matrix $T$ must first add $L$ ``horizontal'' edges in some row of constant imaginary time $t$, and
then propagate to the next row at time $t+1$ by adding $L$ ``vertical'' edges. It is convenient to introduce more elementary operators that add just a single
edge to the lattice. Concretely, ${\sf H}_i$ adds a horizontal edge between sites $i$ and $i+1$ (mod $L$), while ${\sf V}_i$ adds a vertical edge on top of site $i$.
We can thus write
\begin{subequations}
\label{sparse_matrix_factorisation}
\begin{eqnarray}
 T &=& {\sf V} {\sf H} \,, \\
 {\sf H} &=& {\sf H}_L \cdots {\sf H}_2 {\sf H}_1 \,, \\
 {\sf V} &=& {\sf V}_L \cdots {\sf V}_2 {\sf V}_2 \,.
\end{eqnarray}
\end{subequations}

The operators ${\sf H}_i$ and ${\sf V}_i$ must ensure the correct building of the sum $\sum_{A \subseteq E}$ in (\ref{Z_FK}). They can be written
\begin{subequations} \label{HV}
\begin{eqnarray}
 {\sf H}_i = {\sf I} + v {\sf J}_i \,, \\
 {\sf V}_i = v {\sf I} + {\sf D}_i \,,
\end{eqnarray}
\end{subequations}
where ${\sf I}$ denotes the identity operator, while ${\sf J}_i$ and ${\sf D}_i$ will be defined shortly.
Each expression has two terms depending on whether the given edge $e$ belongs to the subset $A$ or not. In the former case, a weight $v$
is applied. The subtle point is to obtain also the non-local weight of $Q$ per completed cluster. To that end, $T$ acts on states $\{s_1,s_2,\ldots,s_L\}$
which are set partitions of $L$ points describing how the sites of a row are interconnected via the parts of the FK clusters living at times prior to $t$.
The join operator ${\sf J}_i$ amalgamates the blocks of the partition corresponding to sites $i$ and $i+1$ (mod $L$). The detach operator
${\sf D}_i$ transforms site $i$ into a singleton, applying a weight $Q$ if it was already a singleton beforehand. The join-detach algebra is defined
by the algebraic rules emanating from these requirements:
\begin{subequations}  \label{join-detach}
\begin{eqnarray}
 {\sf J}_i^2 &=& {\sf J}_i \,, \\
 {\sf D}_i^2 &=& Q {\sf D}_i \,, \\
 {\sf J}_i {\sf D}_j {\sf J}_i &=& {\sf J}_i \mbox{ for } j=i,i+1 \,, \\
 {\sf D}_i {\sf J}_j {\sf D}_i &=& {\sf D}_i \mbox{ for } j=i-1,i \,,
\end{eqnarray}
\end{subequations}
where all indices are considered modulo $L$. Operators associated with sites that are farther apart than in the relations given simply commute.

In two dimensions, the join-detach algebra is closely related to the Temperley-Lieb (TL) algebra \cite{TL71} that we describe next. For a more general
discussion, see \cite{HalversonRam}. The question of how the join-detach algebra must be adapted to accommodate the computation
of correlation functions is deferred to Appendices~\ref{sec:appA1}--\ref{sec:appA2}. For some applications (see Appendix~\ref{sec:appA2} in particular)
we shall also need to consider the transpose of the join-detach algebra, which furnishes another geometrical representation of the TL algebra that we shall call
the {\em split-attach algebra} and describe in some detail in Appendix~\ref{sec:split-attach}.

\subsection{Loops and the Temperley-Lieb algebra}

Another option is to define the transfer matrix in terms of the loops that separate the FK clusters from their duals. We recall that these loops now
live on a tilted square lattice ${\cal M}(G)$. At each vertex of ${\cal M}(G)$ two pieces of loop, labelled $i$ and $i+1$ according to their horizontal
position, can either bounce off a ``vertical'' or a ``horizontal'' edge
of $G$ (or its dual $G^*$), operations that are described in imaginary time by respectively the identity operator ${\sf I}$ and the so-called braid
monoid $e_i$:
\begin{equation}
 {\sf I} =  \ 
 \begin{tikzpicture}[scale=1/4, baseline = {(current bounding box.center)},yscale=-1]
 \draw[black, line width = 1pt] (0,0) .. controls (0.4,0.5) and (0.4,1.5) .. (0,2);
 \draw[black, line width = 1pt] (1,0) .. controls (0.6,0.5) and (0.6,1.5) .. (1,2);
 \end{tikzpicture} \qquad \qquad
 e_i =  \ 
 \begin{tikzpicture}[scale=1/4, baseline = {(current bounding box.center)},yscale=-1]
 \draw[black, line width = 1pt] (0,0) .. controls (0,1) and (1,1) .. (1,0);
 \draw[black, line width = 1pt] (0,2) .. controls (0,1) and (1,1) .. (1,2);
 \end{tikzpicture} \qquad
 \end{equation}
The $e_i$ generate the Temperley-Lieb (TL) algebra which has a long history \cite{TL71} and is deeply associated with work on the Potts
model \cite{Baxter,PPMartin}. 

We note that a horizontal cut through ${\cal M}(G)$, in between two rows of vertices, will intersect the loop pieces in $N = 2L$ points.
If we set ${\sf J}_i = Q^{-1/2} e_{2i}$ and ${\sf D}_i = Q^{1/2} e_{2i-1}$ for any $i=1,2,\ldots,L$, the algebraic relations (\ref{join-detach})
become simply
\begin{subequations}  \label{TLpdef}
\begin{eqnarray}
 e_i^2 &=& n e_i \,, \label{TLpdef-a} \\
 e_i e_{i \pm 1} e_i &=& e_i \,, \\
 \left[ e_i , e_j \right] &=&0 \mbox{ for } \left|i-j \right| \geq 2 \,, \label{TLpdef-c}
\end{eqnarray}
where $n$ is given by (\ref{loop-fugacity}).
These are precisely the defining relation of the TL algebra.

Up to this point we have deliberately been rather loose about specifying the boundary conditions. Indeed, the
TL algebra per se is associated with the Potts model on a strip---i.e., with open boundary conditions (that is, free boundary conditions
for the Potts spins and reflecting boundary conditions for the loops)---and the
generators $e_i$ are defined for $i=1,2,\ldots,N-1$.
In the cylinder geometry---i.e., with periodic boundary conditions---a tempting possibility is to merely
add a last generator ``closing'' the system, $e_{N}$, and  define the labels modulo $N$ in the defining relations (\ref{TLpdef}),
so that in particular $e_{N} e_1 e_{N} = e_1$ and $e_1 e_N e_1 = e_1$.
This natural generalisation however takes one into a sticky mathematical problem: the corresponding algebra is then seen to
be infinite-dimensional, even for finite $N$.
In a nutshell, this occurs  because of through-lines or loops that can wind around the system.
While what must be done with these objects is clear in the Potts model itself, this requires providing 
extra information that is not present in the definition of the ``periodicised'' Temperley-Lieb algebra.
This extra information takes the mathematical form of {\em quotients}. 

To define these quotients more precisely, it is useful to also introduce
a translation generator $u$ that shifts the label
of the $e_i$ generators, giving rise to the following extra relations---in addition to (\ref{TLpdef-a})--(\ref{TLpdef-c})---with integer indices considered modulo $N$ (that is, $i\in\oZ_N$):
\begin{eqnarray}
u e_i u^{-1} &=& e_{i+1}\label{TLpdef-d}\\
u^2 e_{N-1} &=& e_{1} e_2 \dots e_{N-1}\label{TLpdef-e}\ .
\end{eqnarray}
\end{subequations}
The translation operator has the diagrammatic representation
\begin{equation*}
u = 
\begin{tikzpicture}[scale=1/3, baseline = {(current bounding box.center)},yscale=-1]
	\foreach \r in {1,2,5,6}{
	\draw[black, line width = 1pt] (\r,0) .. controls (\r,1) and (\r-1,2) .. (\r-1,3);
	};
	\draw[black, line width = 1pt] (0,0) .. controls (0,1) and (-1,2) .. (-1,3);
	\draw[black, line width = 1pt] (7,0) .. controls (7,1) and (6,2) .. (6,3);
	\filldraw[white] (-.5,0) rectangle (-1.5,3);
	\filldraw[white] (6.5,0) rectangle (7.5,3);
	\node[anchor = north] at (3,1) {$\hdots$};
\end{tikzpicture}.
\end{equation*}
The last relation (\ref{TLpdef-e})
is easily understood in terms of diagrams, for example for $N=4$,
\begin{equation*}
 e_{1}e_{2}e_{3} = \quad
 \begin{tikzpicture}[scale=1/4, baseline = {(current bounding box.center)},yscale=-1]
	\draw[black, line width = 1pt] (0,0) .. controls (0,1) and (1,1) .. (1,0);
	\draw[black, line width = 1pt] (0,3) .. controls (0,2) and (1,2) .. (1,3);
	\draw[black, line width = 1pt] (2,0) -- (2,3);
	\draw[black, line width = 1pt] (3,0) -- (3,3);
	\draw[black, line width = 1pt] (1,3) .. controls (1,4) and (2,4) .. (2,3);
	\draw[black, line width = 1pt] (1,6) .. controls (1,5) and (2,5) .. (2,6);
	\draw[black, line width = 1pt] (0,3) -- (0,6);
	\draw[black, line width = 1pt] (3,3) -- (3,6);
	\draw[black, line width = 1pt] (2,6) .. controls (2,7) and (3,7) .. (3,6);
	\draw[black, line width = 1pt] (2,9) .. controls (2,8) and (3,8) .. (3,9); 
	\draw[black, line width = 1pt] (0,6) -- (0,9);
	\draw[black, line width = 1pt] (1,6) -- (1,9);
\end{tikzpicture}
\quad = \quad
\begin{tikzpicture}[scale=1/4, baseline = {(current bounding box.center)},yscale=-1]
	\draw[black, line width = 1pt] (0,0) .. controls (0,1) and (1,1) .. (1,0);
	\draw[black, line width = 1pt] (2,3) .. controls (2,2) and (3,2) .. (3,3);
	\draw[black, line width = 1pt] (2,0) .. controls (2,1) and (0,2) .. (0,3);
	\draw[black, line width = 1pt] (3,0) .. controls (3,1) and (1,2) .. (1,3);
\end{tikzpicture}
\quad = \quad 
\begin{tikzpicture}[scale=1/4, baseline = {(current bounding box.center)},yscale=-1]
	\foreach \r in {0,...,4}{
		\draw[black, line width = 1pt] (\r,0) .. controls (\r,1) and (\r-1,2) .. (\r-1,3);
		\draw[black, line width = 1pt] (\r,3) .. controls (\r,4) and (\r-1,5) .. (\r-1,6);
	};
	\filldraw[white] (-.5,0) rectangle (-1.5,6);
	\filldraw[white] (3.5,0) rectangle (4.5,6);
	\draw[black, line width = 1pt] (2,6) .. controls (2,7) and (3,7) .. (3,6);
	\draw[black, line width = 1pt] (2,9) .. controls (2,8) and (3,8) .. (3,9);
	\draw[black, line width = 1pt] (0,6) -- (0,9);
	\draw[black, line width = 1pt] (1,6) -- (1,9);
\end{tikzpicture} 
\quad =
u^{2}e_{3}.
\end{equation*}
Note also  that $u^{N}$ is central: it commutes with all the generators $e_i$.
The resulting  algebra is called the \textit{affine Temperley-Lieb} algebra $\ATL{N}(n)$.
In the following we shall draw extensively on known results about its representation theory \cite{MartinSaleur,GrahamLehrer}
and its relation with conformal field theory \cite{GRSV1}.

\subsection{The transfer matrix sectors}

While  $\ATL{N}(n)$ is infinite-dimensional, it is easy to define the finite-dimensional modules which are relevant to us.
First, we fix the number of {\em through-lines}, which are the pieces of loops connecting the bottom and the top of the diagrams.
The number of through-lines is denoted $2j$, with $j \in \mathbb{N}/2$---the factor of $2$ comes about because one can relate
each loop strand to a $\mathfrak{q}$-deformed representation of spin-$1/2$.
Second, we stipulate that whenever $2j$ through-lines wind counterclockwise around
the axis of the cylinder $l$ times, we can unwind them at the price of a complex phase factor
${\rm e}^{2ijlK}$; similarly, for clockwise winding, the phase will be ${\rm e}^{-i 2jlK}$ \cite{MartinSaleur}. This unwinding means more precisely that we equate the words in the algebra corresponding to the winding configurations with a numerical factor (the phase) times the related words without winding. This operation is known to give rise to a generically
irreducible module over $\ATL{N}(n)$, which we denote by
$\AStTL{j}{z^2=\mathrm{e}^{2 iK}}$ and call the
\textit{standard module} \cite{GrahamLehrer}.
A key point is that inside the modules $\AStTL{j}{z^2}$ one has the identity
\begin{equation}
u^{N}=z^{2j} \,,
\end{equation}
meaning that the central element $u^N$ of $\ATL{N}(n)$ is replaced by the complex number $z^{2j}$.

The dimensions of the standard modules $\AStTL{j}{\mathrm{e}^{2 iK}}$   are  given by
%
\begin{equation}\label{eq:dj}
 \hat d_{j}=
 \binom{N}{N/2+j},\qquad j>0\ .
\end{equation}
Note that the dimensions do not depend on $K$, although the representations with
different ${\rm e}^{iK}$ are not isomorphic.

The case $j=0$ is a bit special, due to the absence of through-lines. There is no pseudomomentum, but representations are characterised
by another parameter, related with the weight given to {\em non-contractible} loops (i.e., loops that close around the periodic direction).
Parameterising this weight as $z+z^{-1}$, the corresponding standard module of  $\ATL{N}(n)$ is denoted  $\AStTL{0}{z^2}$ and has  dimension $\binom{N}{N/2}$. 
 These modules are irreducible  for generic $z$. 
  As in the case $j>0$, we indicate only the $z^2$ value, though it does not mean that the two standard modules with $\pm z$ are isomorphic. We will indicate the sign of $z$ when it is necessary.

\subsection{Potts model}
\label{sec:Potts-model}

The fact that we wish to apply $\ATL{N}(n)$ to study the Potts models entails a few minor modifications of the general setup.
First, since the number of sites $N=2L$ is even,
the number of through-lines is also even, so that $j \in \mathbb{N}$. Second, the translation operator in the Potts model shifts the $L$ spins cyclically,
meaning that the TL sites must be shifted by two units. Therefore, we are actually going to use the subalgebra in $\ATL{N}$ generated by the $e_i$'s
and by $u^2$ (instead of $u$ itself), which is why the above notation refers to $z^2$ rather than to $z$ itself. 

The basic ingredient is the finite-dimensional modules of $\ATL{N}$ described above, with the loop weight $n = \sqrt{Q}$ parameterised 
as in (\ref{loop-fugacity}) in order to match (\ref{Z_loop}). However, to account for the particularities of the Potts model, the
algebra that we are  mostly interested  in is a quotient of $\ATL{N}$, known as the Jones-Temperley-Lieb algebra $\rJTL{N}(n)$
 \cite{Jones,ReadSaleur01}. It is obtained by:
\begin{itemize}
 \item[(i)]  replacing non-contractible loops by the same weight $n=\sqrt{Q}$ as for the contractible ones;
 \item [(ii)] identifying diagrams connecting the same sites, even if they are non-isotopic on the cylinder; and
 \item[(iii)] setting $u^N=1$, which allows one to unwind through-lines.
\end{itemize}

Rules (i) and (ii) are only relevant in the case $j=0$, where through-lines are not present.
The first rule leads to $z^2=\q^{\pm 2}$.  In this case, in fact, the affine TL module $\AStTL{0}{\q^2}$ is reducible,
and contains a unique simple submodule isomorphic to $\AStTL{1}{1}$. 
The reason for this is that the general $\ATL{N}(n)$ allows (in diagrammatic terms) to distinguish loop arcs that connect two given sites
on the front or on the back of the cylinder, meaning that a closed loop can be given different weights depending on whether it is 
contractible or not. When this distinction is not needed we must  identify arcs only according to which sites are being
connected, as in rule (ii). Identifying non-isotopic diagrams in this way corresponds to replacing $\AStTL{0}{\q^2}$ by
the (unique) simple quotient $\AStTL{0}{\q^2}/\AStTL{1}{1}\equiv \bAStTL{0}{\q^2}$ of dimension 
\begin{equation}
\overline{d}_0 = \dim \bAStTL{0}{\q^2}= \binom{N}{N/2} - \binom{N}{N/2+1}\ ,
\end{equation}
In technical terms, this quotient is precisely the standard module of $\rJTL{N}(n)$ for $j=0$.

\begin{remark}
The quotient $\bAStTL{0}{\q^2}$ is but one example of representations that appear more generally in $\ATL{N}$ when $\q$
is still generic, but $z$ takes particular values \cite{GrahamLehrer,PerioFusion}.
Indeed, the standard module $\AStTL{j'}{z'}$ has a non-zero homomorphism to $\AStTL{j}{z}$,
\begin{equation}\label{cell-emb}
\StJTL{j'}{z'}\hookrightarrow\StJTL{j}{z} \,,
\end{equation}
if and only if  $j'-j \in \mathbb{N}_0$ and the pairs $(j',z')$ and $(j,z)$ satisfy
\begin{equation}\label{eq:emb-cond}
(z')^2 = (-\q)^{2\epsilon j} \qquad \text{and}\qquad z^2 = (-\q)^{\epsilon 2j'},\qquad \text{for} \quad  \epsilon=\pm1.
\end{equation}
When $\q$ is not a root of unity, there is at most one solution to (\ref{cell-emb}). When there is one, the module $\StJTL{j}{z}$ is not
irreducible, but has a unique proper irreducible submodule isomorphic to $\StJTL{j'}{z'}$. One can then obtain a simple module by taking the quotient
\begin{equation}
\bAStTL{j}{z}\equiv\StJTL{j}{z}/\StJTL{j'}{z'}\label{Wbar-def}
\end{equation}
of dimension
\begin{equation}
\overline{d}_j = \dim \bAStTL{j}{\q^2} = \binom{N}{N/2+j} - \binom{N}{N/2+j'} \,.
\end{equation}

The quotient $\bAStTL{0}{\q^2}$ appearing above is but the simplest example (with $j=0$, $z=\q^2$ and $j'=1$, $z'=1$) of this situation, and it is the {\em only}
such quotient that is relevant for the Potts model at generic $\q$.
\end{remark}

Whenver $j\neq 0$, rule (iii) leads to a quantisation of the momentum:
$K=\pi p/M$, with $M$ a divisor of $j$ (i.e., $M | j$) and with a greatest common divisor $p \wedge M =1$.
The modules encountered so far are thus $ \bAStTL{0}{\q^2}=\bAStTL{0}{\q^{-2}}$, and $\AStTL{j}{\mathrm{e}^{2 i\pi p/M}}$ for $j\neq 0$ and $M|j$. 

On top of this, there is a small subtlety having to do with the relation between through-lines and through-clusters, by which we mean clusters that
propagate along the imaginary time direction. For $j \ge 2$, each of the $2j$ through-lines alternatingly separates a propagating FK cluster and an propagating
dual cluster, implying the existence of precisely $j$ through-clusters.

However, when we wish to impose just one through-cluster, the situation is different.
Since nothing prevents this cluster from wrapping the periodic direction of the cylinder, it will in fact do so with probability one,
implying that through-lines will be absent ($j=0$). On the other hand, there cannot be any non-contractible loops either, since
this would prevent the propagation of the through-cluster. The correct module is thus obtained by giving a vanishing weight
to non-contractible loops. This is easily accomplished by setting $z=\pm i$, leading to $\AStTL{0}{-1}$. 
 
The three types of modules that we have just introduced: 
%
\begin{equation}
 \bAStTL{0}{\q^{\pm 2}}\oplus~\AStTL{j}{\mathrm{e}^{2 i\pi p/M}}{\scriptstyle (M|j,j\ge 2)}\oplus~\AStTL{0}{-1}\label{modules}
\end{equation}
are known to encode the full Potts model partition function on the torus \cite{DFSZ,ReadSaleur01,RichardJacobsen}.
Formal multiplicities for these modules are also known. For $Q$ non-integer, they are real numbers with group-theoretical
significance \cite{tensor2,tensor3}.

The crucial observation that will be made below is that
the modules (\ref{modules}) are also the sufficient objects to describe the four-point correlation functions in the geometrical Potts model. 
An important additional fact is that actually only the modules with {\em even values} of $j$ are necessary for the description of
four-point functions. By contrast, any $j \ge 2$ contributes\footnote{Note that the value $j=1$ is perfectly allowed in algebraic terms, and is crucial for  the description of the statistics of cluster hulls. It does not, however, appear in the torus partition function of the Potts model, nor in the connectivity correlations functions in the bulk.}  to the partition function on the torus, as has been verified in details for finite
systems \cite{JacobsenSalas-torus}.

This last result was not totally obvious a priori. Indeed, one must in general be careful with geometrical questions in models
such as the Potts model, where the set of observables is seemingly not limited, if one sways far enough from locality.
It is well-known, for instance, that correlations involving several independent paths along clusters---the case of two such paths defines
the celebrated backbone exponents---cannot be described using $\ATL{N}$, and the corresponding exponents have never been identified
using Coulomb gas techniques. Indeed, the numerical measurements of \cite{JacobsenZinn1,JacobsenZinn2,DengBloteNienhuis} appear to convincingly
rule out any such identification for this whole class of so-called monochromatic path-crossing exponents.
Similar remarks can be made about other seemingly reasonable observables, such as the shortest-path exponent \cite{ZhouYangDengZiff},
to mention but one example.
Fortunately, then, the matters seem to be (relatively) simpler for the four-point correlation functions.

\subsection{Summary of notations}

For the reader's convenience we summarise here some of the notations used in this paper. They are, as far as possible, the same as those used in  \cite{GRS1,GRS2,GRS3,GJSV,GJRSV,BGJSV,GRSV1,PerioFusion}.

\begin{itemize}

\item$\AStTL{j}{z^2=e^{2iK}}$ --- the standard modules over $\ATL{N}$,

\item$\StJTL{j}{z^2}$ --- the same, with $P=e^{2iK}$,

\item$\bAStTL{0}{\q^2}$ --- the standard module over $\rJTL{N}$ for $j=0$,

\item $\left[j, e^{2iK}\right]$ or $\IrrJTL{j}{z^2}$  --- simple modules over $\ATL{N}(n)$.

\end{itemize}

Moreover, when discussing the transfer matrices for the Potts model correlation functions (see Appendix~\ref{sec:appA}, and
section~\ref{sec:momentum_sectors} in particular)
we shall sometimes need a lighter notation $V_{\ell,k,m}$ for the sector with $\ell$ propagating clusters, an integer momentum variable
$k=0,1,\ldots,\ell-1$ for the through-lines (if any), and a lattice momentum $m=0,1,\ldots,L-1$ which is the precursor of the conformal
spin for a system of finite size $\ell$. The notations $V_{\ell,k}$ or $V_{\ell}$ with some of the variables omitted mean that these take
indiscriminate values. The correspondence with the standard modules is then:

\begin{itemize}

\item $V_0$ is the sector with no through-lines, and non-contractible loops have weight $\sqrt{Q}$: $V_0=\bAStTL{0}{z^2=\q^2}$,

\item $V_1$ is the sector with no through-lines, and non-contractible loops have weight zero: $V_1=\AStTL{0}{z^2=-1}$,

\item $V_{\ell,k}$ is the sector with $j=\ell \ge 2$ {\em pairs} of through-lines and phases $z^2=e^{2i\pi k/j}$: $V_{\ell,k}=\AStTL{j}{z^2=e^{2i\pi k/j}}$. 

\end{itemize}

\section{Four-point functions in the $s$-channel}
\label{sec:four-point}

\subsection{Generalities}
\label{sec:generalities}

We consider a general four-point function of primary operators in a CFT, which we write in the following convenient form in the plane:
\begin{equation}
\langle\Phi_1(z_1,\bar{z}_1)\Phi_2(z_2,\bar{z}_2)\Phi_3(z_3,\bar{z}_3)\Phi_4(z_4,\bar{z}_4)\rangle=\prod_{i<j} z_{ij}^{\delta_{ij}} \bar{z}_{ij}^{\bar{\delta}_{ij}} \G(z,\bar{z}) \,, \label{GenFptf}
\end{equation}
where we have denoted $z_{ij} \equiv z_i - z_j$, with the exponents
\begin{subequations}
\begin{eqnarray}
\delta_{12} &=& 0 \,, \\
\delta_{13} &=& -2h_1 \,, \\
\delta_{14} &=& 0 \,, \\
\delta_{23} &=& h_1-h_2-h_3+h_4 \,, \\
\delta_{24} &=& -h_1-h_2+h_3-h_4 \,, \\
\delta_{34} &=& h_1+h_2-h_3-h_4 \,.
\end{eqnarray}
\end{subequations}
The antiholomorphic exponents $\bar{\delta}_{ij}$ are obtained from the holomorphic ones $\delta_{ij}$ by the replacement $h\to\bar{h}$, and the same convention
henceforth applies to any other quantity. The parameter $z$ denotes the anharmonic ratio
\begin{equation}
z\equiv {z_{12}z_{34}\over z_{13}z_{24}} \,.
\end{equation}
One finds easily that
\begin{equation}
\hbox{Lim}_{\Lambda\to\infty} \Lambda^{2h_3}\bar{\Lambda}^{2\bar{h}_3} \langle 
\Phi_1(z,\bar{z})\Phi_2(0,0)\Phi_3(\Lambda,\bar{\Lambda})\Phi_4(1,1)\rangle=\G(z,\bar{z}) \,. \label{z01inf}
\end{equation}
This function $G(z,\bar{z})$ is what one usually refers to as $ \langle 
\Phi_1(z,\bar{z})\Phi_2(0,0)\Phi_3(\infty,\infty)\Phi_4(1,1)\rangle$. It is known \cite{BPZ} to expand as a sum over conformal blocks
\begin{equation}
\G(z,\bar{z})=\sum_{\Delta,\bar{\Delta}\in {\cal S}} C_{\Phi_1\Phi_2\Phi_{\Delta\bar{\Delta}}}C_{\Phi_{\Delta\bar{\Delta}}\Phi_3\Phi_4}{\cal F}_\Delta^{(s)}(z)\overline{{\cal F}}_{\bar{\Delta}}^{(s)}(\bar{z}) \,,
\end{equation}
where $(\Delta,\bar{\Delta})$ are the conformal weights of the primary fields appearing in the operator product expansion relevant at small $z$.
They define the scaling dimension (eigenvalue of the dilatation operator) $\Delta + \bar{\Delta}$ and the conformal spin (eigenvalue of the rotation
operator) $\Delta - \bar{\Delta}$.

We shall be particularly interested in the limit $z_1\to z_2$ and $z_3\to z_4$: this is called the $s$-channel (borrowing a standard terminology
of particle physics due to Mandelstam). This limit corresponds to taking $z \to 0$ in (\ref{z01inf}), so we can write the expansion
\begin{equation}
\G(z,\bar{z})=\sum_{\Delta,\bar{\Delta}\in {\cal S}} C_{\Phi_1\Phi_2\Phi_{\Delta\bar{\Delta}}}C_{\Phi_{\Delta\bar{\Delta}}\Phi_3\Phi_4}z^{(\Delta-h_1-h_2)}\bar{z}^{\bar{\Delta}-\bar{h}_1-\bar{h}_2}\left[1+O(z,\bar{z})\right] \,. \label{leadingexpan}
\end{equation}
One could of course similarly consider the $t$-channel ($z \to \infty$) and in the $u$-channel ($z \to 1$), by expanding in powers of $\frac{1}{z}$ and
$(z-1)$ respectively. The idea of the  conformal bootstrap programme is that all these expansions determine the same function $\G(z,\bar{z})$ and
hence will impose constraints. The first step in any further discussion is therefore to establish the fields intervening in one of these channels, which
we take here to be the $s$-channel.

The key question we want to address in this paper is thus to determine the set ${\cal S}$ of values of $\Delta,\bar{\Delta}$, which we will tackle
by a combination of algebraic and numerical methods. In its crudest form, the latter involves the brute force numerical determination of a (very) large number of terms appearing in the right-hand side of (\ref{leadingexpan}).

Note that the determination of the set ${\cal S}$  from the knowledge of these terms will only be fully possible in  ``generic'' cases, where none of the $\Delta,\bar{\Delta}$ differ by integers. Otherwise, there will be ambiguities, as a term such as $\Delta+n,\bar{\Delta}+\bar{n}$ (with $n,\bar{n}$ integer) may arise from a genuine primary field, or from a Virasoro descendent of some primary field with weights $\Delta+p,\bar{\Delta}+\bar{p}$, with $p \le n$ and $\bar{p} \le \bar{n}$ (with at least one of the
inequalities being strict). The non-generic case requires $Q$ to take particular values (with $\mathfrak{q}$ being a root of unity); it is clearly more complicated than
the generic case and will typically lead to at least some correlation functions having logarithmic behaviour. A few non-generic cases (not all of them logarithmic)
will be discussed in Appendix~\ref{sec:appB}. But the main text is henceforth dedicated to the generic case, for which we shall determine ${\cal S}$ fully.

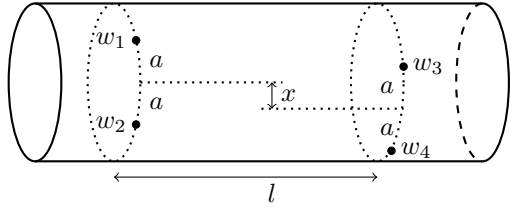
\begin{figure}
\begin{center}
\begin{tikzpicture}[scale=0.7]
 \draw[black,thick] (0,0) arc (0:360:0.5 and 1.5);
 \draw[black,thick] (-0.5,1.5)--(8,1.5);
 \draw[black,thick] (-0.5,-1.5)--(8,-1.5);
 \draw[black,thick,dotted] (1.5,0)--(4.2,0);
 \draw[black,thick,dotted] (1.5,0) arc (0:360:0.5 and 1.5);
 \draw[black,fill] (1.42,0.8) circle (2pt);
 \draw[black,fill] (1.42,-0.8) circle (2pt);
 \draw (1.42,0.8) node[left] {$w_1$};
 \draw (1.42,-0.8) node[left] {$w_2$};
 \draw (1.5,0.4) node[right] {$a$};
 \draw (1.5,-0.4) node[right] {$a$};
 \draw[black,thick,dotted] (6.5,0) arc (0:360:0.5 and 1.5);
 \draw[black,thick] (8.0,-1.5) arc (-90:90:0.5 and 1.5);
 \draw[black,thick,dashed] (8.0,1.5) arc (90:270:0.5 and 1.5);
 \draw[black,thick,dotted] (3.8,-0.5)--(6.5,-0.5);
 \draw[black,fill] (6.50,0.3) circle (2pt);
 \draw[black,fill] (6.27,-1.3) circle (2pt);
 \draw (6.50,0.3) node[right] {$w_3$};
 \draw (6.27,-1.3) node[right] {$w_4$};
 \draw (6.50,-0.1) node[left] {$a$};
 \draw (6.50,-0.9) node[left] {$a$};
 \draw[black,thin,<->] (4,0)--(4,-0.5);
 \draw (4,-0.25) node[right] {$x$};
 \draw[black,thin,<->] (1.0,-1.8)--(6.0,-1.8);
 \draw (4,-2.1) node {$l$};

\end{tikzpicture}
\end{center}
\caption{Four-point functions in the cylinder geometry.}
\label{fig:cylinder}
\end{figure}

Our strategy is to study the expansion (\ref{leadingexpan}) on the cylinder, where we will be able to use, on the numerical side, transfer matrix techniques, and, on the analytic side, algebraic results. The four-point function on the cylinder follows from  (\ref{GenFptf}) via the conformal map
\begin{equation}
 w={L\over 2\pi}\ln z \,.
\end{equation}
Using the fact that the fields are primary, and restricting here to  $i=j=k=l$ for notational simplicity,  we find
\begin{equation}
\langle \Phi(w_1,\bar{w}_1)\Phi(w_2,\bar{w}_2)\Phi(w_3,\bar{w}_3)\Phi(w_4,\bar{w}_4)\rangle_{\rm cyl}=\left({2\pi\over L}\right)^{4(h+\bar{h})} {1\over \left|4\sinh{\pi w_{13}\over L}\sinh{\pi w_{24}\over L}\right|^{2(h+\bar{h})} }~\G(w,\bar{w}) \,,
\end{equation}
where the subscript ``cyl'' on the left-hand side refers to the cylinder geometry, and we have set $w_{ij} \equiv w_i - w_j$ as before. The
expansion variable is now
\begin{equation}
w={\sinh{\pi w_{12}\over L}\sinh{\pi w_{34}\over L}\over \sinh{\pi w_{13}\over L}\sinh{\pi w_{24}\over L}} \,.
\end{equation}
Using (\ref{leadingexpan}) we can write this as 
\begin{eqnarray}
\langle \Phi(w_1,\bar{w}_1)\Phi(w_2,\bar{w}_2)\Phi(w_3,\bar{w}_3)\Phi(w_4,\bar{w}_4)\rangle_{\rm cyl}=\left({2\pi\over L}\right)^{4(h+\bar{h})}{1\over |4\sinh{\pi w_{12}\over L}\sinh{\pi w_{34}\over L}|^{2(h+\bar{h})}}\nonumber\\
\sum_{\Delta,\bar{\Delta}\in {\cal S}} C_{\Phi\Phi\Phi_{\Delta\bar{\Delta}}}C_{\Phi_{\Delta\bar{\Delta}}\Phi\Phi}\left[\left({\sinh{\pi w_{12}\over L}\sinh{\pi w_{34}\over L}\over 
\sinh{\pi w_{13}\over L}\sinh{\pi w_{24}\over L}}\right)^{\Delta}\left({\sinh{\pi \bar{w}_{12}\over L}\sinh{\pi \bar{w}_{34}\over L}\over 
\sinh{\pi \bar{w}_{13}\over L}\sinh{\pi \bar{w}_{24}\over L}}\right)^{\bar{\Delta}}+O(w,\bar{w})\right]\label{CylExpan}
\end{eqnarray}
In practice, to access the $s$-channel properties, we will take the points $w_1,w_2$ on a given slice of imaginary time, and $w_3,w_4$ on another, distant, slice along the cylinder of finite circumference $L$. This geometry is shown in Figure~\ref{fig:cylinder}.
In other words, $w_{12}$ and $w_{34}$ will be fixed, while $w_{13}$ and $w_{24}$ will be large and vary. In this limit, it will then be possible to compare the expansion (\ref{CylExpan}) with the results of transfer matrix calculations, and identify, in particular, the set ${\cal S}$. 

Let us now be more precise. We set 
\begin{subequations}
\label{eq:w1w2w3w4}
\begin{eqnarray}
w_1 &=& ia \,, \\
w_2 &=& -ia \,, \\
w_3 &=& i(a+x)+l \,, \\
w_4 &=& i(-a+x)+l \,,
\end{eqnarray}
\end{subequations}
which means that the points  $w_{1,2}$ and $w_{3,4}$ are a certain distance%
\footnote{Note that $a$ is not the lattice spacing, but some arbitrary parameter. It will always occur in the combination $\frac{2a}{L}$.}
$2a$ apart on the vertical axis, $l$  is the  horizontal distance (imaginary time) between the two groups, and on top of this we have the centre of mass of $w_{3,4}$ shifted by $x$. A short calculation then gives
\begin{eqnarray}
\langle \Phi(w_1,\bar{w}_1)\Phi(w_2,\bar{w}_2)\Phi(w_3,\bar{w}_3)\Phi(w_4,\bar{w}_4)\rangle_{\rm cyl} &=& \left({2\pi\over L}\right)^{4(h+\bar{h})} e^{-8\pi h l/L} \label{CylExpan1} \\
&\times& \left(1-e^{-2\pi (l+ix)/L}\right)^{-4h}\left(1-e^{-2\pi (l-ix)/L}\right)^{-4h}
\nonumber \\
& \times& \G\left({4e^{i\pi}\sin^2 {2\pi a\over L} e^{-2\pi(l+ix)/L}\over (1-e^{-2\pi(l+ix)/L})^2},{4e^{-i\pi} \sin^2 {2\pi a\over L} e^{-2\pi(l-ix)/L}\over (1-e^{-2\pi(l-ix)/L})^2}\right) \,.
\nonumber
\end{eqnarray}
We can then expand this from (\ref{leadingexpan}):
\begin{eqnarray}
\langle \Phi(w_1,\bar{w}_1)\Phi(w_2,\bar{w}_2)\Phi(w_3,\bar{w}_3)\Phi(w_4,\bar{w}_4)\rangle_{\rm cyl} &=& \left({2\pi\over L}\right)^{4(h+\bar{h})} {1\over (4\sin^2 {2\pi a\over L})^{4h}} \sum_{\Delta,\bar{\Delta}\in {\cal S}} C_{\Phi\Phi\Phi_{\Delta\bar{\Delta}}}C_{\Phi_{\Delta\bar{\Delta}}\Phi\Phi} \nonumber\\
&\times& \left(4\sin^2{2\pi a\over L}\right)^{\Delta+\bar{\Delta}}  (-1)^{\Delta-\bar{\Delta}}\xi^\Delta\bar{\xi}^{\bar{\Delta}}[1+O(\xi,\bar{\xi})] \,,
\label{mainexp}
\end{eqnarray}
where we have set 
\begin{equation}
\xi\equiv e^{-2\pi (l+ix)/L},~\bar{\xi}\equiv e^{-2\pi (l-ix)/L} \,.
\end{equation}
The bracket $[1+O(\xi,\bar{\xi})]$ contains now contributions from the conformal blocks and contributions from the hyperbolic functions in the conformal map.

The expansion (\ref{mainexp}) is the crucial tool that we will use systematically in our analysis below.  In the following, we will sometimes use the short-hand notation 
\begin{equation}
A_{\Phi_{\Delta\bar{\Delta}}}\equiv C_{\Phi\Phi\Phi_{\Delta\bar{\Delta}}}C_{\Phi_{\Delta\bar{\Delta}}\Phi\Phi} \,. \label{ampldef}
\end{equation}
We now discuss this in more detail.

\begin{remark}

We also see  that if we exchange $w_1$ and $w_2$ in (\ref{CylExpan}), the leading contributions for a given $\Delta,\bar{\Delta}$ is multiplied  by $(-1)^{\Delta-\bar{\Delta}}$. Hence primary fields with odd integer spin should contribute an opposite weight upon making this exchange.

\end{remark}

%
%

For future reference, the definition of the channels is 
\begin{eqnarray}
\mbox{$s$-channel} &:& z_1\to z_2\nonumber\\
\mbox{$t$-channel}  &:& z_1\to z_4\nonumber\\
\mbox{$u$-channel} &:& z_1\to z_3
\end{eqnarray}
Henceforth, when denoting the probabilities $P_{a_1,a_2,a_3,a_4}$, the four labels specifying the partition ${\cal P}=\{a_1,a_2,a_3,a_4\}$
refer to the points $z_1,z_2,z_3,z_4$ in that order.
Clearly, then, $P_{aaaa}$ should have the same spectrum (and structure constants) in all channels, while, for instance,
$P_{aabb}$ should have the same spectrum (and structure constants) in the $t$- and $u$-channels,
while the spectrum should be different in the $s$-channel. 

\subsection{Exponents}

Contrarily to what is implied in \cite{Ribault}, the exponents of percolation---and more generally the $Q$-state Potts model in the FK cluster
representation---are essentially known (with the exception of certain ``exotic'' exponents, see 
\cite{JacobsenZinn1,JacobsenZinn2,DengBloteNienhuis,ZhouYangDengZiff}).
This knowledge relies on two stages. First, the transfer matrix sectors of the Potts model can be
described in terms of standard modules of the affine TL algebra, as described in section~\ref{sec:Potts-model}.
Second, the continuum limits of these objects are known in the form of spectrum generating functions within the corresponding CFT,
as we now review. This is of course not yet the solution of the $s$-channel conundrum, but since we are able to formulate the four-point functions
in terms of the FK transfer matrix (see section~\ref{sec:algorithm} and Appendix~\ref{sec:appA1}), the results on the generating functions will narrow down
the set of states than can possibly be part of the spectrum ${\cal S}$. Extensive numerical analysis---corroborated by the solvability of a few special cases
(see Appendix~\ref{sec:appB})---will then lead to the results that we give in section~\ref{sec:results}.

The local FK connectivities in the geometrical Potts model and their evolution along the cylinder are described by a transfer matrix or,
in the familiar extreme anisotropic limit, a Hamiltonian. Both transfer matrix and Hamiltonian exhibit the same conformal content---that is,
eigenstates associated with local CFT operators, and the corresponding conformal weights $h,\bar{h}$, together with their multiplicities.
It is convenient to encode the latter into spectrum generating functions. Using for instance the Hamiltonian language and setting 
\begin{equation}
H=-\lambda\sum_{i=1}^{2L} e_i \,,
\end{equation}
with $\lambda$ adjusted so that the sound velocity is unity as usual, we define the generating function of levels (eigenenergies of $H$)
and lattice momentum $P$ as as traces of lattice operators, with the scaling limit \cite{PS}
\begin{equation}
 \hbox{Tr}\, \left[e^{-\beta_R(H-N\varepsilon_0)}e^{-i\beta_I P}\right]\;\xrightarrow{\, \rm scaling}\; \hbox{Tr}\, q^{L_0-c/24}\bar{q}^{\bar{L}_0-c/24} \,. \label{charform}
 \end{equation}
Here $\varepsilon_0$  is the (non-universal) ground state energy per site in the  limit $N\to\infty$.
The scaling limit is defined by taking $N,\beta_R,\beta_I\to\infty$ while keeping the  modular parameters%
\footnote{Here and elsewhere a notation of the type $q(\bar{q})$ means that $q$ is given by the first expression on the right-hand side (the one with a $+$ sign),
and $\bar{q}$ by the second expression (with a $-$ sign).}
$q(\bar{q})=\exp\left[-{2\pi\over N}(\beta_R\pm i\beta_I)\right]$ (with $\beta_{R,I}$ real and $\beta_R>0$) finite. 
The parameters $\beta_R$ and $\beta_I$ define the size of the system in the two principal directions of the torus,
while the trace ensures the periodic boundary conditions in the imaginary time direction. 
We recall that $N=2L$ is the number of sites in the system, often referred to as the length of the spin chain in the
Hamiltonian limit. In other words, only even chains are relevant in our problem.
On the right-hand side of (\ref{charform}), $L_0$ and $\bar{L}_0$ are of course Virasoro generators, while $c$ denotes the central charge.

The generating function (\ref{charform}) calculated  in the modules $\aSt_{j,\mathrm{e}^{2 i K}}$ is \cite{AGR,PS}
\begin{equation}\label{eq_F}
\displaystyle  \mathrm{Tr}_{\aSt_{j,\mathrm{e}^{2iK}}}\left[e^{-\beta_R(H-N\varepsilon_0)}e^{-i\beta_I P}\right] \;\xrightarrow{\, \rm scaling\,}\; F_{j,\mathrm{e}^{2iK}} \equiv\frac{q^{-c/24}\bar{q}^{-c/24}}{P(q) P(\bar{q})} \sum_{e \in \mathbb{Z}} q^{h_{e+{K\over\pi},-j}} \bar{q}^{h_{e+{K\over\pi},j}},
\end{equation}
where
\begin{equation}
\displaystyle P(q) =  \prod_{n=1}^{\infty} (1 - q^n) = q^{-1/24} \eta (q)
\end{equation}
is the (inverse of) the generating function for integer partitions, and $\eta(q)$ is Dedekind's eta function.
Instead of (\ref{loop-fugacity}) we shall find it convenient to parameterise the number of states in the Potts model by
\begin{equation}
 \sqrt{Q}=2\cos \left( {\pi\over m+1} \right) \,, \mbox{ with } m \in [1,\infty] \,,
\end{equation}
so that $\mathfrak{q} = {\rm e}^{\frac{i \pi}{m+1}}$.
Note that to access the generic case ($\mathfrak{q}$ not a root of unity) we do {\em not} restrict $m$ to be integer, as would be the case for the minimal models. The corresponding central charge is then
\begin{equation}
c=1-{6\over m(m+1)} \,,
\end{equation}
and we also use the Kac table parameterisation of conformal weights
\begin{equation}
h_{rs}={[(m+1)r-ms]^2-1\over 4m(m+1)} \,.
\end{equation}
In ``usual'' CFT the labels $(r,s)$ are positive integers, but as for the parameter $m$ we shall here allow them to take more general values,
as is already evident from (\ref{eq_F}).
To make contact with standard references, it is also convenient in the following to introduce the {\em Coulomb gas coupling constant} $g={m\over m+1}$
and the {\em background electric charge} $e_0={1\over m+1}$.
The operator associated with the order parameter has conformal weight $h_{1/2,0}$ \cite{denNIJS,Nienhuis_CG}, the primordial example of an ``unusual''
$h_{rs}$, with $r$ here being non-integer and $s$ non-positive. This operator belongs to the generating function $F_{0,-1}$.

\begin{remark}
\label{rem:switch}
To compare with \cite{Ribault} one must set in their equation (1.1) $\beta^2={m\over m+1}$ (so that ${1\over 2}\leq \beta^2\leq 1$), and $q=Q$.
Moreover, the conventions used in their paper for the exponents are {\em switched} with respect to ours.
In other words, they call $\Delta_{sr}$ what we call $h_{rs}$ (or $\Delta_{rs}$). 
\end{remark}

Restricting now to the cases of interest with momentum $K=\pi p/M$ and $M|j$  gives 
\begin{equation}
\displaystyle F_{j,\mathrm{e}^{2i\pi p/M}} =\frac{q^{-c/24}\bar{q}^{-c/24}}{P(q) P(\bar{q})} \sum_{e \in \mathbb{Z}} q^{h_{e+{p\over M},-j}} \bar{q}^{h_{e+{p\over M},j}} \,,
 \mbox{ with } M|j \mbox{ and } j \in \mathbb{Z} \,. \label{bspec1}
\end{equation}
On top of this we also have to consider the generating function of levels in $\aSt_{0,-1}$,  which reads
\begin{equation}
\displaystyle F_{0,-1} =\frac{q^{-c/24}\bar{q}^{-c/24}}{P(q) P(\bar{q})} \sum_{e \in \mathbb{Z}} q^{h_{e+1/2,0}} \bar{q}^{h_{e+1/2,0}} \,. \label{bspec2}
\end{equation}
Finally we need the generating function for the quotient module $\bAStTL{0}{\q^2}$. The twist $e^{iK}=\q$ corresponds in our notation to ${K\over\pi}=e_0$, so we have first 
\begin{equation}
\displaystyle F_{0,\q^2} =\frac{q^{-c/24}\bar{q}^{-c/24}}{P(q) P(\bar{q})} \sum_{e \in \mathbb{Z}} q^{h_{e+e_0,0}} \bar{q}^{h_{e+e_0,0}} \,.
\end{equation}
The subtraction necessary to obtain the module $\bAStTL{0}{\q^2}=\AStTL{0}{\q^2}/\AStTL{1}{1}$ leads to the expression for the generating function of the corresponding levels:
\begin{equation}
\displaystyle \bar{F}_{0,\q^2} =\frac{q^{-c/24}\bar{q}^{-c/24}}{P(q) P(\bar{q})} \left[\sum_{e \in \mathbb{Z}} q^{h_{e+e_0,0}} \bar{q}^{h_{e+e_0,0}}-\sum_{e \in \mathbb{Z}} q^{h_{e,1}} \bar{q}^{h_{e,-1}}\right] \,.
\end{equation}
The subtraction can actually be implemented term by term. Introducing the characters of the so-called Kac modules---which are Verma modules where a single singular vector at level $r s$ has been removed---,
\begin{equation}
K_{rs}=q^{-c/24} {q^{h_{rs}}-q^{h_{r,-s}}\over P(q)}=q^{-c/24} q^{h_{rs}}{1-q^{rs}\over P(q)} \,,
\end{equation}
we have 
\begin{equation}
\bar{F}_{0,\q^2}=\sum_{r=1}^\infty K_{r1}(q)K_{r1}(\bar{q}) \,. \label{bspec3}
\end{equation}
To summarise, corresponding to the modules (\ref{modules}) we expect (and confirm below) that the set of exponents contributing to the four-point connectivities is encoded into
 \begin{equation}
 F_{j,\mathrm{e}^{2i\pi p/M}}{\scriptstyle( M|j,~j\geq 2)}\oplus~F_{0,-1}\oplus~\bar{F}_{0,\q^2} \,.
 \end{equation}
Moreover, we shall see below that only $j$ even contributes, which cannot be foreseen at this stage.
We note that for generic values of $Q$ or $\q$ (i.e, with $m$ irrational), there are no coincidences of exponents in the different sectors (generating functions)---this is also true on the lattice, where there are no coincidences of eigenvalues. Moreover, in a given sector, no two exponents differ by integers. 

\subsubsection{Numerical validation of the generating functions}

It appears useful at this stage to test the internal coherence of the ingredients brought together this far. On one hand, in section \ref{sec:Potts-model} we have
related the sectors of the Potts model transfer matrix $T$ to certain standard modules, $\aSt_{j,z^2}$ and $\bAStTL{0}{\q^2}$, of the affine TL algebra.
On the other hand, we have just given their corresponding spectrum generating functions, $F_{j,z}$ and $\bar{F}_{0,\q^2}$.
This means that a numerical diagonalisation of $T$ in the various sectors should produce---after a proper extrapolation to the continuum limit $L \to \infty$---the
primaries and descendents (with multiplicities) of these generating functions. We are not aware of a previous careful study that this is indeed so. 

To this end, we first outline in Appendix~\ref{sec:appA3} the extraction of the eigenvalue spectrum of $T$ in the various sectors relevant for the Potts model.
Fixing the values of the momentum and the conformal spin is a non-trivial operation that is expounded in Appendix~\ref{sec:momentum_sectors}.
Once this has been done, we can examine the spectrum of $T$; this is done first for a generic value of $Q$ in Appendix~\ref{sec:specQ12},
and then for a few non-generic values: $Q=4$ in Appendix~\ref{sec:specQ4}, and $Q=2$ in Appendix~\ref{sec:appA_Ising}. When combined, these three
cases permit us to test examples of Verma modules with zero, one, or infinitely many singular vectors.

In all cases we find that the agreement with the expected spectrum generating functions is perfect. In the generic case, we are able to see descendents up
to level 6 for the identity operator, and up to level 3 for other operators. Moreover, the set of primaries fully agree with the expectations from the affine
TL algebra. In the Ising case we are able to follow the first 29 scaling levels and observe descendents up to level 9 for both operators ($I$ and $\epsilon$)
in the even sector, with an agreement better than $10^{-4}$ for almost all scaling dimensions. Moreover, the degeneracy observed for each ``completed'' level
is in perfect accord with the spectrum generating functions.

The techniques used in the numerical analysis may be of independent interest and can be consulted in the appendices (see also
Appendix~\ref{sec:practical-remarks} for a few practical remarks).

\subsection{The numerical algorithm}
\label{sec:algorithm}

The geometrical setup for four-point functions is shown in Figure~\ref{fig:cylinder}. As stated earlier, our lattice discretisation consists in embedding
a periodic square lattice $G=(V,E)$ of width $L$ Potts spins into the cylinder, with the edges $E$ being either horizontal or vertical with respect to the
figure (axial geometry). We possess two different strategies for obtaining numerical results for the correlation functions.

The first strategy gives access
to the most general FK correlation functions, namely the 15 different $P_{a_1,a_2,a_3,a_4}$. It is practically feasible up to size $L=7$, after a
considerable numerical effort. It applies for both generic and non-generic values of $Q$, meaning that in the latter case it can determine the
indecomposable structure of correlation functions.

The second strategy applies to a smaller set of correlation functions, namely the four order-parameter correlators $G_{aaaa}$, $G_{aabb}$,
$G_{abba}$ and $G_{abab}$. Its advantage is that it gives access to larger sizes, in practice up to $L=11$, at a much smaller computational expenditure than
the first method. However, it applies only to generic values of $Q$ and, at least in its present form, cannot determine the Jordan block structure
at non-generic $Q$-values.

We now briefly outline the two methods, while relegating all technical details to Appendices~\ref{sec:appA1}--\ref{sec:appA2}.

\subsubsection{First method}
\label{sec:1st_method}

It is shown in the appendix that all $P_{a_1,a_2,a_3,a_4}$ can be computed, for fixed values of the distances $a$, $x$ and $l$,
via a suitable modification of the FK representation of the transfer matrix
in which certain clusters (viz., the ones touching one or more of the points $w_1,w_2,w_3,w_4$) carry specific marks. The spectrum of this modified
transfer matrix is contained within that of the original one, namely the one described in section~\ref{sec:Potts-model} in terms of affine TL representations.
The spectrum can be proven to be real, so we can order the {\em distinct} eigenvalues as $\Lambda_0 > \Lambda_1 > \cdots > \Lambda_i > \cdots$.
The correlation function then takes the following form, for generic values of $Q$,
\begin{equation}
 P_{a_1,a_2,a_3,a_4} = \sum_i A_i \left( \frac{\Lambda_i}{\Lambda_0} \right)^l \,, \label{PAeval}
\end{equation}
where the amplitudes $A_i = A_i(a,x,L)$ are to be determined.  The corresponding expression for non-generic values has the same form,
but with the replacement
\begin{equation}
 A_i \longrightarrow \sum_{j=0}^{r_i-1} A_i^{(j)} l^j \,, \label{generalised_amplitudes}
\end{equation}
whenever the eigenvalue $\Lambda_i$ is associated with a Jordan block of rank $r_i$. In the latter case the generalised 
amplitudes $A_i^{(j)} = A_i^{(j)}(a,x,L)$ are again independent of $l$.

\begin{remark}
\label{rem:degeneracy}

There is of course an exact degenerescence of the scaling dimension of a CFT operator with non-zero spin $\Delta-\bar{\Delta}$ and
that of its conjugate (i.e., obtained by the exchange $\xi \to \bar{\xi}$). This is prefigured in the lattice discretisation by the exact degenerescence
of the eigenvalues corresponding to eigenstates of $T$ with opposite non-zero lattice momenta $\pm m$ (cf.\ Appendix~\ref{sec:momentum_sectors}).
Because of the regrouping of degenerate eigenvalues in (\ref{PAeval}) it should thus be remembered to divide the amplitude $A_i$ of such states by a factor
of two when comparing to the CFT predictions (see Appendix~\ref{sec:appB} for many examples of this phenomenon).

\end{remark}

For finite $L$ the set of eigenvalues of $T$---and hence the number ${\cal N} \equiv \sum_i r_i$ of (generalised) amplitudes to be determined---is finite.
It follows that the form (\ref{PAeval}) is an exact expression, not merely an asymptotic expansion.
Therefore, to determine the amplitudes $A_i$---or the generalised amplitudes
$A_i^{(j)}$ for the cases with Jordan blocks---for given separations $a$, $x$ and size $L$, it suffices in principle to numerically determine
the spectrum $\{ \Lambda_i \}$, compute the correlation functions $P_{a_1,a_2,a_3,a_4}$ for ${\cal N}$ different values $l$,
and to invert the linear system (\ref{PAeval}).

In practice, of course, things are more complicated, and several remarks must be made (see Appendix~\ref{sec:practical-remarks}). The most important
of those is that the magnitude of the terms in (\ref{PAeval}) decreases exponentially fast, in particular when ${\cal N}$, and hence $l$, is
large. Therefore both $\{ \Lambda_i \}$ and $P_{a_1,a_2,a_3,a_4}$ must be computed to an exceedingly high numerical precision.
For instance, our most demanding computation (see section~\ref{sec:big-computation} for details) required a 4000-digit numerical precision.

Another remark is that when ${\cal N} \gg 1$ we often wish to determine only the first ``few'' amplitudes (corresponding to $i \le$ some $i_{\rm max}$).
This can be done by using the expression (\ref{PAeval}), truncated to the first $i_{\rm max}$ terms, as an asymptotic expression, i.e., by 
solving for $A_i$ the linear system provided by the numerically computed $P_{a_1,a_2,a_3,a_4}$
with $l = l_{\rm min}, l_{\rm min}+1,\ldots,l_{\rm min}+i_{\rm max}$, where $l_{\rm min}$ is taken sufficiently large.
One then has to carefully check that the desired $A_i$ are stable, within the desired numerical precision, to small changes in $l_{\rm min}$.

\subsubsection{Second method}
\label{sec:2nd_method}

Our other method applies to the computation of the order-parameter correlators $G_{a_1,a_2,a_3,a_4}$, defined in (\ref{order-param-corr}).
This requires another variant of the FK transfer matrix with marked clusters, as described in details in Appendix~\ref{sec:appA2}. The number
of different marks allowed must be chosen as the number of different symbols among $a_1,a_2,a_3,a_4$, and the dimension of the transfer
matrix grows with this number. In practice we have employed two different marks, in order to gain access to the four correlators
$G_{aaaa}$, $G_{aabb}$, $G_{abba}$ and $G_{abab}$.

The spin operator ${\cal O}_a(\sigma_k)$, defined in (\ref{spin-op}), can be expressed within
this basis and has essentially the effect of attributing the label $a$ to the spin situated at vertex $k \in V$.
Our geometrical setup is such that vertices $w_1$ and $w_2$ belong to the same time slice, while $w_3$ and $w_4$ belong to another time slice,
with the relative positions within these two time slices being specified by Figure~\ref{fig:cylinder}. We henceforth denote the spin operators
simply by ${\cal O}_{a_k}$, for $k=1,2,3,4$, and keep implicit their point of insertion in the relevant time slices.

Let $\langle v_i |$ and $| v_i \rangle$ denote the left and right eigenvectors of $T$. In the case of simple eigenvalues we then have
\begin{subequations}
\begin{eqnarray}
 \langle v_i | T &=& \Lambda_i \langle v_i | \,, \\
 T | v_i \rangle &=& \Lambda_i | v_i \rangle \,.
\end{eqnarray}
\end{subequations}
But even for the generic values of $Q$ that we consider here, some of the eigenvalues are degenerate, due to symmetries of the lattice and of the
order parameter symbols $a_k$. In that case we denote the multiplicity of $\Lambda_i$ by $d_i$, and we endow the 
corresponding eigenvectors with an extra label, $| v_{i,j} \rangle$,
where $j=1,2,\ldots,d_i$.

The left and right eigenvectors can be obtained efficiently within an iterative diagonalisation
scheme, such as the Arnoldi method (see again Appendix~\ref{sec:practical-remarks} for details).
Left and right vectors will obviously be orthogonal if they correspond to different eigenvalues, but the Arnoldi
method does not guarantee orthogonality within the degenerate subspaces. It is however
possible to perform an additional diagonalisation step that will ensure that the orthogonality holds with respect to both labels:
\begin{equation}
 \langle v_{i,j} | v_{i',j'} \rangle \propto \delta_{i,i'} \delta_{j,j'} \,. \label{orthogonality}
\end{equation}

\begin{remark}

Some readers are likely to be well acquainted with the representation theory of the TL algebra, in which
``geometrical'' scalar products (see e.g.\ \cite{JS_combi,dGJP}) are introduced between basis states (often called link patterns in the context
of the loop representation). These scalar products ``count'' loops and clusters formed by the gluing of the
states, leading to $Q$-dependent results. We must therefore stress that the scalar products appearing in this
section are simply the standard Euclidean scalar products between ordinary vectors.
Moreover, all eigenvectors turn out to be real, so there is no issue of complex conjugation.

\end{remark}

We now claim that the amplitudes $A_i$ for the generic case without Jordan blocks can be obtained as
\begin{equation}
 A_i = \sum_{j=1}^{d_i} \frac{ \langle v_0 | {\cal O}_{a_3} {\cal O}_{a_4} | v_{i,j} \rangle \, \langle v_{i,j} | {\cal O}_{a_1} {\cal O}_{a_2} | v_0 \rangle}
 {\langle v_0 | v_0 \rangle \, \langle v_{i,j} | v_{i,j} \rangle} \,. \label{scalar_prod_method}
\end{equation}
It is crucial for the validity of this result that the orthogonalisation in degenerate subspaces has been performed (see Appendix~\ref{sec:appA2}).

We have performed extensive checks that the first and second methods give the same results, in situations where they are both applicable.
The advantage of the second method is that it is numerically much more efficient, and hence gives access to larger sizes $L$.
The sources for this gain in efficiency are explained in Appendix~\ref{sec:appA2}.

Formula (\ref{scalar_prod_method}) has a nice geometrical interpretation that makes direct contact with Figure~\ref{fig:cylinder}.
Indeed the numerator of the formula (read from right to left) and the figure (read from left to right) are completely analogous:
\begin{enumerate}
 \item The propagation from the free boundary condition at imaginary time $t \to -\infty$ to the time slice containing $w_1$ and $w_2$ corresponds to the production of
 the ground state $| v_0 \rangle$.
 \item The first two operators are then inserted by ${\cal O}_{a_1} {\cal O}_{a_2}$.
 \item The piece $| v_{i,j} \rangle \, \langle v_{i,j} |$ corresponds to the projection on a definite state appearing in the $s$-channel of the four-point function.
 \item This is followed by the insertion of the two remaining operators, at $w_3$ and $w_4$.
 \item The projection on $\langle v_0 |$ matches the propagation to the free boundary condition at the other extreme of the cylinder ($t \to +\infty$).
\end{enumerate}
We stress that the validity of (\ref{scalar_prod_method}) depends crucially on the orthogonality (\ref{orthogonality}).

\subsubsection{Continuum limit}
\label{sec:num_cont_limit}

To extract the continuum limit ($L \to \infty$) of the amplitudes it is important to be able to associate each $A_i$ with a definite field in the continuum limit.%
\footnote{But see remark~\ref{rem:degeneracy} above.}
For instance, one question that one might want to answer is what would be the amplitude contributing to a given primary $\Phi_{\Delta \bar{\Delta}}$,
i.e., to identify the $A_i$ that will converge to the amplitude of $\Phi_{\Delta \bar{\Delta}}$ in the expansion
(\ref{mainexp}).

Several remarks are in order in this respect. 
Obviously, it is the ratio between two amplitudes---rather than each amplitude taken individually---that is universal and hence related to CFT.
Therefore we assume tacitly in what follows that each amplitude of interest is measured via its ratio to the one that gives the leading
contribution to the considered correlation function. Moreover, the conformal mapping to the cylinder implies that we should correct the
raw lattice result by a conformal factor, namely the powers of $\sin \frac{2 \pi a}{L}$ appearing in (\ref{mainexp}).
Once this has been done, our main claim is that we are capable of analysing the numerical results so as to establish the convergence
\begin{equation}
 A_i \to A_{\Phi_{\Delta \bar{\Delta}}} \mbox{ as } L \to \infty \,,
\end{equation}
where the conformal amplitude has been defined in (\ref{ampldef}). More generally, we can obtain the corresponding results for subdominant
contributions from the conformal blocks, corresponding to the amplitude multiplying a term of the type $\xi^{\cal N} \bar{\xi}^{\bar{\cal N}}$ in
the square bracket of (\ref{mainexp}). This is interpreted as the (total) amplitude of the descendents at level ${\cal N},\bar{\cal N}$ of the
primary $\Phi_{\Delta \bar{\Delta}}$. The challenge involved in making this identification is to make sure that we possess enough information
about the lattice model to unambiguously associate a given field in the continuum limit with its ``corresponding'' eigenvalue of the transfer
matrix in finite size $L$.

In the lattice model, each $A_i$ is unambiguously associated with the eigenvalue $\Lambda_i$. A careful study of the transfer matrix
(see Appendix~\ref{sec:momentum_sectors}) enables us to attribute to each eigenvalue three labels $\ell,k,m$, formally restricting to
a representation denoted $V_{\ell,k,m}$. The first label $\ell$ gives the number of 
propagating FK clusters, so in the notation of the standard modules $\AStTL{j}{z^2}$ we have $j = \ell$ for all $\ell \neq 1$, while $\ell = 1$
corresponds to $j = 0$. The second label $k$ is directly related with the momentum of through-lines, via $z^2=e^{2i\pi k/j}$. And finally
the third label $m$ is the lattice momentum that gives directly the conformal spin, $h - \bar{h} = m$, at least if $L$ is large enough to
accommodate the desired spin.

While certainly very helpful, the three labels $\ell,k,m$ are not quite enough to determine which conformal block to associate with $A_i$,
nor at which level ${\cal N},\bar{\cal N}$. Roughly speaking, the trouble is that the $k$'th smallest scaling dimension in the continuum limit
does not necessarily correspond to the $k$'th largest eigenvalue of $T$ in finite size $L$. While this is certainly true for $L \gg 1$, there are
numerous crossovers in finite size, and these have to be monitored carefully in order to make to correct identifications. How we overcame
this delicate problem is described in Appendix~\ref{sec:specT_CFTlimit}.

Finally, once the finite-size approximation $A_i$ to a given CFT amplitude has been determined, for several different sizes $L$, 
the numerical value of the latter is determined by finite-size extrapolation techniques. This is again discussed in Appendix~\ref{sec:specT_CFTlimit}.
A large number of concrete applications of the entire method can be found in section~\ref{sec:results} and (with more details provided)
in Appendix~\ref{sec:appB}.

\section{Results}
\label{sec:results}

\subsection{Checks}

Our approach, being based on properties of the lattice model, requires a careful control of the continuum limit. There are several aspects to this. The most obvious one is that, since we are studying  four-point functions of a CFT, we should, ideally, have all ratios $|z_{ij}| \gg 1$ (where all distances are measured in units of  the lattice spacing). On the cylinder, we have chosen to take points far apart along the cylinder axis, but placed, pairwise, on identical imaginary time slices. Since the cylinder widths $L$ are limited for technical reasons, this means that  $|w_{12}|,|w_{34}|$ will be limited, in fact, to a few lattice spacings.%
\footnote{We could take the two points within each pair to reside on different time slices, of course,
but this would only allow us to get real parts of $w_{ij}$ bigger, with the imaginary parts similarly limited.} 


\medskip

We first observe that the dependence $A(x)$ of the amplitudes on the shift $x$ in the space-like direction (see Figure~\ref{fig:cylinder}) 
between the two groups of points is in fact trivial. Taking into account that all amplitudes have been normalised as ratios with respect
to the leading one, as well as remark~\ref{rem:degeneracy}, it is seen from (\ref{mainexp}) that
\begin{equation}
 A(x) = \cos \left( \frac{2 \pi s x}{L} \right) A(0) \,,
 \label{amp_dep_x}
\end{equation}
where $s = \Delta - \bar{\Delta}$ denotes the conformal spin (which coincides with the lattice momentum, $m = s$). 

An alternative means of deriving (\ref{amp_dep_x}) goes through the inspection of (\ref{scalar_prod_method}).
Imagine evaluating the first scalar product in the numerator in a geometry where the cylinder has been rotated by the amount $-x$.
This rotation will re-align the second pair of operators ${\cal O}_{a_3} {\cal O}_{a_4}$ with the first pair ${\cal O}_{a_1} {\cal O}_{a_2}$,
like in the computation of $A(0)$. The ground state $\langle v_0 |$ is obviously rotationally invariant, but an intermediate state
$| v_{i,j} \rangle$ of lattice momentum $m \neq 0$ is not, and will therefore pick up a corresponding phase factor under the rotation.
Summing this over the degenerate contributions $\pm m$ reproduces (\ref{amp_dep_x}).

We have checked numerically on explicit examples that (\ref{amp_dep_x}) holds true exactly in finite size. As a matter of fact, determining numerically
the dependence $A(x)$ is a convenient means of establishing the lattice momentum $m$ of a given state, complementary to the techniques
explained in Section~\ref{sec:momentum_sectors}.

\medskip

A maybe more subtle aspect is that the lattice observables are not in general pure scaling fields. This means that the conformal field whose four-point functions we want to study, is identified on the lattice as the Potts spin operator only up to additional corrections  (``higher (or excited) spin operators''), whose contributions become negligible only when all distances are once again much larger than the lattice spacing: put otherwise, measured four-point functions on the lattice are a mixture of four-point functions of pure scaling fields. 

For our approach to be useful, it is necessary to perform many tests in order to control these potential drawbacks. As discussed extensively in Appendix~\ref{sec:appB}, we have checked that, for the sizes we were able to access:

\begin{itemize}

\item{} The mixture of excited spin operators can be neglected (see Appendix~\ref{sec:appB1}); and

\item{} The values of the extrapolated amplitudes $A_{\Phi_{\Delta\bar{\Delta}}}$---see (\ref{ampldef})---as well as those of the first few (in practice, a handful)
 subdominant contributions to each conformal tower, extracted via the method outlined in section~\ref{sec:algorithm} are in fine agreement with
 their exact CFT values in three exactly solvable cases ($Q=0,2,4$), which are treated respectively in Appendices~\ref{sec:appB-Q0}, \ref{sec:appB-Q2} and
 \ref{sec:appB-Q4}. In the most favourable cases the relative deviations are as small as $10^{-4}$.

\end{itemize}

Moreover, even for operators higher in the spectrum ${\cal S}$ where the lattice determination of  amplitudes may not have fully converged,  our approach, combined with the algebraic understanding of transfer matrix sectors, indicates unambiguously which coupling constants will remain non-zero in the scaling limit, even if error bars on their extrapolated values are not negligible.

\subsection{The $s$-channel spectrum of $P_{abab}$ or $P_{abba}$:  $F_{j,\mathrm{e}^{2i\pi p/M}}{\scriptstyle( M|j,~j\geq 2)}$, $j$ even}

\begin{figure}
\definecolor{hazygray}{gray}{0.9}
\definecolor{darkergray}{gray}{0.83}
\begin{center}
\begin{tikzpicture}[scale=0.55]
 \draw[hazygray,fill,rotate=175,yshift=-63.0] (0,1.4) arc(0:360:4.0 and 0.3);
 \draw[hazygray,fill,rotate=175,yshift=-23.0] (-0.2,1.4) arc(0:360:3.6 and 0.3);
 \draw[black,thick] (0,0) arc (0:360:0.5 and 1.5);
 \draw[black,thick] (-0.5,1.5)--(8,1.5);
 \draw[black,thick] (-0.5,-1.5)--(8,-1.5);
 \draw[black,thick,dotted] (1.5,0) arc (0:360:0.5 and 1.5);
 \draw[black,fill] (1.42,0.8) circle (2pt);
 \draw[black,fill] (1.42,-0.8) circle (2pt);
 \draw (1.42,0.8) node[left] {$w_1$};
 \draw (1.42,-0.8) node[left] {$w_2$};
 \draw[black,thick,dotted] (6.5,0) arc (0:360:0.5 and 1.5);
 \draw[black,thick] (8.0,-1.5) arc (-90:90:0.5 and 1.5);
 \draw[black,thick,dashed] (8.0,1.5) arc (90:270:0.5 and 1.5);
 \draw[black,fill] (6.50,0.3) circle (2pt);
 \draw[black,fill] (6.27,-1.3) circle (2pt);
 \draw (6.50,0.3) node[right] {$w_3$};
 \draw (6.27,-1.3) node[right] {$w_4$};
 \draw (4,-1.5) node[below] {$P_{abab}$};
\end{tikzpicture}
\qquad \qquad
\begin{tikzpicture}[scale=0.55]
 \draw[darkergray,fill] (3,-1.5)--(5,1.5)--(4,1.5)--(2,-1.5)--cycle;
 \draw[hazygray,fill,rotate=159,yshift=-73.0] (0,1.4) arc(0:360:3.6 and 0.3);
 \draw[hazygray,fill] (1.3,-0.4) -- (3,-1.5) -- (2,-1.5) -- (1.3,-1.0) .. controls (1.1,-0.8) and (0.8,-0.6) .. (1.3,-0.4);
 \draw[hazygray,fill] (4,1.5)--(5,1.5)--(7,0.5) .. controls (7.2,0.4) and (6.7,0.0) .. (6.0,0.1) -- cycle;
 \draw[black,thick] (0,0) arc (0:360:0.5 and 1.5);
 \draw[black,thick] (-0.5,1.5)--(8,1.5);
 \draw[black,thick] (-0.5,-1.5)--(8,-1.5);
 \draw[black,thick,dotted] (1.5,0) arc (0:360:0.5 and 1.5);
 \draw[black,fill] (1.42,0.8) circle (2pt);
 \draw[black,fill] (1.42,-0.8) circle (2pt);
 \draw (1.42,0.8) node[left] {$w_1$};
 \draw (1.42,-0.8) node[left] {$w_2$};
 \draw[black,thick,dotted] (6.5,0) arc (0:360:0.5 and 1.5);
 \draw[black,thick] (8.0,-1.5) arc (-90:90:0.5 and 1.5);
 \draw[black,thick,dashed] (8.0,1.5) arc (90:270:0.5 and 1.5);
 \draw[black,fill] (6.50,0.3) circle (2pt);
 \draw[black,fill] (6.27,-1.3) circle (2pt);
 \draw (6.50,0.3) node[right] {$w_3$};
 \draw (6.27,-1.3) node[right] {$w_4$};
 \draw (4,-1.5) node[below] {$P_{abba}$};
\end{tikzpicture}
\end{center}
\caption{FK cluster configurations that contribute to the correlation functions $P_{abab}$ and $P_{abba}$.}
\label{fig:2clusters}
\end{figure}
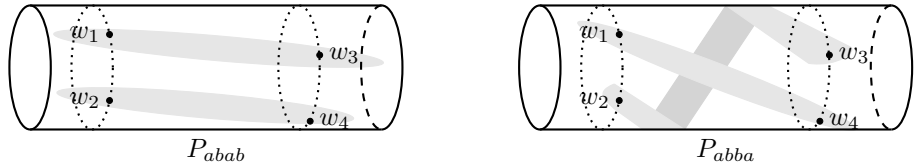

We first study the cases where (at least) two distinct clusters are forced to propagate between the two distinguished time slices in
Figure~\ref{fig:cylinder}. Specifically, $P_{abab}$
 is the probability that points $w_1,w_3$ and $w_2,w_4$
respectively belong to the same clusters. This quantity is also called $P_2$ in \cite{Ribault}. Similarly, $P_{abba}$ is the probability
that $w_1,w_4$ and $w_2,w_3$ belong to the same clusters. These two correlation functions are depicted in Figure~\ref{fig:2clusters}.

\subsubsection{Results in finite size}


It is evident from Figure~\ref{fig:2clusters} that the leading term in this sector should be the term corresponding to the propagation of two different clusters, that is, four cluster boundaries.
The affine TL modules $\AStTL{j}{z^2}$ (or their continuum counterparts $F_{j,z^2}$) correspond to $2j$ through-lines, so the propagation of two clusters
must have a contribution with $j=2$. The corresponding generating function of levels has two sectors, depending on whether a pair of boundaries going around the
system picks up a phase $z^2=1$ or $z^2=-1$. No other choice is possible since for two pairs of boundaries (picking up a phase $z^4$), we do not want a phase.%
\footnote{The reader might wonder why we have used the parameter $z$ since only $z^2$ seems to appear in the discussions. This is in part to conform with the literature, although $z$ itself may also have meaning for other questions in the Potts model. For instance, the single shift operator $u$ appplied to
the system amounts to performing a duality transformation.}
Hence we expect the contribution of modules
\begin{equation}
\AStTL{2}{1}\oplus\AStTL{2}{-1} \,. \label{2modcontr}
\end{equation}
We have first checked that {\em all} the eigenvalues associated with these two modules contribute to the probabilities $P_{abab}$ and $P_{abba}$ for all finite sizes. 

This is however not all. Sectors with a higher number of clusters than the two imposed by the choice of indices might also be thought to contribute to these
correlation functions. One could think of several mechanisms for such contributions. First, there might be more clusters, distinct from the two imposed by
the boundary conditions, that ``by chance'' connect the two time slices. Second, the two clusters might have more complicated topologies,
with for instance one of them (say, the one containing points $w_1$ and $w_3$ in the left part of Figure~\ref{fig:2clusters}) starting out at one insertion
point (here $w_1$), and wrapping all around the other cluster (containing points $w_4$ and $w_2$), before arriving at its terminal point (here $w_3$).

We have found it difficult to provide a convincing argument that certain subclasses of configurations will necessarily lead to further contributions
to the correlation functions, in terms of the modules $\AStTL{j}{z^2}$; we think there are underlying symmetry and branching rules considerations that may answer this riddle on general grounds, and that we do not yet control. 
Fortunately the numerical results are completely clear. We find that, for all finite sizes, all eigenvalues associated with the
modules $\AStTL{j}{\mathrm{e}^{2 i\pi p/M}},M|j$, with $j$ even, and only those, contribute to the probabilities $P_{abab}$ and $P_{abba}$.
For $j=4$ for instance, this allows contributions from the momentum sectors
$z^2=1,\exp(i\pi/4), \exp(i\pi/2),\exp(3i\pi/4)$,
and thus the following modules, in addition to those of (\ref{2modcontr}),
\begin{equation}
\AStTL{4}{1}\oplus \AStTL{4}{i}\oplus \AStTL{4}{-1}\oplus \AStTL{4}{-i} \,.
\end{equation}
Note that for a given width $L$, the maximum value of $j$ is bounded from above, $j \le L$. We have checked that, as $L$ increases, higher values of $j$ start contributing to the probabilities, provided that the separation $2a$ between the insertion points is sufficiently large. More precisely, we have observed that:
\begin{itemize}
 \item For $L=5$ and separation $2a=2$, all the eigenvalues with $j=2,4$ and none of the eigenvalues with $j=0,1,3$ contribute to the probabilities.
 \item Still for $L=5$, but diminishing to separation $2a=1$, the contributions from $j=4$ disappear.
 \item For $L=7$ and separation $2a=3$, the two probabilities get contributions from all
 eigenvalues with $j=2,4,6$ and none of the eigenvalues with $j=0,1,3,5$.
 \item 
 Still for $L=7$, but diminishing to separation $2a=2$, the contributions from $j=6$ disappear. Diminishing further to $2a=1$, the contributions from $j=4$ disappear as well.
\end{itemize}

The above result is corroborated by a closer study of the spectrum of the transfer matrix described in section~\ref{sec:2nd_method}, namely the one
that produces the correlation functions of order parameter operator. Its eigenvalues are precisely those corresponding to the modules
(\ref{modules}) with $j$ even, while those corresponding to $j$ odd are not observed at all.

Motivated by the above list of observations, we conjecture that in fact a given sector $j \in 2 \mathbb{N}$ contributes only when $2a \ge j/2$.

\subsubsection{Non-zero coupling to the sector $j=6$}
\label{sec:big-computation}

The computation with $L=7$ and $2a=3$ establishing that the sector $j=6$ does contribute to the correlation functions $P_{\cal P}$
is the most demanding among all of those
made for this paper. For the benefit of readers interested in computational aspects (and those wanting to scrupulously assess the validity
of our conclusions) we wish to describe it in some more detail---other readers may wish to skip this
section and resume the reading below.

This computation was performed for the generic value $Q=\frac32$. There is a total of
$3\,932$ distinct eigenvalues in the sectors with $j=0,2,4,6$, corresponding to all possible momenta. Using the methods of Appendix
\ref{sec:momentum_sectors} we can classify them in sectors $V_{\ell,k,m}$ corresponding to $\ell$ propagating clusters with cluster momentum $k$
and lattice momentum $m$. Ordering all the eigenvalues as
$\Lambda_1 > \Lambda_2 > \cdots > \Lambda_{3932}$, with $\Lambda_1$ being the dominant eigenvalue in $V_0$ (i.e., the ground state),
the dominant eigenvalues in the sectors $V_1$, $V_2$, $V_4$ and $V_6$---which obviously have vanishing momenta, $k=m=0$---are
respectively $\Lambda_2$, $\Lambda_5$, $\Lambda_{205}$ and $\Lambda_{2390}$.

Suppose first that we considered some correlation function coupling to all of these eigenvalues, and we wished to isolate the amplitude
corresponding to $\Lambda_{2390}$ by using the first method of Appendix~\ref{sec:appA1}. The ratio
$r = \Lambda_{2390}/\Lambda_1 \simeq 3.451 \cdot 10^{-4}$, and since we need to determine 2\,390 coefficients $A_i$ in (\ref{PAeval})
we will need the same number of equations, obtained by choosing the distance $l = l_{\rm min}+1,\ldots,l_{\rm min}+2390$.
We would need (at the very least) to take $l_{\rm min} = 100$ in order to be in the asymptotic regime.
Then, since $r^{2490} \simeq 2.6 \cdot 10^{-8821}$, we see that the terms entering (\ref{PAeval}) would differ by almost nine thousand
orders of magnitude, so allowing some margin for numerical instabilities we would have to compute the correlation function for
(at least) 2\,500 different values of $l$ to a numerical precision of (at least) 10\,000 digits. This task is hopelessly impossible, given that
the transfer matrix of Appendix~\ref{sec:appA1} is of dimension $\sim 10^6$ in this case.

To do better, we need to consider a particular well-chosen combination of correlation functions, designed so that it decouples from a sufficient
number of low-lying states in the spectrum. The quantity
\begin{equation}
 P^* = P_{aaaa} + P_{aabb} + \frac{1}{Q-1} \left( P_{abab} + P_{abba} \right)
\end{equation}
is a such a good combination. On symmetry grounds, it decouples from the sectors with odd momenta $k$.
The term $P_{aaaa}$ is rather easily checked to pick up contributions from the sectors $V_1$, $V_2$
and $V_4$ (and maybe higher values of $\ell$), whereas $P_{aabb}$ couples in addition to $V_0$.
Meanwhile, $P_{abab}$ and $P_{abba}$ get contributions only from $V_2$ and $V_4$
(and maybe higher values of $\ell$); it is indeed clear that since these terms impose two long clusters
(see Figure~\ref{fig:2clusters}) they cannot couple to $V_0$ and $V_1$.
The surprising property of $P^*$ is now that, with the above choice of the two coefficients in its definition, all contributions from $V_1$ and $V_2$
disappear from the combination. In other words, $P^*$ couples to $V_0$, $V_{40}$, $V_{42}$, and maybe $V_{\ell,k}$
with higher values of $\ell$ and even $k$. We have ${\rm dim}\, V_0 = 232$, ${\rm dim} \, V_{40} = 190$ and ${\rm dim} \, V_{42} = 182$.
Therefore we shall be able to settle whether there is a non-zero amplitude for the 6-cluster sector provided we can look beyond the
first $604$ eigenvalues.

To that end, we have computed $P^*$ for $l=100,101,\ldots,900$ to be on the safe side. Noting that $r^{900} \simeq 1.3 \cdot 10^{-3116}$
we have performed the computations to a numerical precision of $4\,000$ digits. This required about $3 \times 10^5$ hours of single-processor
CPU time. The conclusion is that we have unambiguously established that $P^*$ picks up non-zero contributions from the first few eigenvalues
in each of the sectors $V_{60}$, $V_{62}$ and $V_{64}$, with amplitudes in the range $\sim 10^{-10}$. While these numbers may seem small,
they follow the clear trend (observed in all cases) that the non-zero amplitudes decay exponentially with the index of the corresponding eigenvalue.
Moreover, these amplitudes are numerically stable towards changing $l_{\rm min}$ throughout the range $l_{\rm min} \in [100,200]$. We have
also checked that the absolute values of  contributions which are genuinely supposed to be zero
(such as those from sectors $V_{61}$, $V_{63}$ and $V_{65}$) come out numerically
as $\ll 10^{-500}$, which is fully compatible with the above estimates of the required numerical precision.

\subsubsection{Exponents}

Introducing   our usual notation $F_{j,z^2}$,  the spectrum in the sector with $j=2$ propagating clusters is encoded in the generating functions
\begin{equation}
F_{2,1}\oplus F_{2,-1}
\end{equation}
This corresponds to the conformal weights given by (\ref{bspec1}):
\begin{subequations}
\begin{eqnarray}
(h_{e,-2};h_{e,2}) \,, & & e\in \mathbb{Z} \,, \\
(h_{e+1/2,-2};h_{e+1/2,2}) \,, & & e\in \mathbb{Z} \,.
\end{eqnarray}
\end{subequations}
Note that for the first part of the spectrum, $h-\bar{h}$ is an even integer, while it is an odd integer for the first part.  We will denote these two contributions  by $2S$ and $2A$ respectively, where $S$ stands for symmetric and $A$ for antisymmetric.
Going back to an earlier remark, this means that we should have, for primary fields
\begin{subequations}
\begin{eqnarray}
\Delta-\bar{\Delta} &=& \hbox{even  in S part} \,, \\
\Delta-\bar{\Delta} &=& \hbox{odd  in A part} \,.
\end{eqnarray}
\end{subequations}
We thus have, for the $2S$ part
\begin{subequations}
\begin{eqnarray}
 h_{e,2} &=& {[(m+1)e-2m]^2-1\over 4m(m+1)} \,, \\
h_{e,-2} &=& h_{e,2}+2e \,,
\end{eqnarray}
\end{subequations}
%
%
while for the $2A$  part 
\begin{subequations}
\begin{eqnarray}
h_{e+1/2,2} &=& {[(m+1)e+{1-3m\over 2}]^2-1\over 4m(m+1)} \,, \\
h_{e+1/2,-2} &=& h_{e+1/2,2}+2e+1 \,.
\end{eqnarray}
\end{subequations}

Similarly, in the sector with $j=4$ propagating clusters we have 
\begin{equation}
F_{4,1}\oplus F_{4,i}\oplus F_{4,-1}\oplus F_{4,-i}
\end{equation}
The corresponding conformal weights from (\ref{bspec1}) are
\begin{subequations}
\begin{eqnarray}
(h_{e,-4};h_{e,4}) \,, & & e\in\mathbb{Z} \,, \\
(h_{e+1/4,-4};h_{e+1/4,4}) \,, & & e\in\mathbb{Z} \,, \\
(h_{e+1/2,-4};h_{e+1/2,4}) \,, & & e\in\mathbb{Z} \,, \\
(h_{e+3/4,-4};h_{e+3/4,4}) \,, & & e\in\mathbb{Z} \,,
\end{eqnarray}
\end{subequations}
but the fourth set is identical to the second one. As the number of clusters increases, so does the number of allowed sectors.
Our finite-sizes observations are clearly in favour of the extension of this pattern, invariably with all even values of $j$.

Clearly, we find  exponents $(h_{r,s},h_{r,-s})$ in the $s$-channel with $r\in \mathbb{Z}$ and $s\in2\mathbb{Z}$, and $r\in \mathbb{Z}+1/2,s\in 2\mathbb{Z}$. We will   refer to these exponents  as sets $S_{\mathbb{Z},2\mathbb{Z}}$ and  $S_{\mathbb{Z}+1/2,2\mathbb{Z}}$, in analogy with \cite{Ribault}.
These sub-spectra arise from the modules $\AStTL{j}{1}$ and $\AStTL{j}{-1}$ with $j$ even. But we find that the $s$-channel, in finite size at least,  contains  many more exponents, arising from phases $z^2\neq 1,-1$ and, in terms of exponents, corresponding to rational values of the first label with higher denominators, such as those with first label $e+1/4$. 

The next key question is whether some sort of simplification might occur in the scaling limit---for instance,
whether some sectors that contribute to the probabilities in finite size might do so with amplitudes that go to zero as $L\to\infty$. 
We have seen absolutely no evidence of this. To make the point as clear as possible,
we illustrate it on the case of the antisymmetric combination  $P_{abab}-P_{abba}$.

\subsubsection{Amplitudes and the antisymmetric combination $P_{abab}-P_{abba}$}

The  antisymmetry of the combination implies that only modules with $z^{j}=-1$ contribute, which translates into primaries with 
$h-\bar{h}$ an odd number---what we have called earlier the $j$ even, A sectors.

Let us now focus on how this contributes to $P_{abab}-P_{abba}$. We have the first fields at spin $|h-\bar{h}|=1,3$ with weights  
$(h_{1/2,\mp2},h_{1/2,\pm2})$ and $(h_{3/2,-\mp 2},h_{3/2,\pm2})$ in the sector ${\cal S}_{\mathbb{Z}+{1\over 2},2\mathbb{Z}}$. But according to our earlier analysis we also expect contributions from, in particular, $(h_{1/4,\mp4},h_{1/4,\pm4})$. 
%
%
%
%

To make things concrete, we can take for instance $Q=1/2$  (so $m$ is irrational), in which case  we find
\begin{subequations}
\begin{eqnarray}
(h_{1/2,-2},h_{1/2,2}) &=& (1.156405\cdots,0.156405\cdots) \,, \\
(h_{3/2,-2},h_{3/2,2}) &=& (2.969378\cdots,-0.030621\cdots) \,, \\
(h_{1/4,-4},h_{1/4,4}) &=& (2.925269\cdots,1.925269\cdots) \,.
\end{eqnarray}
\end{subequations}
We observe that  the field with $(h_{1/2,-2}+2,h_{1/2,2})$ has total dimension larger than $(h_{3/2,-2},h_{3/2,2})$. Therefore, at momentum 3, the field $(h_{3/2,-2},h_{3/2,2})$ will be the first contribution, and so will be $(h_{1/2,-2},h_{1/2,2})$ at momentum one. It is therefore very easy to identify the corresponding contributions to the four-point function:  %
\begin{eqnarray}
P_{aabb}-P_{abba}\propto (z\bar{z})^{-2h_{1/2,0}}\left( A_{\Phi_{h_{1/2,-2},h_{1/2,2}}}z^{h_{1/2,-2}}\bar{z}^{h_{1/2,2}}+
A_{\Phi_{h_{3/2,-2},h_{3/2,2}}}z^{h_{3/2,-2}}\bar{z}^{h_{3/2,2}}+
\ldots\right.\nonumber\\
\left.A_{\Phi_{h_{1/4,-4},h_{1/4,4}}}z^{h_{1/4,-4}}\bar{z}^{h_{1/4,4}}+\ldots\right) \,.
\end{eqnarray}
Since $m$ is irrational, there is no mixing in the conformal mapping, and we have on the cylinder
\begin{eqnarray}
P_{aabb}-P_{abba}\propto A_{\Phi_{h_{1/2,-2},h_{1/2,2}}}\left(4\sin^2{2\pi a\over L}\right)^{h_{1/2,-2}+h_{1/2,2}}\xi^{h_{1/2,-2}}\bar{\xi}^{h_{1/2,2}}+\ldots\nonumber\\
+A_{\Phi_{h_{3/2,-2},h_{3/2,2}}}\left(4\sin^2{2\pi a\over L}\right)^{h_{3/2,-2}+h_{3/2,2}}\xi^{h_{3/2,-2}}\bar{\xi}^{h_{3/2,2}}+\ldots\nonumber\\
+A_{\Phi_{h_{1/4,-4},h_{1/4,4}}}\left(4\sin^2{2\pi a\over L}\right)^{h_{1/4,-4}+h_{1/4,4}}\xi^{h_{1/4,-4}}\bar{\xi}^{h_{1/4,4}}+\ldots
\end{eqnarray}
To restate the obvious, what we do then is measure the combination of probabilities on the left, identify the various terms on the right
(via the exponential $l$-dependence of $\xi, \bar{\xi}$), and account for the geometrical factors (the powers of $4\sin^2 {2\pi a\over L}$) to extract,
for a given sizes $L$, an estimate of the amplitudes. 

\subsubsection{The $(1/4,\mp 4)$ amplitude}

\begin{figure}[ht]
\begin{center}
    \includegraphics[width=12cm]{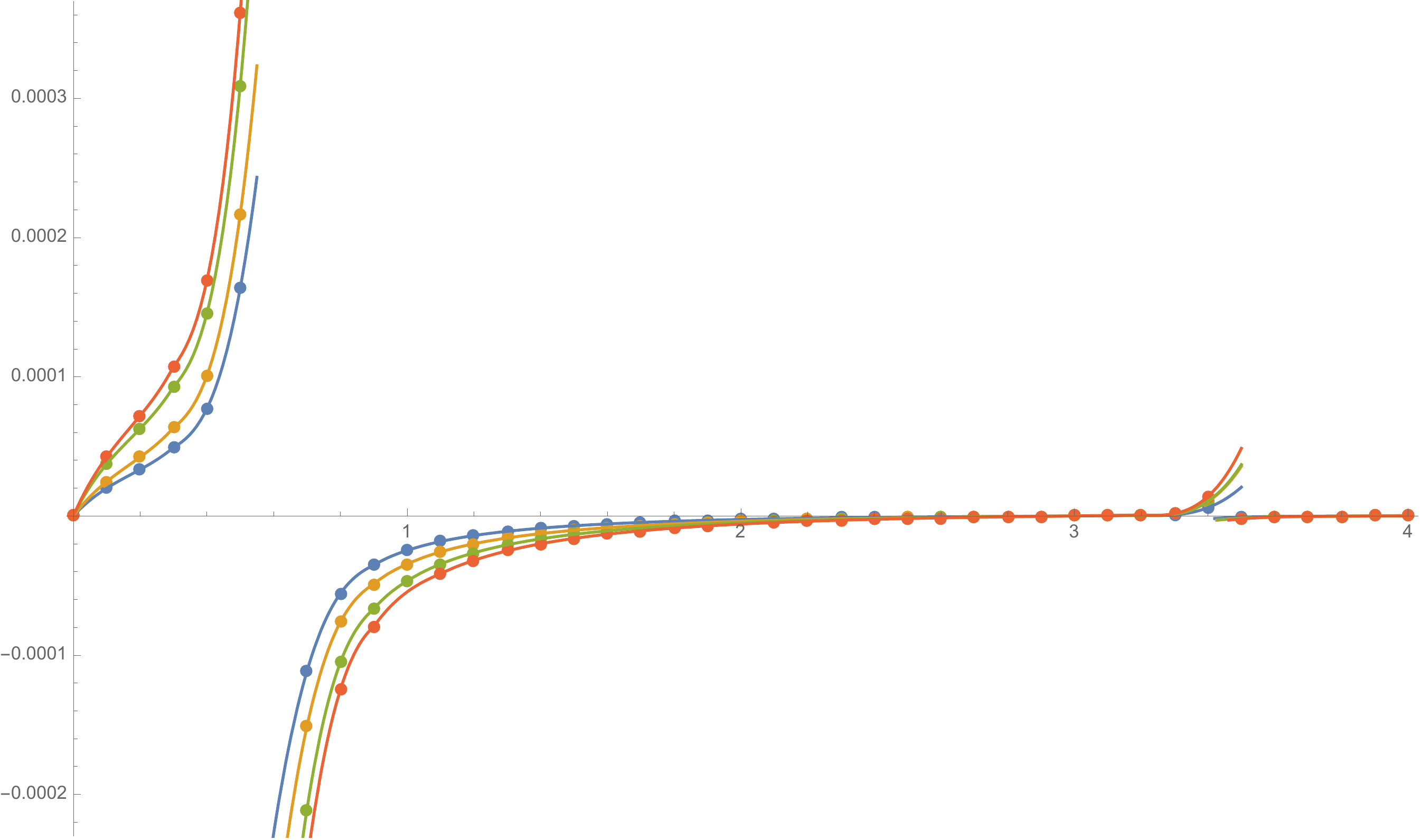}
     \caption{The ratio $A_{\Phi_{h_{1/4,-4},h_{1/4,4}}}/A_{\Phi_{h_{1/2,-2},h_{1/2,2}}}$ as a function of $Q$ for $L=5,6,7,8$ (blue, orange, green and red dots respectively). This ratio is generically non zero. It exhibits (in finite size) simple poles at $Q=4 \cos^2{\pi\over 8}$, $Q=4\cos^2{4\pi\over 8}$, and vanishes exactly (in finite size) for $Q=0,3,4$. }\label{quarteramp}
\end{center}
\end{figure}

We give in figure (\ref{quarteramp}) the results for the ratio $A_{\Phi_{h_{1/4,-4},h_{1/4,4}}}/A_{\Phi_{h_{1/2,-2},h_{1/2,2}}}$ as a function of $Q$ for various sizes $L$. While this amplitude is small  (amplitudes typically decay very fast with the dimension of the associated primaries), it is clearly not zero in general, nor does it show any indication of going to zero as $L$ increases. 
We note however that, for all finite sizes, $A_{\Phi_{h_{1/4,-4},h_{1/4,4}}}=0$ for $Q=0,3,4$. This is well expected, as discussed in the
Appendices \ref{sec:appB-Q0} and \ref{sec:appB-Q4} in particular. We find on the other hand that $A_{\Phi_{h_{1/4,-4},h_{1/4,4}}}\neq 0$ for $Q=2$
(cf.\ Appendix~\ref{sec:appB-Q2}).

\begin{figure}[ht]
\begin{center}
    \includegraphics[width=4cm]{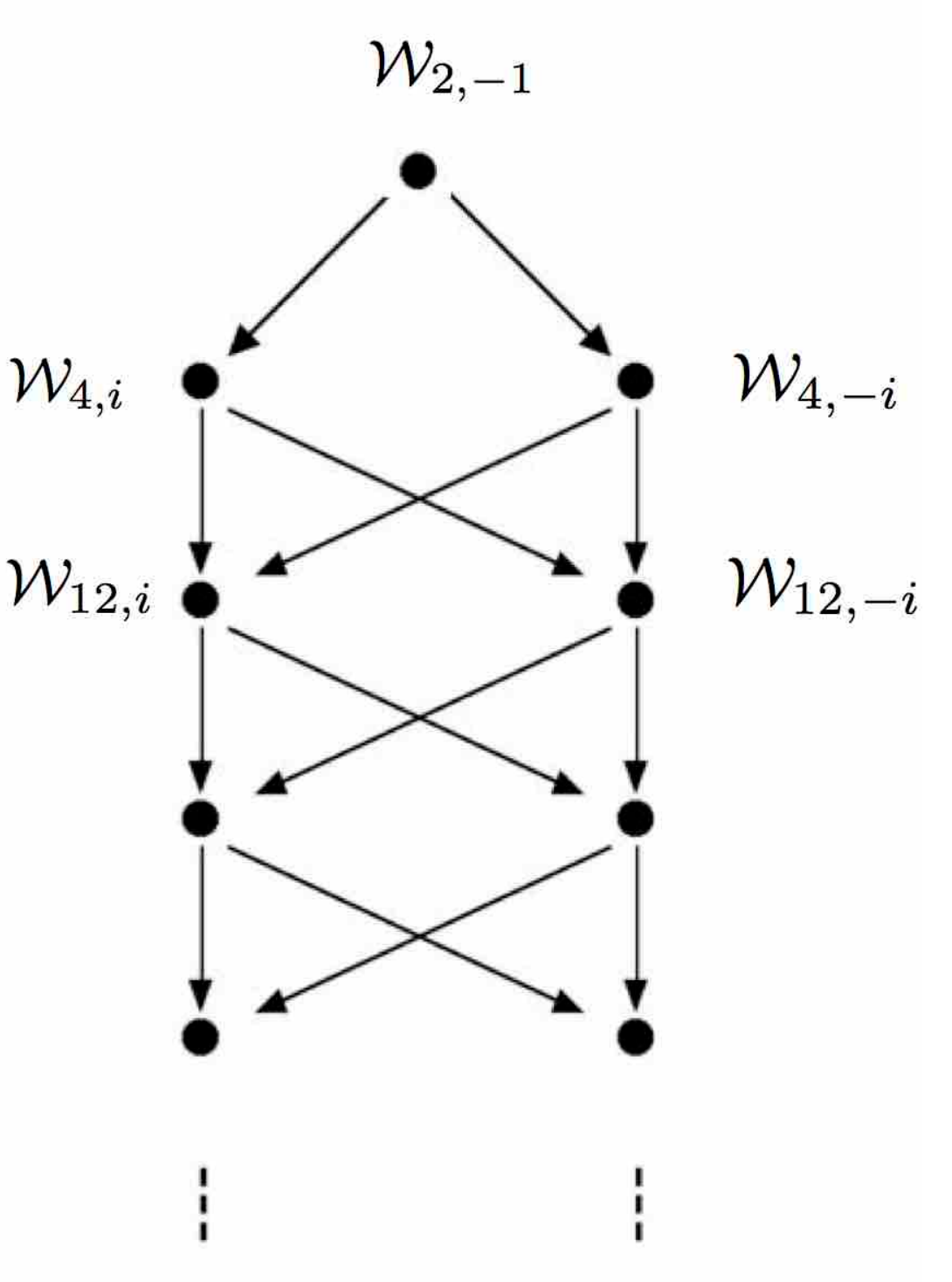}
     \caption{Sub-module structure of $\AStTL{2}{-1}$ for $\q=e^{3i\pi/8}$.  Note the appearance of sub-modules isomorphic to $\AStTL{4}{\pm i}$, which lead to glueing of standard modules into bigger, indecomposable modules, and Jordan cells for the transfer matrix.}\label{figmodjor}
\end{center}
\end{figure}

While the amplitude is small in general, it is found to become large---nay divergent---for two special values:
\begin{subequations}
\begin{eqnarray}
 Q &=& 4\cos^2{\pi\over 8} = 3.414213\cdots \,, \\
 Q &=& 4\cos^2{3\pi\over 8} = 0.585786\cdots \,.
\end{eqnarray}
\end{subequations}
There are several ways to understand this. We will discuss a CFT analysis  in the conclusion. From  the lattice  point of view, 
the divergence  arises because   the transfer matrix exhibits a Jordan cell in the lowest level of $\AStTL{4}{\pm i}$. This Jordan cell  arises from representation theory of the Jones algebra for $\q=e^{i\pi/8},\q=e^{3i\pi/8}$. To illustrate this, take for instance the case $\q=e^{3i\pi/8}$. The module $\AStTL{2}{-1}$ becomes reducible for this value of $\q$, and admits a sequence of submodules as represented in Figure~\ref{figmodjor}. The presence of submodules $\AStTL{4}{i},\AStTL{4}{-i}$ (in particular) suggests\footnote{While the structure of modules $\AStTL{j}{z^2}$ in degenerate cases is well under control, what happens here is the glueing of two standard modules for $\q$ a root of unity. The understanding of which modules glue with which ones for a given transfer matrix is a bit more complicated, and involves more representation theory; see  \cite{GRSV1} for a discussion of this point.} that  excited states in $\AStTL{2}{-1}$ (a module with two propagating clusters) mix with states  in $\AStTL{4}{i},\AStTL{4}{-i}$ within the module involving four propagating clusters. This mixture leads to Jordan cells in the transfer matrix. As shown in (\ref{generalised_amplitudes})---and further
discussed in Appendix~\ref{sec:appB-Q0} in the case $Q=0$---a Jordan cell in turns translates into a contribution to the correlation function that is linear in (imaginary) time on the cylinder. This, finally, corresponds formally to an infinite amplitude.

\subsection{The $s$-channel spectrum of $P_{aaaa}$:   $F_{j,\mathrm{e}^{2i\pi p/M}}{\scriptstyle( M|j,~j\geq 2)}$, $j,jp/M$ even and $F_{0,-1}$ 
}

We now turn to $P_{aaaa}$: this is the probability that all four points belong to the same cluster. It is called $P_0$ in \cite{Ribault}. For all finite sizes, we find that the modules $\AStTL{j}{e^{2i\pi p/M}}$ with $M|j,~j\geq 2$ and $j$ even contribute when $jp/M$ is even: this corresponds to the sectors with an even number of clusters propagating, and values of $z$ obeying $z^j=1$, what we have called earlier the $j$ even, S sectors. Geometrically, these contributions arise from configurations where for instance the points $1,3$ and $2,4$ are joined by two clusters which are only connected outside of the interval between their two (imaginary) time slices. On top of this, we also have the contribution where the four points belong to a single cluster arising between their two (imaginary) time slices. As discussed earlier, having a cluster propagating along the cylinder does not imply that there are boundaries around the cluster. The corresponding module of the Jones algebra is thus not a module with $j=1$: rather, it occurs as $\AStTL{0}{-1}$, i.e., as a module with no through-lines, but for which non-contractible loops (which would cut the connection between $1,2$ and $3,4$) are forbidden. 

Like for $P_{abab}$ and $P_{abba}$ we find that all eigenvalues in these modules do contribute in finite size, and that none of the amplitudes seem to vanish as $L\to \infty$. This suggests that the spectrum of critical exponents is given by 
$F_{j,\mathrm{e}^{2i\pi p/M}}{\scriptstyle( M|j,~j\geq 2)}$, $j,jp/M$ even and $F_{0,-1}$. 

In the two clusters ($j=2$) sector, this leads to 
\begin{equation}
(h_{e,-2};h_{e,2}),~~ e\in\mathbb{Z}
\end{equation}
while in the  four clusters  ($j=4$) sector we find
\begin{eqnarray}
(h_{e,-4};h_{e,4}),~~ e\in\mathbb{Z}\nonumber\\
(h_{e+1/2,-4};h_{e+1/2,4}),~~ e\in\mathbb{Z}
\end{eqnarray}
These two contributions occur as well in $S_{\mathbb{Z}+1/2,2\mathbb{Z}}$. New contributions appear for higher even values 
of $j$. For instance we find also 
\begin{eqnarray}
(h_{e,6};h_{e,-6}),~~ e\in\mathbb{Z}\nonumber\\
(h_{e\pm1/3,6};h_{e\pm1/3,-6}),~~ e\in\mathbb{Z}
\end{eqnarray}
 On top of this we have  the `one-cluster sector', which is described by $F_{0,-1}$ (i.e., non-contractible loops are killed). This corresponds to the set of conformal weights
\begin{equation}
(h_{e+1/2,0};h_{e+1/2,0}),~~e\in\mathbb{Z} 
\end{equation}
%
%
%
%
%
which is also in $S_{\mathbb{Z}+1/2,2\mathbb{Z}}$.

\subsection{The $s$-channel spectrum of $P_{aabb}$: $F_{j,\mathrm{e}^{2i\pi p/M}}{\scriptstyle( M|j,~j\geq 2)}$, $j,jp/M$ even, $F_{0,-1}$ 
and $\bar{F}_{0,\q^2}$}

The quantity $P_{aabb}$ is the probability for two ``short clusters'' (as opposed to the ``long clusters'' shown in Figure~\ref{fig:2clusters}): points $1,2$ belonging to one cluster, points $3,4$  to the other. It is called $P_1$ in \cite{Ribault}. We find that all the 
 eigenvalues occurring in $P_{aaaa}$ also contribute to $P_{aabb}$. On top of these, we also find  the eigenvalues from the module $\bAStTL{0}{\q^2}=\bAStTL{0}{\q^{-2}}$. This module corresponds to a sector with  no (forced) propagating cluster, which is obtained simply by giving non-contractible loops their bulk weight. As usual now, none of the corresponding amplitudes seem to vanish in the limit $L\to\infty$.

%
%
The operator content from  $\bAStTL{0}{\q^{\pm2}}$ involves diagonal primaries, with weights  
%
\begin{equation}
(h_{e+e_0,0};h_{e+e_0,0}),~~e\in\mathbb{Z} \,,
\end{equation}
where 
\begin{equation}
h_{e+e_0,0}={[(m+1)(e+1)-1]^2-1\over 4m(m+1)} \,.
\end{equation}
%
Of course this is the same set as the set 
\begin{equation}
(h_{e,1};h_{e,1}),~~e\in\mathbb{Z} \,,
\end{equation}
after a shift of the electric charge. We will denote this set as $\mathcal{S}^d_{\mathbb{Z},1}$.

\subsection{Summary}

We can now summarise our spectra in the $s$-channel
\begin{align}
\setlength{\arraycolsep}{6mm}
\renewcommand{\arraystretch}{1.2}
 \begin{array}{c|ll}
   & \mbox{$s$-channel} & \mbox{Parities}
   \\
   \hline
 P_{aaaa} &  \mathcal{S}_1\equiv\mathcal{S}_{\mathbb{Z}+1/2,0}\cup\{\mathcal{S}_{\mathbb{Z}+{p\over M},j}\} & j \in 2 \mathbb{Z}, \ jp/M\hbox{ even}    \\
 P_{aabb} &\mathcal{S}_2\equiv \mathcal{S}^d_{\mathbb{Z},1}\cup \mathcal{S}_{\mathbb{Z}+1/2,0}\cup\{\mathcal{S}_{\mathbb{Z}+{p\over M},j}\} & j \in 2 \mathbb{Z}, \ jp/M\hbox{ even}     \\
P_{abab/abba} & \mathcal{S}_3\equiv \{\mathcal{S}_{\mathbb{Z}+{p\over M},j} \}& j \in 2 \mathbb{Z}, \ jp/M\hbox{ integer}    \\
   \hline
 \end{array}
\end{align}
where we have allowed $j$ to take positive or negative values, since the sets of exponents are invariant under $j\to -j$. Recall that e.g. the set $\mathcal{S}_{\mathbb{Z},2\mathbb{Z}}$ refers to {\sl pairs} of exponents $(h_{r,s},h_{r,-s})$ with $r\in \mathbb{Z},s\in2\mathbb{Z}$, while $\mathcal{S}^d_{\mathbb{Z},1}$ denotes pairs $(h_{r,1},h_{r,1})$, with $r\in\mathbb{Z}$.   Recall also that $p,M$ are coprime integers, and that the value $p=0$ in particular is allowed. The case ${p\over M}={1\over 2}$ appears already in \cite{Ribault}.  

Note that these are the generic results, i.e., those valid for $m$ irrational.  Some contributions  vanish for special values of 
$Q$, such as $Q=0$, $Q=2$ and $Q=4$ (see Appendix~\ref{sec:appB}) and in some cases Jordan blocks appear.

The spectra in the other channels follow from simple geometrical considerations:
\begin{align}
\setlength{\arraycolsep}{6mm}
\renewcommand{\arraystretch}{1.2}
 \begin{array}{c|llc}
   & \mbox{$t$-channel} & \mbox{$u$-channel}
   \\
   \hline
 P_{aaaa}&  \mathcal{S}_1 & \mathcal{S}_1 
   \\
 P_{aabb} &\mathcal{S}_3  & \mathcal{S}_3    \\
P_{abab} & \mathcal{S}_3 &  \mathcal{S}_2 &    \\
P_{abba} & \mathcal{S}_2 &  \mathcal{S}_3 &    \\
   \hline
 \end{array}
\end{align}

An important property of our spectra  in the case of $P_{aabb}$ is that only states with positive conformal weights propagate along the cylinder: no  ``effective central charge'' appears, despite the non-unitarity of the CFT. This is contrast with what would be observed, for instance, in the case of minimal models corresponding to $m+1\equiv {p\over p'}$, $p,p'$ integer, where the effective ground state with $c_{\rm eff}=1-{6\over pp'}$ would appear. It is our understanding that a similar phenomenon takes place in the conjectured expressions of \cite{Ribault}. 

\section{Comparison with results in \cite{Ribault}}
\label{sec:comp-ribault}

\begin{figure}[ht]
\begin{center}
    \includegraphics[width=10cm]{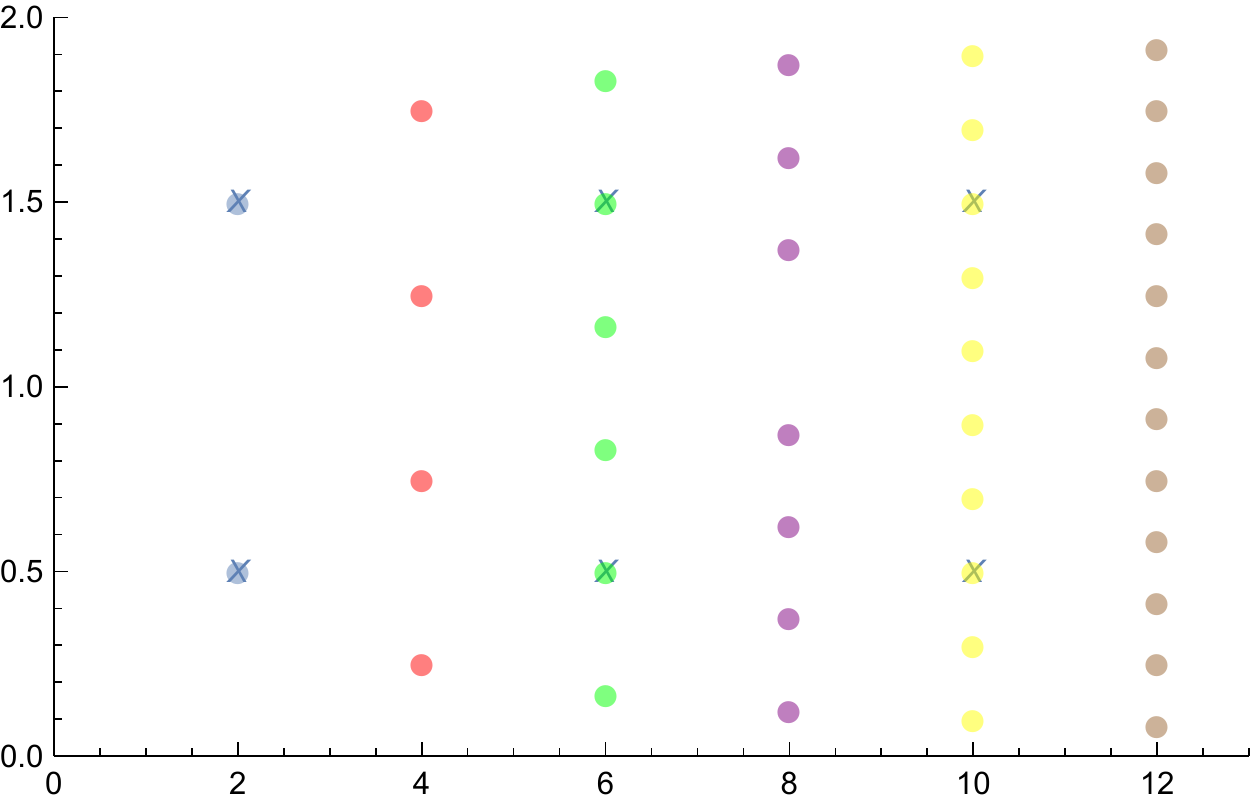}
     \caption{A sample of the full spectrum in the s-channel for $P_{abab}-P_{abba}$  represented by the pairs $(r,s)$ of the $h_{r,s}$ exponents ($r$ is on the y-axis, and $s$ on the x-axis. The spectrum considered in \cite{Ribault}, depicted as crosses, is seen to be a tiny subset of the full spectrum: the projections of the dots on the y-axis in fact should cover it densely (represented here are  exponents for $M=2,4,6,8,10,12$ only).}\label{figexpmiss}
\end{center}
\end{figure}

\begin{figure}[ht]
\begin{center}
    \includegraphics[width=3cm]{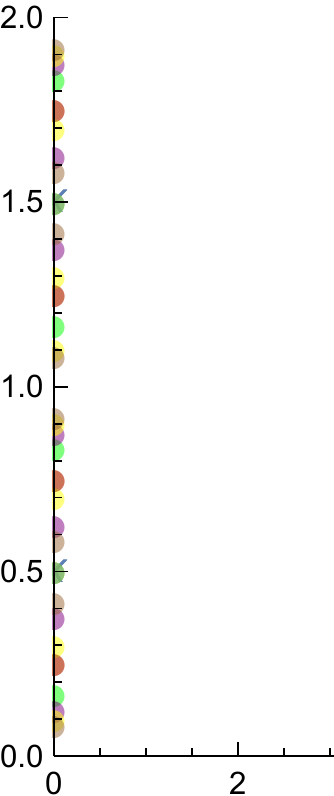}
     \caption{The projection on the vertical axis of the exponents represented in Figure \ref{figexpmiss}}\label{figexpmiss1}
\end{center}
\end{figure}

The comparison with the proposal in \cite{Ribault}  requires some discussion, since the authors in this reference did not, in particular, provide conjectured results for $P_{aaaa}$. The simplest quantity  to consider is $P_{abab}-P_{abba}\equiv P_2-P_3$ in the notations of that reference. Indeed, from eq. (3.2) in \cite{Ribault} 
\begin{equation}
R_\sigma=\lambda(P_0+\mu P_\sigma)
\end{equation}
we see that, in their notations,  $R_2-R_3=\lambda\mu(P_2-P_3)$. The spectrum in the $s$-channel of $P_2-P_3=P_{abab}-P_{abba}$, according to our analysis, is made of the fields $(h_{e+p/M,-j};h_{e+p/M,j})$ for $e\in \mathbb{Z}$, with $j$ even and   $pj/M$  an odd integer. Note that all these fields have $h-\bar{h}$ odd. In \cite{Ribault}, meanwhile, the  spectrum is $S_{\mathbb{Z}+1/2,2\mathbb{Z}}$ (after switching indices  in\cite{Ribault} to make their conventions the same as ours),  restricted like for us to odd spin $h-\bar{h}$.  So for instance the field with weights $(h_{1/4,\mp 4};h_{1/4,\pm 4})$ for which we have seen that the amplitude is generically non-zero, is absent in the solution proposed in \cite{Ribault}. This suggests that their solution is, generically,  not the correct one, and that {\bf an infinity of fields is missing in their proposal}. We illustrate this qualitatively in Figures \ref{figexpmiss},\ref{figexpmiss1}.

Meanwhile, it is fascinating  to compare results for amplitudes that are predicted in \cite{Ribault} and which are also found to occur in our analysis. A good example of this is the first  amplitudes for the sector with $j=2$, namely $A_{\Phi_{h_{3/2,-2},h_{3/2,2}}}$ and $A_{\Phi_{h_{1/2,-2},h_{1/2,2}}}$. The bootstrap in \cite{Ribault} produces amplitudes which are in fact simply related with those of Liouville field theory at $c<1$, and thus admit analytical expressions
\cite{EstienneIkhlef,RibaultMigliaccio}. In particular, their conjecture is%

\begin{equation}
{A_{\Phi_{h_{3/2,-2},h_{3/2,2}}}\over A_{\Phi_{h_{1/2,-2},h_{1/2,2}}}}=
 2^{-\frac{4}{\beta^2}} \frac{\Gamma(\frac32+\frac{1}{4\beta^2})} {\Gamma(\frac{1}{4\beta^2})}
\frac{\Gamma(\frac32+\frac{3}{4\beta^2})}{\Gamma(1+\frac{3}{4\beta^2})}
\frac{\Gamma(-1-\frac{1}{4\beta^2})}{\Gamma(\frac32-\frac{1}{4\beta^2})}
\frac{\Gamma(2-\frac{3}{4\beta^2})}{\Gamma(\frac32-\frac{3}{4\beta^2})}\label{exactconj}
\end{equation}
where $\beta^2={m\over m+1}$, $\sqrt{Q}=2\cos{\pi\over m+1}$, $m\in[1,\infty]$. This conjecture reproduces results which are believed to be exact at $Q=0,3,4$---the result for $Q=0$ is discussed in our Appendix \ref{sec:appB-Q0}; the result for $Q=4$ follows from a work by A.\ Zamolodchikov (as discussed in \cite{Ribault}), and the  result for $Q=3$ is unpublished work of R.\ Santachiara.

\begin{figure}[ht]
\begin{center}
    \includegraphics[width=10cm]{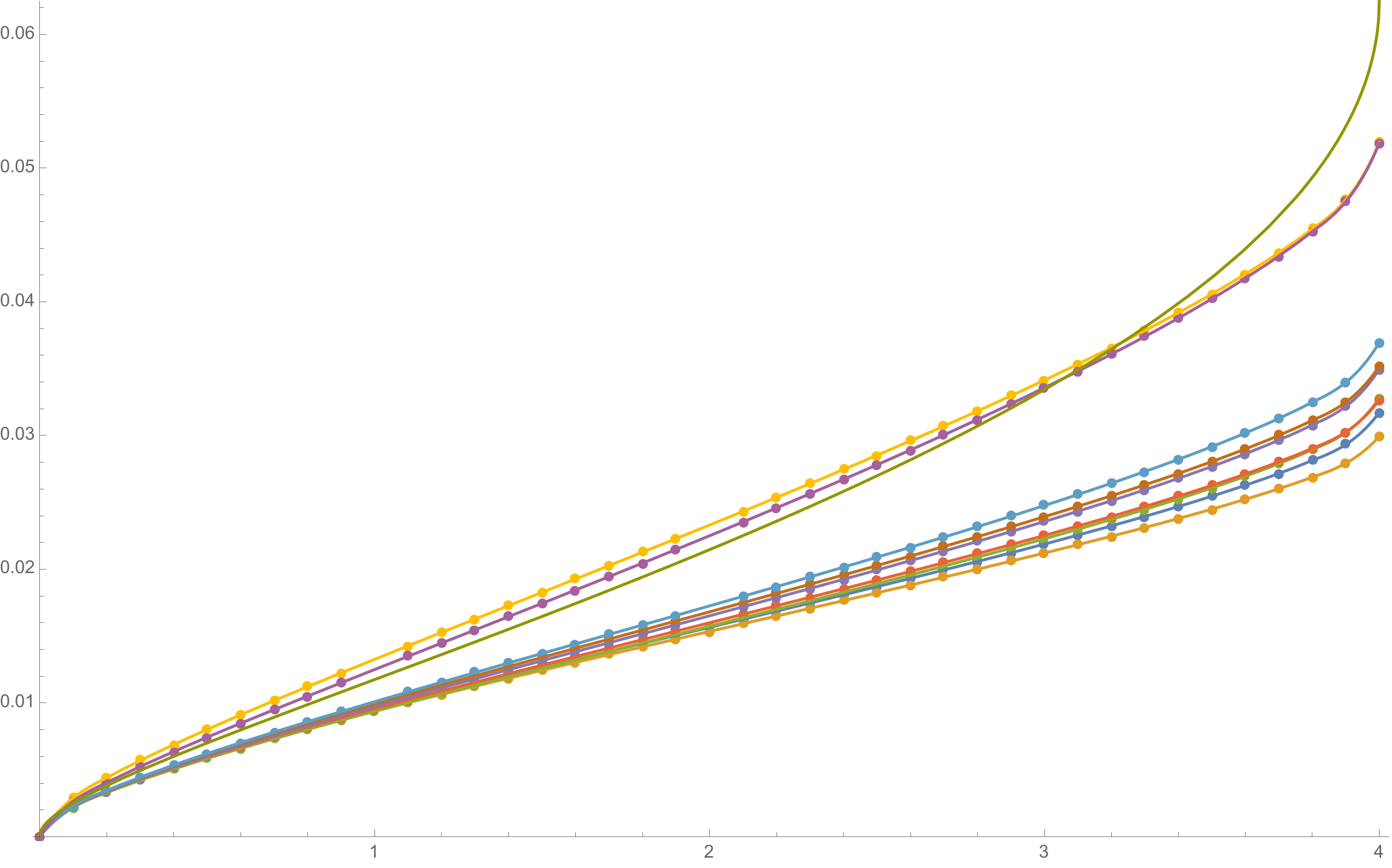}
     \caption{Results for the ratio $A_{\Phi_{h_{3/2,-2},h_{3/2,2}}}/ A_{\Phi_{h_{1/2,-2},h_{1/2,2}}}$, as a function of $Q$, for sizes $L=5,6,7,8,9,10,11$ corresponding to colours blue, orange, green, red, purple, mauve, clear blue. The points in purple are obtained by extrapolating data for $L=5,7,9,11$ and those in yellow by extrapolating data for $L=6,8,10$. The green curve is the conjecture (\ref{exactconj}).}\label{figconjsyl}
\end{center}
\end{figure}

Numerical results for this ratio are given in figure \ref{figconjsyl}. They are intriguingly close ---after reasonable extrapolation---to the formula (\ref{exactconj}). The agreement is worse near $Q=4$, but as commented elsewhere in this paper, this discrepancy can possibly be attributed to the presence of a marginal operator affecting corrections to scaling. We do not know whether (\ref{exactconj}) might actually be exact, or whether it is just very close to the exact result. Numerics, at this stage, do not really allow us to settle this issue.

Meanwhile, the uncertainty of the numerical determination shown in Figure~\ref{figconjsyl} can be estimated from the difference between the extrapolations
through even and odd system sizes $L$. Given that this uncertainty is (for most values of $Q$) comparable to the distance to the conjectured result (\ref{exactconj})
is certainly a strong motivation for further improving the numerical algorithm and gain access to a few more sizes. This could maybe be achieved if one could
impose the sector and momentum constraints within our scalar product method (see Appendix~\ref{sec:appA2}).

%
%
%
%
%
%
%

\section{Conclusion}
\label{sec:conclusion}

We believe that the numerical and algebraic evidence presented in this paper invalidates the results in \cite{Ribault}. This is 
a very intriguing conclusion, since, in particular, the authors of \cite{Ribault} presented Monte Carlo simulations of four-point functions in the plane that were in good agreement with their bootstrap prediction. It is possible that the conjecture in \cite{Ribault}, while not the correct answer to the problem of describing geometrical correlations in the Potts model, is indeed a solution to the bootstrap, and moreover  captures  numerically  the essential features of the four-point functions,  failing only at an accuracy, or for values of the cross-ratio $z$, not accessible using the Monte-Carlo approach. 
If this is the case, this raises several questions, in particular about the number of possible solutions to the bootstrap,%
\footnote{Recall that there are cases where several solutions to the bootstrap are known to exist, for instance the Liouville theory at $c=1$ and the Runkel-Watts limit of minimal models \cite{RunkelWatts,RibaultSantachiara}.}
and what, if anything, is truly described by the proposal in \cite{Ribault}. 

To shed more light on this issue, an obvious route is to build four-point functions following the methodology in \cite{Ribault} but based on our spectra. This is quite challenging technically, because of the large number of primary fields with dimensions of the same order of magnitude we would have to involve. Another, more fundamental aspect worth mentioning is that, in our spectrum, many of the conformal weights have degenerate values, with singular conformal blocks. It is not clear whether the regularisation procedure used in the bootstrap approach \cite{Ribault} is actually the relevant one for the $Q$-state Potts model. This, we believe, could be answered by  numerical studies in the spirit of the present paper and \cite{KooSaleur}. 

A particularly intriguing fact is that we found  numerically a ratio   $A_{\Phi_{h_{3/2,-2},h_{3/2,2}}}/ A_{\Phi_{h_{1/2,-2},h_{1/2,2}}}$ which is not incompatible with the proposal in \cite{Ribault}; see Figure~\ref{figconjsyl}. It could be that the solution to the bootstrap relevant for the Potts model  involves for this  ratio a value close to the one in \cite{Ribault} and yet different, over the whole range $Q\in [0,4]$. It could also be that the amplitudes in \cite{Ribault}---which, to the best of our understanding, are actually given by standard formulae for Liouville at $c<1$, naively extended to the case of fields with $h\neq \bar{h}$---are exact, but that  something has to be added.  

Adding ``something'' to the spectrum in \cite{Ribault} is definitely necessary if one wishes to avoid correlation functions with many singularities as $Q$ is varied. To see why  this is the case,
we consider the contributions to the antisymmetric combination of probabilities:
including now higher order terms in the conformal blocks
\begin{eqnarray}
P_{aabb}-P_{abba} &\propto& (z\bar{z})^{-2h_{1/2,0}}\left( A_{\Phi_{h_{1/2,-2},h_{1/2,2}}}{\cal F}_{h_{1/2,-2}}^{(s)}\overline{{\cal F}}_{h_{1/2,2}}^{(s)}\right.+\nonumber\\
& & \left.A_{\Phi_{h_{3/2,-2},h_{3/2,2}}}{\cal F}_{h_{3/2,-2}}^{(s)}\overline{{\cal F}}_{h_{3/2,2}}^{(s)}+
A_{\Phi_{h_{1/4,-4},h_{1/4.4}}}z^{h_{1/4,-4}}\bar{z}^{h_{1/4,4}}+\ldots\right) \,. \label{basicform}
\end{eqnarray}
Note that we used here conformal blocks where the dependency $ z^{-2h_{1/2,0}}$ (resp. $\bar{z}^{-2h_{1/2,0}}$) has been factored out.

When $Q\to 4\cos^2{3\pi\over 8}\equiv Q^*$, we find that $h_{3/2,2}\to h_{1,2}$, a degenerate value. The conformal block in the  four-point function coming from 
\begin{equation}
(h_{3/2,-2},h_{3/2,2})=(h_{3/2,-2},h_{1,2})=\left(-{1\over32}+3,-{1\over 32}\right)
\end{equation}
has a null-state at level 2 for the $\bar{z}$ components, with weights
\begin{equation}
(h_{3/2,-2},h_{1,-2})=\left(-{1\over32}+3,-{1\over 32}+2\right) \,.
\end{equation}
The appearance of the null-state  means that the conformal block $\overline{{\cal F}}_{h_{3/2,2}}$ has a pole of the form ${1\over Q-Q^*}$ multiplying the term $z^{h_{3/2,-2}}\bar{z}^{h_{1,-2}}$. 
Setting
\begin{equation}
{\cal F}_{h_{3/2,-2}}^{(s)}\overline{{\cal F}}_{h_{3/2,2}}^{(s)}\approx \ldots+{r^*\over  Q-Q^*}z^{h_{3/2,-2}}\bar{z}^{h_{3/2,2}+2}+\ldots \,,
\end{equation}
we see that  the amplitude of the singular term in the bracket in (\ref{basicform}) is 
\begin{equation}
A_{\Phi_{h_{3/2,-2},h_{3/2,2}}}{r^*\over  Q-Q^*} z^{h_{3/2,-2}}\bar{z}^{h_{3/2,2}+2} \,.
\end{equation}
Meanwhile we  observe that the weights for the singular term  coincide with the weights from the $1/4$ sector:
\begin{equation}
(h_{1/4,-4},h_{1/4,4})=\left({-1\over 32}+3,{-1\over 32}+2\right) \,.
\end{equation}
Recall that we have found   numerically that the amplitude of this field also has a simple pole when $Q\to Q^*$:
\begin{equation}
A_{\Phi_{h_{1/4,-4},h_{1/4,4}}}\approx {R\over Q-Q^*} \,,
\end{equation}
so the amplitude of the second singular term in (\ref{basicform}) is 
\begin{equation}
{R\over Q-Q^*}z^{h_{1/4,-4}}\bar{z}^{h_{4,1/4}} \,.
\end{equation}
For technical reasons, we normalise all quantities by $A_{\Phi_{h_{1/2,-2},h_{1/2,2}}}$ (this amplitude is not expected to be singular), so we set 
\begin{equation}
{A_{\Phi_{h_{1/4,-4},h_{1/4,4}}}\over A_{\Phi_{h_{1/2,-2},h_{1/2,2}}}} \approx {r\over Q-Q^*} \,.
\end{equation}
We find numerically  that on ``resonance'', there is a Jordan cell of rank two mixing the two terms, but no singularity. 
This means that we should have the condition
\begin{equation}
r+{A_{\Phi_{h_{3/2,-2},h_{3/2,2}}}\over A_{\Phi_{h_{1/2,-2},h_{1/2,2}}}}   r^*=0 \,.
\end{equation}
To put things differently, the appearance of a null-state in the conformal block $\overline{{\cal F}}_{h_{3/2,2}}$ leads to a divergence in the four-point function (assuming the formula for $A_{\Phi_{h_{3/2,-2},h_{3/2,2}}}$ given earlier is correct). To cancel this divergence, a contribution $A_{\Phi_{h_{1/4,-4},h_{1/4,4}}}$ is {\bf necessary}. Moreover, this contribution must exhibit a simple pole, as we have observed numerically. It is possible that this picture generalises, with singularities in the proposal of \cite{Ribault} exactly cancelled out by the additional terms we find in our lattice analysis. This will be discussed elsewhere \cite{JRSS}.

To conclude this paper, we re-iterate the remark that the  eigenvalues contributing to the probabilities are (a subset of) those appearing in the Potts model partition function \cite{DFSZ}. While this would be a well expected fact  for a  model defined locally such as the Ising model or any kind of height model, this is  not so obvious in our case. Indeed, in  a model where correlations are defined non-locally there is no clear connection between the partition function and at least some of the observables. To give a simple example, we know well that the probability that two points are connected with a cluster allowing two independent paths involves exponents not present in the partition function \cite{JacobsenZinn1,DengBloteNienhuis}.
The fact that no such exponents are needed for the $P_{a_1 a_2 a_3 a_4}$ suggests that these are rather close to ``ordinary'' observables, and that we may be able to understand them in terms of fully local operators acting on the space of states. One of the main ``elementary'' mysteries in this description is why only sectors with an even number of clusters contribute. We believe that thinking more deeply about algebraic aspects of the problem on the lattice will shed some light on this question.


\paragraph{Acknowledgements.}
We thank the authors of \cite{Ribault} and especially S.\ Ribault for inspiring discussions.
We also thank C.R.\ Scullard for discussions about numerical aspects and especially for the use of his unpublished {\sc C++} implementation of the {\sc Arpack} version
of the Arnoldi algorithm. This work was supported by the ERC Advanced Grant NuQFT.

\vskip2cm

\appendix

\section{Computing four-point functions on the cylinder}
\label{sec:appA}

In this appendix we gather all the algebraic and numerical technology that enables us to compute four-point functions in the FK cluster
model on a cylinder.

The geometrical setup is shown in Figure~\ref{fig:cylinder}. For the sake of simplicity we shall define the model on
an axially oriented square lattice of width $L$ spins. The four points are inserted in two different time slices, separated by $l$ lattice spacings,
with points $w_1,w_2$ in the first slice and points $w_3,w_4$ in the second slice.

The algebraic tool is the transfer matrix $T$ based on the join-detach algebra, which we have already briefly
reviewed in section~\ref{sec:join-detach}. To adapt it to the computation of correlation functions the crucial point
is the representation of the algebra, i.e., the choice of the set of states on which $T$ acts.
While the partition function $Z$ can be simply computed by taking these states to be set partitions of $L$ points
(see \cite{BloteNightingale82,SalasSokal01} for details) we shall need to endow these partitions with various kinds of marks.

We are interested in two different situations that present some subtle differences. First we describe how to compute
directly the 15 different FK cluster correlation functions $P_{a_1,a_2,a_3,a_4}$, defined by (\ref{P_corr_def}), that lie at the
heart of the numerical method outlined in section~\ref{sec:1st_method}. Second, we go into the details of the alternative
method of section~\ref{sec:2nd_method} which is based on computing scalar products involving the order parameter
operators (\ref{spin-op}).

\subsection{Computation of $P_{a_1,a_2,a_3,a_4}$}
\label{sec:appA1}

Let $w_k$ (with $k=1,2,3,4$) be four marked points on the cylinder, as shown in Figure~\ref{fig:cylinder}.
Let ${\cal P} = \{a_1,a_2,a_3,a_4\}$ be a set partition of the four points, defined by the corresponding integer
labels $a_k$. We wish to construct a transfer matrix that builds the statistical weight $W_{\cal P}$ corresponding
to a correlation function in which the marked points $k$ and $\ell$ belong to identical (resp.\ different) FK
clusters if $a_k = a_\ell$ (resp.\ $a_k \neq a_\ell$). For instance, the choice ${\cal P} = \{1,2,1,2\}$ means that
$w_1$ and $w_3$ belong to the same cluster, while $w_2$ and $w_4$ also belong to a common cluster
which is different from the first one, as shown in the left panel of Figure~\ref{fig:2clusters}.

On the cylinder there are $B_4 = 15$ different correlation functions, where the Bell number $B_N$ denotes
the number of partitions of an $N$-element set. On the strip there would be only $C_4 = 14$ correlation
functions, where $C_N$ denote the Catalan numbers.

We consider the square lattice of width $L$ wrapped on a cylinder. The sites within each row are labelled
by $i = 0,1,\ldots,L-1$, with the labels considered modulo $L$ by the periodic boundary conditions. To make contact
with Figure~\ref{fig:cylinder}, we let
$w_1$ and $w_2$ correspond to sites $0$ and $2a$ (with $a \in \mathbb{N}/2$) in the row labelled by the
transfer matrix ``time'' $t_1 = 0$, while $w_3$ and $w_4$ correspond to sites $x$ and $x+2a$ in the row $t_2 = l$.
A cluster containing (at least) one of the points $w_k$ is called a {\em marked cluster}.

To allow the marked clusters to wind around their insertion points, we consider a larger piece of the lattice,
going from row $t_0 = -M$ to row $t_3 = l+M$. We impose free boundary conditions on the two extremites of
the {\em finite} cylinder defined by $t \in [t_0,t_3]$.
The desired correlation functions (or probabilities) are then given by the limit
\begin{equation}
 P_{\cal P}(l) = \lim_{M \to \infty} \frac{W_{\cal P}}{\sum_{r=1}^{15} W_{{\cal P}_r}} \,. \label{norm_proba}
\end{equation}
For practical purposes, to obtain $P_{\cal P}(l)$ to a given numerical
precision, it suffices to take $M$ sufficiently large, so that the results for $M$ and $M+1$ coincide to the chosen
precision. A more precise criterion for the choice of $M$ can be given once the spectrum of the transfer matrix is known
(see section~\ref{sec:M_bdry_cond}).

The transfer matrix acts as usual on states $\{s_1,s_2,\ldots,s_L\}$ which are certain set partitions of $L$ points,
but now endowed with suitable markings.
We have $s_i = s_j$ if and only if sites $i$ and $j$ are seen to be in the same FK cluster at a given time $t$---by this we
mean that $i$ and $j$ are connected by a piece of FK cluster on the partly constructed cylinder $t \in [t_0,t]$.
The symbols $s_i$ can take the values $1,2,\ldots,L$ for unmarked clusters, and $\bar{1},\bar{2},\bar{3},\bar{4}$
for marked clusters containing one of the points $w_k$. If a marked cluster contains more than one marked point,
the chosen symbol is the lowest one, in order to avoid any redundancy in the correspondence between cluster connectivities
and states. Unmarked clusters are indistinguishable.

As shown in (\ref{HV}) the transfer matrix can be written as a product of elementary operators
${\sf H}_i = {\sf I} + v {\sf J}_i$ and ${\sf V_i} = v {\sf I} + {\sf D}_i$
that add respectively a horizontal edge between sites $i$ and $i+1$ (mod $L$), and a vertical edge on top of site $i$.
We have $v = {\rm e}^K - 1$, with $K$ the Potts coupling, and the critical value on the square lattice is $v_{\rm c} = \sqrt{Q}$.
The join and detach operators, ${\sf J}_i$ and ${\sf D}_i$ satisfy the join-detach algebra
with relations (\ref{join-detach}). We now describe a modified representation of this algebra that properly
takes into account the possibility of having marked clusters.

A clusters is said to be {\em left behind} at site $i$ if ${\sf D}_i$ detaches a singleton in the set partition.
When this happens, ${\sf D}_i$ applies a weight $Q$, 
irrespective of whether the cluster being left behind is unmarked or marked.

\begin{remark}

In some applications it is natural to give weight $1$ to marked clusters. In particular, this is mandatory
in the limit $Q \to 0$, since otherwise all correlation functions would be identically zero. However, for the time being we choose to
give weight $Q$ to any cluster, including the marked ones, since then $\sum_{r=1}^{15} W_{{\cal P}_r} = Z(Q,v)$, the partition function on a
cylinder of circumference $L$ and length $l+2M$. The quantity $Z(Q,v)$ can easily be obtained by independent
means \cite{ChangSalasShrock}, and the sum rule then provides a valuable check of the correctness of the algorithm.

\end{remark}

To apply the initial condition, we start from the all-singleton state $\{1,2,\ldots,L\}$ at time $t_0$.
The final condition at time $t_3$ is to project out any state in which the desired connections between marked clusters
have not been achieved. To be precise, if there exists two distinct labels $\bar{k} \neq \bar{\ell}$ in the connectivity
state such that $a_k = a_\ell$, then the state must be discarded. Any state that survives this projection is attributed
a weight $Q$ per cluster (again irrespectively of whether it is unmarked or marked). The weighted sum over retained states produces
the sought-for $W_{\cal P}$.

Marked symbols are introduced in the time evolution by joining site $i=0$ (resp.\ $i=2a$) to the marked
symbol $\bar{1}$ (resp.\ $\bar{2}$) at time $t=t_1$, and by joining site $i=x$ (resp.\ $i=x+2a$) to the marked
symbol $\bar{3}$ (resp.\ $\bar{4}$) at time $t=t_2$. 

To impose the desired four-point connectivity, we modify the action of the join operators ${\sf J}_i$ as follows:
\begin{itemize}
 \item When ${\sf J}_i$ joins an unmarked cluster to a marked cluster with symbol $\bar{k}$, the result is a marked cluster with
  the same label $\bar{k}$.
 \item ${\sf J}_i$ can join two marked clusters with labels $\bar{k}$ and $\bar{\ell}$ only if $a_k = a_\ell$.
  In that case, the result is a marked cluster with the lowest label ${\rm min}(\bar{k},\bar{\ell})$.
\end{itemize}
The detach operators ${\sf D}_i$ need a more subtle modification. We describe the marked points $w_1$ and $w_2$ as {\em unseen}
(i.e., not yet visited by the transfer process) when the time $t < t_1$. Similarly the marked points $w_3$ and $w_4$ are unseen
when $t < t_2$. The required modification is:
\begin{itemize}
 \item The cluster $\bar{k}$ is allowed to be {\em left behind} only if the label $a_k$ is different from the labels of any
 other marked cluster in the connectivity state, and also different from the labels of all yet unseen marked points.
\end{itemize}

\subsubsection{Example}

Consider the case ${\cal P} = \{ a_k \} = \{1,2,1,2\}$ of two propagating clusters. At times $t_1 < t < t_2$, the
points $w_1$ and $w_2$ have been seen, so the connectivity state can contain the symbols $\bar{1}$ and $\bar{2}$.
In fact, both symbols {\em must} be present, because the ${\sf J}_i$ operator cannot join $\bar{1}$ and $\bar{2}$ (since we
have $a_1 \neq a_2$); moreover none of them can be left behind, because points $P_3$ and $P_4$ are unseen.
Indeed, $a_1 = a_3$ then prevents $\bar{1}$ from being left behind, and $a_2 = a_4$ prevents $\bar{2}$ from being
left behind.

At later times $t_2 < t < t_3$, all points have been seen. Symbols $\bar{1}$ and $\bar{3}$ can join, because $a_1 = a_3$
(and similarly for $\bar{2}$ and $\bar{4}$). But before this happens, none of the clusters $\bar{1}$ and $\bar{3}$ can be
left behind (by the rule on ${\sf D}_i$). When $\bar{1}$ and $\bar{3}$ join, the resulting cluster carries the symbol $\bar{1}$
(by the second rule on ${\sf J}_i$). After this happens, $\bar{1}$ can be left behind (by the rule on ${\sf D}_i$, because
$\bar{3}$ is no longer used in the connectivity state).

\subsubsection{Checks}

In addition to extensive checks for small systems, where all configurations can be drawn by hand, we have checked that:
\begin{enumerate}
 \item The unnormalised sum $\sum_{r=1}^{15} W_{{\cal P}_r} = Z(Q,v_{\rm c})$ for any values of $Q$, $L$, $l$, and $M$.
 \item The probabilities converge to any desired numerical precision upon taking $M \gg L,l$ large enough.
\end{enumerate}
Moreover, for the situation with shift $x=0$ (see Figure~\ref{fig:cylinder}) it is non-trivial from the point of view of the transfer
algorithm that the following lattice symmetries hold true:
\begin{enumerate}
 \setcounter{enumi}{2}
 \item The four probabilities in which three points are in the same cluster are all equal.
 \item The six probabilities in which two points are in the same cluster and the other two are in two distinct clusters, are equal two by two.
\end{enumerate}
A more restricted set of lattice symmetries holds true also for $x \neq 0$; for instance the four probabilities in which three points are in the same cluster,
are equal two by two. We have checked this as well.

\subsubsection{Sample results}

To help the readers desirous of writing their own implementation, we here give
some sample results for $\mathbb{P}_{\cal P}(l)$ for a system with $Q=1$ and size $L=5$. We take distance $l=3$ between the time slices
and place the points in each slice as nearest neighbours ($2a=1$), with no offset between the two slices ($x=0$). The set of 15
probabilities (up to the symmetries described above) are then:
\begin{subequations}
\begin{eqnarray}
 P_{aaaa} &=& 0.4137474261084028402728075794444609019425125830604 \,, \\
 P_{aaab} &=& 0.0527654414874488570020983215187468064670272213592 \,, \\
 P_{aabb} &=& 0.1710044985220526406478058107746167574154887333139 \,, \\
 P_{abab} &=& 0.0008428892255516309347686777944043427144862186950 \,, \\
 P_{abba} &=& 0.0000809519640647414138510576682844254047786037826 \,, \\
 P_{aabc} &=& 0.0728732174729894992103081083804105937132582003470 \,, \\
 P_{abac} &=& 0.0082903229000495425925235985388971176349864439965 \,, \\
 P_{abca} &=& 0.0061033620128809220144422552541186396032488302690 \,, \\
 P_{abcd} &=& 0.0287286634582927910878256638963936447516380264857 \,.
\end{eqnarray}
\end{subequations}
It was necessary to take $M=200$ to achieve 50 correct digits. Note that our code is written
so that the number of digits of numerical precision can be adjusted to any desired value.

As a final check we consider the limit $l \gg L$. Then we expect
\begin{subequations}
\begin{eqnarray}
 P_{abcd} &=& (1-p)^2 \,, \\
 P_{aabc} &=& p(1-p) \,, \\
 P_{aabb} &=& p^2 \,,
\end{eqnarray}
\end{subequations}
where $p$ is the finite probability that two nearest neighbours are in the same cluster. All other $P_{{\cal P}_r}(l)$ will be negligible in that limit.
By taking $L=5$ and $l=384$ (with still $M=200$) we have verified that this is indeed the case, and we estimate
\begin{equation}
 p = 0.76315602507834269413 \,.
\end{equation}
%

\subsubsection{Case of $Q \to 0$}

The limit $Q \to 0$ (with $v = \sqrt{Q} \to 0$ as well) is somewhat special from the point of view of normalisations, since then $Z(Q,v) = 0$.
The partition function which is used to normalise the weights $W_{\cal P}$ and turn them into probabilities, as in (\ref{norm_proba}), is
then taken as $\widetilde{Z} = W_{aaaa}$, the number of spanning trees containing all four marked points. Moreover, any marked cluster is assigned a weight $1$
(instead of $Q$ in the general case), whereas unmarked clusters still have the weight $Q=0$, which implies that such clusters are disallowed.

\subsection{Computation of $G_{a_1,a_2,a_3,a_4}$}
\label{sec:appA2}

The second numerical method presented in section~\ref{sec:2nd_method} provides a means of computing one by one the amplitudes appearing
in the order parameter correlation functions $G_{a_1,a_2,a_3,a_4}$ defined in (\ref{order-param-corr}). This method is of a very different nature
than the one just presented (i.e., for the computation of $P_{a_1,a_2,a_3,a_4}$), since it does not compute the four-point correlator
directly on a cylinder of finite length, but rather goes directly for the asymptotic quantities $A_i$. Thus, the appearance of $|v_0 \rangle$ and $\langle v_0 |$ in (\ref{scalar_prod_method})
amounts effectively to taking the limit $M \to \infty$ of the distance to the boundary conditions, and the piece
$| v_{i,j} \rangle \, \langle v_{i,j} |$ projects directly on an intermediate state in the $s$-channel, which is equivalent to taking the
limit $l \to \infty$.

As a prerequisite to this approach---and in order recover in particular the same results as with the first method of section~\ref{sec:1st_method}---it is however
necessary to produce yet another representation of the join-detach algebra. In particular, we shall need the action of the order parameter operators
${\cal O}_{a}$ to be well-defined for generic values of $Q \in \mathbb{R}$. By (\ref{spin-op}), this hinges on giving a well-defined
meaning---and in particular, a meaning that extends to non-integer $Q$---to the operator $\delta_{\sigma_k,a}$ that fixes the value
of the spin at point $w_k$ to be equal to the label $a$.
Note that such extensions of order-parameter related quantities have played an important role in several recent pieces of work
involving the present authors \cite{VJ_spin_FK,DJS_spin1,DJS_spin2,tensor1,tensor2,tensor3}.

As stated around (\ref{sumrules}) we are here only interested in correlation functions $G_{a_1,a_2,a_3,a_4}$ employing at most two
distinct symbols, $a$ and $b$. We shall therefore take $T$ to act on basis states which are set partitions of $L$ points with two marked values,
denoted $a$ and $b$. It is crucial that these values are now different by definition. This provides a difference with the computation of
$P_{a_1,a_2,a_3,a_4}$, where two differently marked clusters could be joined under some circumstances. More precisely, the rules are now the following:
\begin{itemize}
 \item The operator $\delta_{\sigma_k,a}$ (the non-trivial part of ${\cal O}_a$) acts at the point $w_k$ by transforming an unmarked cluster touching
 that site into a marked cluster of label $a$.
 If the cluster is already marked with label $b$, ${\cal O}_a$ acts as the identity times $\delta_{a,b}$.
 \item The join operator ${\sf J}_i$ acts as usual on two unmarked clusters. When acting on an unmarked cluster and a marked cluster
 of label $a$, the result is a marked cluster of label $a$. Finally, when acting on two marked clusters of labels $a$ and $b$, ${\sf J}_i$
 acts as the identity times $\delta_{a,b}$.
 \item The detach operator ${\sf D}_i$ transforms site $i$ into an unmarked singleton. If $i$ was already a singleton beforehand, a
 weight $Q$ is applied if the corresponding cluster is unmarked, and $1$ if it is marked.
\end{itemize}

To compute the scalar products in (\ref{scalar_prod_method}) it is convenient to produce all the intervening vectors in the same space. In particular,
the ground state eigenvector $| v_0 \rangle$ is written within the space of set partitions with two marked values, although it is easily seen that its
component along any state containing marked clusters is zero.

\subsubsection{Orthogonalisation}

It has already been stated in the main text that the scalar product method relies on the left and right eigenvectors being orthogonal, even within
degenerate subspaces. Standard numerical methods for non-symmetric matrices, such as the Arnoldi algorithm, do not immediately produce the
eigenvectors in this form. Rather, within each degeneral subspace corresponding to the eigenvalue $\Lambda_i$, the scalar product of eigenvectors
come out as
\begin{equation}
 \langle v_{i,j} | v_{i',j'} \rangle = \alpha^{(i)}_{j,j'} \,,
\end{equation}
where the nombers $\alpha^{(i)}_{j,j'}$ can be viewed as the elements of some matrix $\alpha^{(i)}$. However, if we replace the right eigenvectors by the linear combinations
\begin{equation}
 | \widetilde{v}_{i',j'} \rangle = \sum_k \beta^{(i)}_{k j'} | v_{i,k} \rangle
\end{equation}
it is easy to see that we obtain the orthonormality $\langle v_{i,j} | \widetilde{v}_{i',j'} \rangle = \delta_{j,j'}$ provided we take
$\beta^{(i)} = \left( \alpha^{(i)} \right)^{-1}$. In practice, the size of degenerate subspaces is rather small (of dimension $1,2,4$ or $8$ in the problem at hand),
and so any elementary method of producing the inverse matrix $\left( \alpha^{(i)} \right)^{-1}$ will solve the problem conveniently.

\subsubsection{Checks}

We have made extensive checks that the $A_i$ obtained from the scalar product method are identical to those
obtained from the more involved first method based on (\ref{PAeval}), provided one takes into account the linear
relations (\ref{sumrules}), and forms the sum over orthogonalised degenerated subspaces in (\ref{scalar_prod_method}).

We have verified that the scalar product method also gives the correct amplitudes of $G_{aaaa}$ in the simpler case where the states of $T$
are constrained to have only one marked value $a$. It seems likely---although we have not actually tried this---that it will also extend to
the most general $G_{a_1,a_2,a_3,a_4}$ provided the number of marked values is (at least) equal to the number of different symbols $a_k$
in the correlation fucntion.

\subsection{Spectrum of $T$}
\label{sec:appA3}

At many occasions throughout this work we need to obtain the eigenvalues $\Lambda_i$ of the transfer matrix. This is obviously a much easier problem
than obtaining the correlation functions, and has been discussed in many places, so we can be rather brief.

\subsubsection{First representation}

The $\Lambda_i$ are related to the asymptotic decay of the two-point functions, so in analogy with section~\ref{sec:appA1} they can be obtained
by using the representation of the join-detach algebra in the regime $t_1 < t < t_2$, where the boundary conditions $t_1 \to -\infty$ and $t_2 \to \infty$
have been pushed to the extremities of the cylinder. More precisely, we are interested in the sector with $\ell$ propagating FK clusters, so we
can build on the same representation as in section~\ref{sec:appA1}, but with states employing $\ell$ distinct marked symbols. The action of the
elementary operators ${\sf J}_i$ and ${\sf D}_i$ must ensure that $\ell$ distinct clusters propagate along the cylinder, and the rules match those of
section~\ref{sec:appA1} for $t_1 < t < t_2$:
\begin{itemize}
 \item ${\sf J}_i$ cannot join two marked clusters corresponding to different labels.
 \item ${\sf D}_i$ cannot leave behind any marked cluster.
 \item The different labels are treated as indistinguishable.
\end{itemize}
The last rule means that if the first site in the cluster with mark $\bar{k}$ is denoted $i_k$,
we quotient the set of states in order to impose $i_1 < i_2 < \ldots < i_\ell$.

We have checked that the spectrum of this transfer matrix coincides with that of the loop model (\ref{Z_loop}) in the affine TL representations (\ref{modules}),
where as usual $j = \ell$ for $j \ge 2$, while the first and the third terms in the direct sum correspond to $\ell=0$ and $\ell=1$ respectively.

\subsubsection{Second representation}

It is also of interest to express $T$ in the representation of section~\ref{sec:appA2}. There are now precisely two different marks, and the rules read:
\begin{itemize}
 \item ${\sf J}_i$ cannot join two marked clusters corresponding to different labels.
 \item ${\sf D}_i$ can leave behind a marked cluster (with weight 1).
 \item The different labels are treated as distinguishable.
\end{itemize}
We find that the spectrum of this transfer matrix reproduces precisely (\ref{modules}), 
but only for $j=0,2,4,\ldots$, including the structure $\bAStTL{0}{\q^{\pm 2}}\oplus~\AStTL{0}{-1}$ with two direct summands for $j=0$
and the absence of (at least the first few) odd terms ($j=1,3$).
At present it is not obvious to us what is the precise decomposition of this representation in terms of simple affine TL modules,
but we take the observations just mentioned as a first sign that our main result on the $s$-channel
of four-point functions might have a natural interpretation within this representation of the join-detach algebra.
We therefore suggest that it might be
worthwhile to study more precisely the algebra obtained from generators ${\cal O}_a(\sigma_i)$, ${\sf J}_i$ and ${\cal D}_i$, which contains the
usual join-detach algebra as a sub-algebra.

\subsection{Momentum sectors of $T$}
\label{sec:momentum_sectors}

As discussed in section~\ref{sec:num_cont_limit}, it is important to be able to associate the eigenvalues $\Lambda_i$ of $T$ with the
conformal properties that emerge in the continuum limit. To this end we shall need to attach to each $\Lambda_i$ more information than
just the number of propagating FK clusters that it corresponds to.

It is most practical for our purposes to consider here the loop model, in its representation of standard modules
$\AStTL{j}{z^2=\mathrm{e}^{2 iK}}$ within the affine TL algebra $\ATL{N}(n)$.
We denote in this section by $T_{N,2 \ell}$ the corresponding transfer matrix on link patterns with $N=2L$ strands
(recall that $L$ is the number of Potts spins in a row) and $2 \ell$ through-lines.
Each through-line picks up a factor $z$ (resp.\ $z^{-1}$) when it traverses the
periodic boundary condition towards the right (resp.\ left).

Our goal is to make apparant two more quantum numbers: the momentum $k$ of the through-lines, and the lattice momentum $m$.
The former comes directly from the quantisation of $z$, and the latter is equal, in finite size, to the conformal spin: $m = h - \bar{h}$.
We shall obtain each momentum sector by transforming $T_{N,2\ell}$ into an appropriate matrix of smaller dimension.

To this end we proceed as follows. The dimension of $W_{j,z}$ is ${N \choose N/2-j}$.
Within this space we choose one representative state
for each orbit of the cyclic group $C_L$ generated by $u^2$, where $u$ is the affine TL shift operator. Note that $[T_{N,2\ell},u^2] = 0$
for the Potts-model transfer matrix. The orbit length corresponding to a representative state $|s\rangle$ is denoted $g_s$, so the
orbit can be written
\begin{equation}
 |s\rangle, u^2 |s\rangle, u^4 |s\rangle, \ldots, u^{2(g_s-1)} |s\rangle \,.
\end{equation}
Obviously $g_s$ is a divisor of $L$, and the weighted sum over representative states (with each $|s\rangle$ being weighted by its orbit length $g_s$)
equals $\mbox{dim}\, W_{j,z}$.

\begin{example}
For $L=3$ and $\ell=0$ the dimension of $W_{j,z}$ is ${6 \choose 3} = 20$ (we do not need to take the quotient yet),
and there are 8 distinct orbits (namely 6 with $g_s = 3$, and 2 with $g_s = 1$). For $L = 8$ and $\ell=4$ the dimension of $W_{j,z}$ is ${16 \choose 4} = 1820$,
and there are 224 orbits with $g_s = 8$, 6 orbits with $g_s = 4$, and 2 orbits with $g_s = 2$.
\end{example}

The motion within orbits gives rise to another momentum variable, distinct from the twist $z$ of the affine TL algebra.
Recall that the twist variable is $z = {\rm e}^{i \pi k/\ell}$ for each through-line traversing the seam towards the right, and since the total phase factor for
a turn of all $2 \ell$ through-lines must be $z^{2 \ell} = 1$ we have $k = 0,1,\ldots,\ell-1$.
Similarly we now wish to construct the sector of lattice momentum $\omega_m \equiv (\omega)^m$, where $\omega \equiv {\rm e}^{2 \pi i / L}$,
by attributing a weight $\omega_m$ to each translation of one Potts spin (hence two TL loop strands) within the orbits of $C_L$. Since the total
phase factor for a rotation through $L$ spins must be $(\omega_m)^L = 1$ we have $m = 0,1,\ldots,L-1$. We stress that $k$ is related to the
number of FK clusters $\ell$, whereas $m$ is related to the system size $L$. 


We now wish to construct a restriction $T_{N,2\ell,m}$ to given values of the momentum labels $k \in \mathbb{Z}_\ell$
and $m \in \mathbb{Z}_L$. This means that the spectrum of $T_{N,2\ell}(n,z)$ must be decomposed as the union of the
spectra of $T_{N,2\ell,m}$, including multiplicities:
\begin{equation}
 \mbox{spec} \, T_{N,2 \ell}(n,z) = \bigcup_{m=0}^{L-1} \mbox{spec} \, T_{N,2 \ell,m}(n,z,\omega_m) \,.
 \label{decomp-spectrum-m}
\end{equation}
It is practical for us to denote the set of eigenvalues of $T_{N,2\ell,m}(n,z,\omega_m)$, with specified values of
the momenta $z = {\rm e}^{i \pi k/\ell}$ and $\omega_m = {\rm e}^{2 i \pi m/L}$, simply as $V_{\ell,k,m}$.
We denote the revelant dimensions as $D_\ell = \mbox{dim}\, W_{j,z}$ and $d_{\ell,k,m} = \mbox{dim}\, V_{\ell,k,m}$,
where $d_{\ell,k,m}$ will be determined below. We should of course have
\begin{equation}
 D_\ell = \sum_{m=0}^{L-1} d_{\ell,k,m}
 \label{sumrule-D-d}
\end{equation}
for any $\ell=0,1,\ldots,L$ and $k \in \mathbb{Z}_\ell$, in accordance with (\ref{decomp-spectrum-m}).


Our approach is to construct, for each $m$, a $D_\ell \times d_{\ell,k,m}$ matrix $S_\text{in}$, and a $d_{\ell,k,m} \times D_\ell$ matrix $S_\text{out}(z,\omega)$, such that
\begin{equation}
 T_{N,2\ell,m}(n,z,\omega_m) = S_\text{out}(z,\omega) T_{N,2\ell}(n,z) S_\text{in} \,. \label{Sout-Sin}
\end{equation}
In the matrix $S_\text{in}$, each state $|s\rangle$ in the restricted space is mapped to the corresponding representative state in the full space,
with a Boltzmann weight equal to $g_s$. In other words, if the representative state $|s\rangle$ is ordered as the $j$'th basis state in the restricted space
and as the $i$'th basis state in the full space, then $(S_\text{in})_{ij} = g_s$. All other matrix elements are zero.

In the matrix $S_\text{out}(z,\omega)$, each state $|t\rangle$ in the full space is identified as $|t\rangle = u^{2 k} |s\rangle$, where $|s\rangle$ is the representative state
corresponding to $|t\rangle$ (i.e., with $|s\rangle$ living in the restricted space) and $k$ the number of double shifts (by convention, towards the left)
necessary to bring $|t\rangle$ into the representative form. Notice that under these shifts, it is possible that a number of through-lines $p$ will
cross the periodic boundary condition (towards the left). We must keep in mind that the convention for $\ATL{N}(n)$ is that each through-line that crosses
the periodic boundary condition towards the right (resp.\ left) acquires the weight $z$ (resp.\ $z^{-1}$). Therefore, the
Boltzmann associated with bringing $|t\rangle$ into the representative form $|s\rangle$ is $(\omega_m)^k z^{-p}/g_s$.
Thus, if $|t\rangle$ is ordered as the $j$'th basis state in the full space and its
representative $|s\rangle$ is the $i$'th basis state in the restricted space, then
\begin{equation}
 (S_\text{out}(z,\omega))_{ij} = \frac{(\omega_m)^k}{g_s z^p}= \frac{\omega^{m \cdot k}}{g_s z^p} \,.
\end{equation}
All other matrix elements are zero.

A crucial point is that the orbit lengths must be compatible with the momentum that we impose. To be precise, the restricted states
corresponding to given values of the labels $k,m$ are those with orbit lengths $g_s$ satisfying
\begin{equation}
 (k-m) g_s = 0 \mbox{ mod } L \,. \label{orbit-length-criterion}
\end{equation}
The dimension $d_{\ell,k,m}$ is precisely determined by the number of restricted states satisfying this constraint.

\begin{example}
Let us consider again the case $L=8$ and $\ell=4$. Any $V_{\ell,k,m}$ with $k-m \in 4 \mathbb{Z}$ comprises all $224+6+2=232$
representative states (corresponding to $g_s = 8, 4, 2$), so that the corresponding $d_{\ell,k,m} = 232$.
When $k-m \in 4 \mathbb{Z}+2$ only the states with $g_s = 8,4$ are allowed, so that $d_{\ell,k,m} = 230$.
And finally, when $k-m \in 2 \mathbb{Z}+1$, only the states with $g_s = 8$ are allowed, so that $d_{\ell,k,m} = 224$.
To obtain the sumrule (\ref{sumrule-D-d}) we notice that for any $k \in \mathbb{Z}_\ell$, the eight values of $m \in \mathbb{Z}_L$
corresponds to $8$ cases where $g_s = 8$ is allowed, $4$ cases where $g_s = 4$ is also allowed, and $2$ cases where $g_s = 2$
is also allowed, i.e., we recognise the values of the $g_s$ themselves in this count. Thus
$8 \times 224 + 6 \times 4 + 2 \times 2 = 1820$, as it should. By the same reasoning, (\ref{sumrule-D-d}) is verified for arbitrary
values of $L$ and $\ell$.
\end{example}

We have written an algorithm that produces the basis change matrices $S_\text{in}$ and $S_\text{out}(z,\omega)$ very efficiently. The crux is obviously to
deal with the ordering of the states and the identification of the relevant orbits.
We stress that the construction with basis change matrices $S_\text{in}$ and $S_\text{out}(z,\omega)$ is perfectly compatible with
iterative diagonalisation schemes (such as the Arnoldi method), which are particularly efficient for dealing with transfer matrices that allow for
a sparse matrix factorisation.

Finally we note that the spaces $V_{\ell,k,m}$ are invariant upon simultaneously changing
the signs of both momenta (modulo $\ell$ and $L$, respectively). Thus
\begin{equation}
 V_{\ell,k,m} = V_{\ell,\ell-k,L-m} \,.
 \label{V_symmetry}
\end{equation}
We have checked this exhaustively for $L=5$ and $L=6$.

\subsubsection{Checks}

To check the algorithm, we have performed extensive tests on the case $N = 2L = 10$.
We have first checked that $[u^2,T_{N,2\ell}(n,z)] = 0$ for $\ell=0,1,\ldots,L$
and any values of $n = \sqrt{Q}$ and $z$.

The sector decomposition was checked in details for a generic value $Q=\frac{3}{2}$. Our general result is that the union of $V_{\ell,k,m}$
where $k=0,1,\ldots,\ell-1$ and $m=0,1,\ldots,L-1$ produces, for each $\ell$, precisely the spectrum of the FK transfer matrix 
$T_{N,2\ell}$ with the correct multiplicities for degenerate eigenvalues.
The advantage of this construction is thus twofold: it suffices to diagonalise much smaller matrices, and at the same
time we can associate the quantum numbers $(\ell,k,m)$ with each eigenvalue.

\begin{itemize}
 \item For $\ell = 0$ we formally set $k=0$. Without taking the $\bar{\cal W}_{j,z^2}$ quoitient, we have
 compared the 42 eigenvalues obtained from numerically diagonalising the FK transfer matrix with those obtained
 from the momentum sector decomposition of the full $\ATL{N}(n)$ module. This allows us to assign the correct momentum label to each
 FK eigenvalue. We find that $V_{0,0,0}$ has $10$ eigenvalues, while $V_{0,0,1}$ and $V_{0,0,2}$ each have $8$. Taking into account
 the $\pm m$ degeneracies, this gives all the required $10 + 2 \times(8 + 8) = 42$ eigenvalues indeed.
 \item For $\ell = 1$ we also formally set $k=0$. We find that $V_{1,0,0}$ has $52$ eigenvalues, but only $21$ distinct eigenvalues;
 of these $5$ are fourfold degenerate and the remaining $16$ are twofold degenerate. Each of $V_{1,0,1}$ and $V_{1,0,2}$ contains
 $50$ eigenvalues, but only $25$ are distinct; each of these are twofold degenerate. This accounts for the
 required $52 + 2 \times(50 + 50) = 252$ eigenvalues of the affine TL module.
 The overall twofold degeneracy is explained by the fact, that since there are no winding loops (recall $d=0$) and the square lattice is
 selfdual, the $\ATL{N}(n)$ module in fact decomposes in two isomorphic representations each of dimension $\frac12 {2L \choose L} = 126$.
 However, this still leaves $5$ degenerate eigenvalues in $V_{1,0,0}$. These degeneracies are
 due to the choice of the square lattice, which is not only symmetric under rotations but also under reflections (i.e., the symmetry
 is the dihedral group $D_5$, which is larger than the cyclic group $C_5$).
 \item For $\ell=2$, we have $V_{2,k,m}$ with $k=0,1$ and $m=0,\pm 1,\pm 2$. Each of these contains $24$ eigenvalues, which accounts
 for all $2 \times 5 \times 24 = 240 = 2 \times {10 \choose 3}$ eigenvalues in $\ATL{N}(n)$.
 However, it turns out that $V_{2,0,0}$ (resp.\ $V_{2,1,0}$) has only $18$ (resp.\ $12$) distinct eigenvalues.
 These degeneracies occur only in the sectors with vanishing lattice momentum ($m=0$), which is indeed compatible with the above
 remark about dihedral symmetry.
 \item For $\ell=3$, we have $V_{3,k,m}$ with $k=0,1,2$ and $m=0,\pm 1,\pm 2$. Each of these contains $9$ eigenvalues, which accounts
 for all $3 \times 5 \times 9 = 135 = 3 \times {10 \choose 2}$ eigenvalues in $\ATL{N}(n)$.
 In this case $V_{3,0,0}$ has only $7$ distinct eigenvalues. Moreover, we observe that $V_{3,1,0}$ is equal to $V_{3,2,0}$, which is
 compatible with the general symmetry (\ref{V_symmetry}).
 \item For $\ell=4$, we have $V_{4,k,m}$ with $k=0,1,2,3$ and $m=0,\pm1,\pm2$. Each of these contains $2$ eigenvalues, which accounts
 for all $4 \times 5 \times 2 = 40 = 4 \times {10 \choose 1}$ eigenvalues in $\ATL{N}(n)$.
 In this case $V_{4,2,0}$ has only $1$ distinct eigenvalue.
 Moreover, we observe that $V_{4,1,0}$ is equal to $V_{4,3,0}$, again in agreement with (\ref{V_symmetry}).
 \item The case $\ell=5$ is obviously very degenerate, as will generally be the case when $\ell=L$.
\end{itemize}

\subsection{Spectrum of $T$ and the CFT limit}
\label{sec:specT_CFTlimit}

We wish to verify the CFT interpretation of the spectrum of the FK model transfer matrix $T$ in the sectors $V_{\ell,k,m}$.
As above, $\ell$ denotes the number of propagating clusters, $k$ is the twist parameter for the affine TL through-lines
(with allowed values $k=0$ for $\ell = 0$, and $k \in \mathbb{Z}_\ell$ for $\ell \ge 1$), and $m$ is the lattice momentum
(with allowed values $m \in \mathbb{Z}_L$ for a periodic strip of width $L$ Potts spins). Moreover we have the symmetry (\ref{V_symmetry}).

As we shall see, the conformal interpretation of the labels $k,m$ is as follows.
The twist label $k$ fixes the first Kac label in $\phi_{r,s}$ to be $r = k/\ell + e$, where $e$ is an integer.
The momentum label $m$ detemines the conformal spin, $s = h - \bar{h} = m \mbox{ mod } L$.
Below we give detailed evidence that if the true value of the conformal spin is too large to be accommodated in a given size $L$
(that is, $|h-\bar{h}| > L/2$), there will in general be an appropriate eigenvalue with $m = (h - \bar{h}) \mbox{ mod } L$
that nicely fits into the finite-size scaling formulae.

In the sequel we consider strips of widths $L \ge 5$.
For technical reasons, it is feasible to diagonalise $T$ with respect to all quantum numbers $(\ell,k,m)$
only for $L \le 7$, but such sizes are insufficient to numerically determine the scaling dimensions $h + \bar{b}$
for a significant number of low-lying excitations. However, we are able diagonalise $T$ in each sector $V_{\ell} \equiv \sum_{k,m} V_{\ell,k,m}$
for higher values of $L$, by simply imposing the number of through-lines (and for $\ell = 1$, also setting to zero the fugacity of non-contractible loops),
without having to compute the basis change matrices $S_{\rm in}$ and $S_{\rm out}$
appearing in (\ref{Sout-Sin}). Concretely, we have done so for $\ell = 0$ and $L \le 14$, for $\ell = 1$ and $L \le 13$, and for $\ell = 2$ and $L \le 12$.

For a fixed value of $\ell$, the eigenvalues are labelled $\Lambda_\ell^{(i)}(L)$, where we have supposed the ordering $\Lambda_\ell^{(1)} > \Lambda_\ell^{(2)} > \cdots$.
Note in particular that we disregard multiplicities (the eigenvalues with $m \neq 0$ are two-fold degenerate, as are those with $m=0$ and $k \neq 0$).
We shall refer to $i$ as the ``rank'' of the eigenvalue within $V_\ell$. From the eigenvalues we can compute the effective scaling dimension $x = h + \bar{h}$ via
\begin{equation}
 x_{\ell}^{(i)}(L) = - \frac{L}{2 \pi} \log \left( \frac{\Lambda_{\ell}^{(i)}(L)}{\Lambda_0^{(1)}(L)} \right) \,. \label{eff_scaling_dim}
\end{equation}

The practical problem is that the rank $i = i_L$ of a given scaling level in $V_{\ell,k,m}$ will depend on $L$. We would of course expect $i$ to stabilise at some constant value $i_\infty$
for $L$ sufficiently large, for the level that becomes the $i_\infty$'st lowest-lying excitation in the conformal limit. The trouble is that for all but the few lowest excitations, $i_\infty$
turns out to be comparable to or larger than the values of $L$ that can be accessed numerically. As explained above, we however know the values of $i_L$ that are compatible with given
$(k,m)$ for $L \le 7$. Fortunately, by plotting $x_{\ell}^{(i)}(L)$ against $1/L$, and fitting the available sizes to a polynomial in $1/L$, we can gradually reconstruct the sequence
$i_L$ by adding one size at a time, by carefully monitoring the quality of the fit and checking whether the extrapolation
\begin{equation}
 x_{\ell}^{(i_\infty)} = \lim_{L \to \infty} x_{\ell}^{(i)}(L)
\end{equation}
produces a reasonable value. In this way we obtain excellent fits that extend in general from $L_{\rm min} = {\rm max}(5,2m)$ to the largest accessible size $L_{\rm max}$.
The outcome of this procedure can be verified by checking that, for any given $L$, each value of $i$ is attributed to one and only one scaling level. Moreover, the extrapolated
scaling dimension $x_{\ell}^{(i_\infty)}$ should make sense in the CFT, being in particular compatible with the values of $(k,m)$ that we have thus inferred.

For $m > 3$ we cannot proceed in this way, because $i_L$ is only defined for $L \ge L_{\rm min} = 2m$ and we only know the sector labels for $L \le 7$. In this case, we can usually work the
other way around, either guessing the possible values of $i_L$ from those which are not used by other levels, or by carefully monitoring the fits starting at the highest values of
$L$ (where $i_L$ can be supposed to be constant, or almost constant) and gradually proceeding to include also lower sizes and in the same time determining the corresponding $i_L$.

\subsubsection{A generic case $Q=\frac12$}
\label{sec:specQ12}

In Table~\ref{Tab_exp_V0} we report the outcome of these computations within $V_0$, where we have taken an arbitrary and generic value $Q = \frac12$. We show the sequence $(i_5,i_6,\ldots)$ for each
level, the corresponding representation $V_{\ell,k,m}$, and the numerically extrapolated scaling dimension $x_\ell \equiv x_\ell^{(i_\infty)}$. Comparing this to the possible
analytical values%
\footnote{The analytical value is obviously an exact number, but in cases where it is not an integer the Table only shows its decimal representation, truncated to the number of digits to
which it agrees with the numerical result. This allows the reader to quickly assess the accuracy of the numerical work.}
allows us to identify the corresponding scaling field in the CFT limit. In the latter we use the notation $L_{-n} L_{-\bar{n}}$ to denote any highest-weight
representation at level $(n,\bar{n})$ in the holomorphic (resp.\ antiholomorphic) Verma module. For instance, $L_{-2}$ denotes here what is usually called either $L_{-2}$ or $(L_{-1})^2$, or a linear combination of those terms.
In a generic module we should then expect a multiplicity corresponding to
$p(n) \times p(\bar{n})$, where $p(n)$ denotes the number of integer partition of $n$. This count can however be smaller due to the presence of null vectors.

\begin{table}
\begin{center}
\begin{tabular}{|c|cccccccccc|ll|l|}
\cline{1-14}
  $V_{\ell,k,m}$
    & \multicolumn{10}{|c|}{$(i_5,i_6,\ldots,i_{14})$} & \multicolumn{2}{|c|}{$x_\ell$} & Identification \\  \cline{2-13}
   & 5 & 6 & 7 & 8 & 9 & 10 & 11 & 12 & 13 & 14 & Numerics & Exact & of scaling field \\
  \hline
 $V_{000}$ & 1 & 1 & 1 & 1 & 1 & 1 & 1 & 1 & 1 & 1 & 0              & 0              & $\phi_{1,1} \times \phi_{1,1} \equiv I$ \\
 $V_{000}$ & 2 & 2 & 2 & 2 & 2 & 2 & 2 & 2 & 2 & 2 & 1.438925 & 1.438918 & $\phi_{2,1} \times \phi_{2,1} \equiv \varepsilon$ \\
 $V_{002}$ & 3 & 3 & 3 & 3 & 3 & 3 & 3 & 3 & 3 & 3 & 2.00007   & 2              & $L_{-2} I$ \\
 $V_{001}$ & 4 & 4 & 4 & 4 & 4 & 4 & 4 & 4 & 4 & 4 & 2.4394     & 2.4389     & $L_{-1} \varepsilon$ \\
 $V_{003}$ & -- & 5 & 5 & 5 & 5 & 5 & 5 & 5 & 5 & 5 & 3.0009    & 3              & $L_{-3} I$ \\ \hline
 $V_{002}$ & 5 & 6 & 6 & 6 & 6 & 6 & 6 & 6 & 6 & 6 & 3.434      & 3.439       & $L_{-2} \varepsilon$ \\
 $V_{000}$ & 6 & 7 & 8 & 9 & 8 & 7 & 7 & 7 & 7 & 7 & 3.4391    & 3.4389     & $L_{-1} \bar{L}_{-1} \varepsilon$ \\
 $V_{004}$ & -- & -- & -- & 7 & 7 & 8 & 8 & 8 & 8 & 8 & 4.01       & 4              & $L_{-4} I$ \\
 $V_{003}$ & -- & -- & 7 & 8 & 9 & 9 & 9 & 9 & 9 & 10 & 4.41     & 4.44         & $L_{-3} \varepsilon$ \\
 $V_{004}$ & -- & -- & -- & 10 & 11 & 12 & 10 & 10 & 10 & 9 & 3.99986 & 4  & $L_{-4} I$ \\ \hline
 $V_{000}$ & 8 & 11 & 12 & 12 & 13 & 13 & 13 & 12 & 11 & 11 & 4.0008 & 4 & $L_{-2} \bar{L}_{-2} I$ \\
 $V_{005}$ & -- & -- & -- & -- & -- & 10 & 11 & 11 & 12 & 13 & 4.988 & 5 & $L_{-5} I$ \\
 $V_{001}$ & 7 & 9 & 10 & 11 & 12 & 14 & 15 & 14 & 13 & 12 & 4.4381 & 4.439 & $L_{-2} \bar{L}_{-1} \varepsilon$ \\
 $V_{004}$ & -- & -- & -- & 7 & 10 & 11 & 12 & 13 & 14 & 14 & 5.40 & 5.44 & $L_{-4} \varepsilon$ \\
 $V_{003}$ & -- & -- & 14 & 15 & 17 & 16 & 16 & 17 & 16 & 15 & 4.4397 & 4.4389 & $L_{-3} \varepsilon$ \\ \hline
 $V_{000}$ & 11 & 14 & 16 & 17 & 18 & 18 & 18 & 19 & 18 & 16 & 4.50369 & 4.50378 & $\phi_{3,1} \times \phi_{3,1} \equiv \varepsilon'$ \\
 $V_{005}$ & -- & -- & -- & -- & -- & 15 & 17 & 18 & 20 & 18 & 4.9983 & 5 & $L_{-5} I$ \\
 $V_{001}$ & 10 & 13 & 17 & 18 & 20 & 22 & 21 & 21 & 22 & 23 & 4.998 & 5 & $L_{-3} \bar{L}_{-2} I$ \\
 $V_{000}$ & 12 & 18 & 23 & 28 & 32 & 33 & 33 & 34 & 36 & ? & 5.984 & 6 & $L_{-3} \bar{L}_{-3} I$ \\
  \hline
\end{tabular}
\end{center}
\caption{Conformal spectrum in the sector $V_0$, for $Q=\frac12$}
  \label{Tab_exp_V0}
\end{table}

Since the method just outlined for identifying the exponents may be of more general interest, we now give a detailed example of its application.

\begin{figure}[htp] \centering{
\includegraphics[scale=0.6]{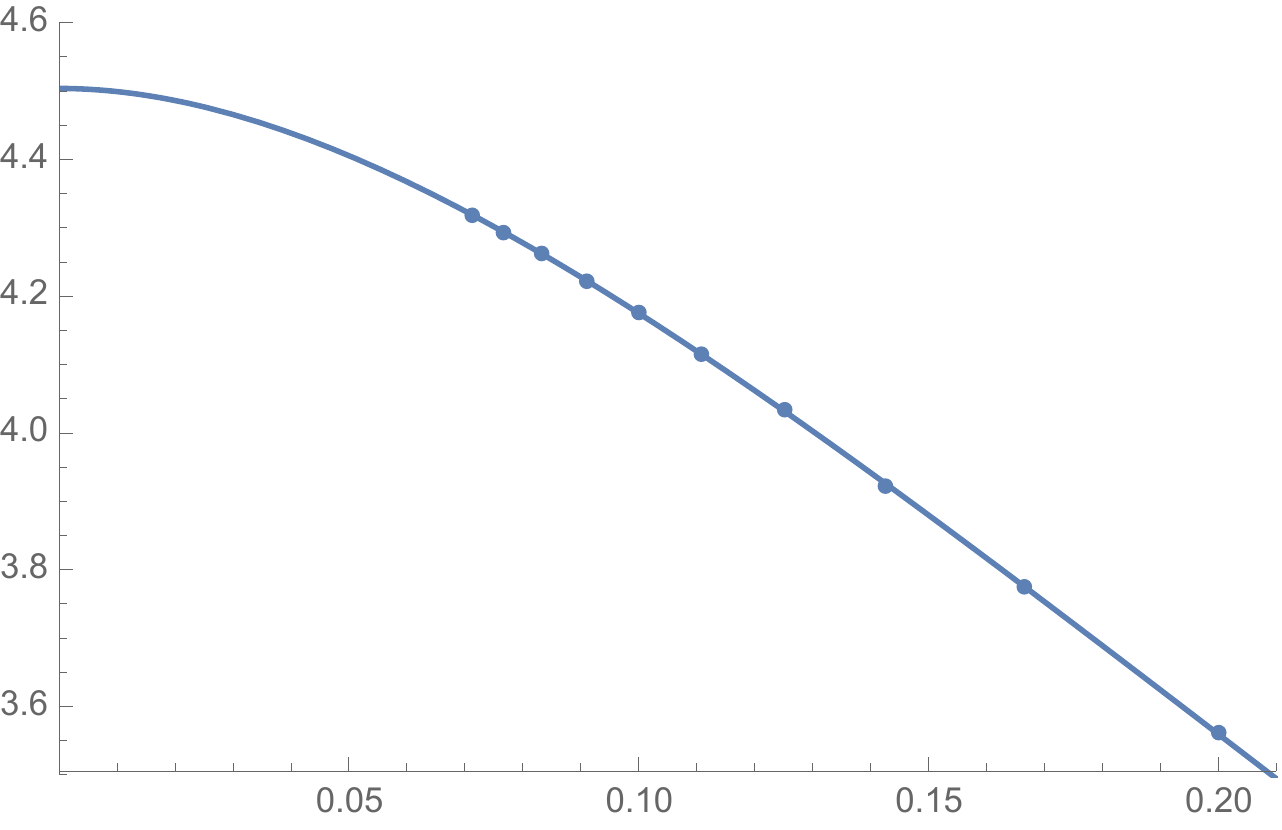}}
\caption{Effective scaling dimension, plotted against $1/L$, of the eigenvalues in $V_{0}$ (for $Q=\frac12$)
 that correspond to the CFT scaling level $\phi_{3,1} \times \phi_{3,1}$, along with a polynomial extrapolation of order $9$.}
\label{fig:phi31}
\end{figure}

\begin{example}

We focus on the scaling level which corresponds to the eigenvalue of rank $i_{14} = 16$ for $L=14$, i.e., line 16 in Table~\ref{Tab_exp_V0}.
Assume that the first 15 lines of the table have already been determined. We then try to determine $i_{13}$, which a priori can be any of
the values not used in the first 15 lines: $i_{13} = 15, 17, 18, 19,\ldots$. To this end, plot the effective scaling dimensions (\ref{eff_scaling_dim})
for $L=13, 14$ against $1/L$ along with a linear fit in $1/L$, for all possible values of $i_{13}$. It is quickly seen that only the choice $i_{13} = 16$
will lead to an exponent of the expected value $x_\ell \approx 5$.  Fixing this choice, we next seek to determine $i_{12}$, which can take
any of the unused values $i_{12} = 15,16,18,19,\ldots$. Plot now the effective scaling dimensions for $L=12,13,14$ against $1/L$ along with
a quadratic fit in $1/L$. This fit clearly singles out $i_{12} = 19$ and narrows down the value of the exponent to $x_\ell \approx 4.5$. Going on
in this way, including one by one the smaller sizes, and increasing the order of the fits, one determines all the ranks $i_L$ in line 16 of the table,
down to the smallest size $L_{\rm min} = 5$. The final fit, using all ten sizes $L=L_{\rm min},\ldots,L_{\rm max}$ and a polynomial extrapolation
of order $9$ is shown in Figure~\ref{fig:phi31}. It leads to a very precise estimate $x_\ell \approx 4.50369$, which is in excellent
agreement with the exact scaling dimension $4.50378\cdots$ of the field $\phi_{3,1} \times \phi_{3,1}$. Moreover, we can check that for the three
smallest sizes, the eigenvalues with $(i_5,i_6,i_7) = (11,14,16)$ all have lattice momentum $m=0$, so the CFT field must indeed have
conformal spin $h - \bar{h} = 0$.

\end{example}

The contents of Table~\ref{Tab_exp_V0} can be quickly summarised by saying that we have numerically observed within $V_0$ three different primaries,
\begin{equation}
 \phi_{1,1} \times \phi_{1,1} \equiv I \,, \qquad
 \phi_{2,1} \times \phi_{2,1} \equiv \varepsilon \,, \qquad
 \phi_{3,1} \times \phi_{3,1} \equiv \varepsilon' \,,
\end{equation}
along with their corresponding {\em Kac modules} (where the generic Verma module of $\phi_{r,s}$ has been quotiented by one singular vector
at level $r s$). More precisely:
\begin{itemize}
 \item For the identity operator $I$ we have observed all descendents up to and including the total level
 $N \equiv n + \bar{n} = 5$, and some of the descendents at level $6$. This operator is seen to be degenerate at level 1.
 \item For the energy operator $\varepsilon$ we have observed all descendents up to and including level $N=3$, and some
 of the descendents at level $4$. This operator is degenerate at level 2, as witnessed by the fact that we only observe
 a single state (and not $p(2)=2$ states) at level $(n,\bar{n}) = (2,0)$.
 \item For the second energy operator $\varepsilon'$ we have observed only the primary at level $N=0$, due to the high scaling
 dimension $x = 4.503782\cdots$.
\end{itemize}

Table~\ref{Tab_exp_V1} gives the results of a similar investigation for the sector $V_1$. We here observe numerically two different primaries,
\begin{equation}
 \phi_{1/2,0} \times \phi_{1/2,0} \equiv \sigma \,, \qquad
 \phi_{3/2,0} \times \phi_{3/2,0} \equiv \sigma' \,,
\end{equation}
along with their corresponding {\em Verma modules}.
More precisely:
\begin{itemize}
 \item For the magnetisation operator $\sigma$ we have observed all descendents up to and including the total level
 $N = 3$, and some of the descendents at level $4$.
 \item For the second magnetisation operator $\sigma'$ we have observed all descendents up to and including level $N=2$.
 \item We have not observed the expected third magnetisation operator, $\phi_{5/2,0} \times \phi_{5/2,0}$, due to its high
 scaling dimension $x = 4.960593\cdots$.
\end{itemize}

\begin{remark}

We stress that no degenerate states appear in these magnetisation operators, and accordingly their Verma modules are generic.
This is in a nutshell why the determination of four-point functions in the bulk case is so difficult: the absence of degenerate states
implies that we cannot write down differential equations satisfied by the correlation functions.
 
\end{remark}

\begin{table}
\begin{center}
\begin{tabular}{|c|ccccccccc|ll|l|}
\cline{1-13}
  $V_{\ell,k,m}$
    & \multicolumn{9}{|c|}{$(i_5,i_6,\ldots,i_{13})$} & \multicolumn{2}{|c|}{$x_\ell$} & Identification \\  \cline{2-12}
   & 5 & 6 & 7 & 8 & 9 & 10 & 11 & 12 & 13 & Numerics & Exact & of scaling field \\
  \hline
 $V_{100}$ & 1 & 1 & 1 & 1 & 1 & 1 & 1 & 1 & 1 & 0.082745 & 0.082757 & $\phi_{1/2,0} \times \phi_{1/2,0} \equiv \sigma$ \\
 $V_{101}$ & 2 & 2 & 2 & 2 & 2 & 2 & 2 & 2 & 2 & 1.08278 & 1.08276 & $L_{-1} \sigma$ \\
 $V_{100}$ & 4 & 3 & 3 & 3 & 3 & 3 & 3 & 3 & 3 & 1.7091 & 1.7087 & $\phi_{3/2,0} \times \phi_{3/2,0} \equiv \sigma'$ \\ 
 $V_{102}$ & 3 & 4 & 4 & 4 & 4 & 4 & 4 & 4 & 4 & 2.0819 & 2.08276 & $L_{-2} \sigma$ \\
 $V_{102}$ & 5 & 6 & 5 & 5 & 5 & 5 & 5 & 5 & 5 & 2.0830 & 2.0828 & $L_{-2} \sigma$ \\ \hline
 $V_{100}$ & 6 & 7 & 6 & 6 & 6 & 6 & 6 & 6 & 6 & 2.08255 & 2.08276 & $L_{-1} \bar{L}_{-1} \sigma$ \\
 $V_{101}$ & 7 & 8 & 8 & 9 & 8 & 8 & 7 & 7 & 7 & 2.704 & 2.079 & $L_{-1} \sigma'$ \\
 $V_{103}$ & -- & 5 & 7 & 7 & 7 & 7 & 8 & 8 & 8 & 3.094 & 3.083 & $L_{-3} \sigma$ \\
 $V_{103}$ & -- & 9 & 9 & 10 & 10 & 9 & 9 & 9 & 9 & 3.072 & 3.083 & $L_{-3} \sigma$ \\
 $V_{101}$ & 9 & 11 & 10 & 11 & 11 & 11 & 10 & 10 & 10 & 3.080 & 3.083 & $L_{-2} \bar{L}_{-1} \sigma$ \\ \hline
 $V_{103}$ & -- & 12 & 11 & 12 & 12 & 12 &11 & 11 & 11 & 3.075 & 3.083 & $L_{-3} \sigma$ \\
 $V_{101}$ & 12 & 14 & 14 & 14 & 13 & 14 & 13 & 12 & 12 & 3.081 & 3.083 & $L_{-2} \bar{L}_{-1} \sigma$ \\
 $V_{104}$ & -- & -- & -- & 8 & 9 & 10 & 12 & 13 & 13 & 4.078 & 4.083 & $L_{-4} \sigma$ \\
 $V_{102}$ & 8 & 10 & 12 & 13 & 14 & 15 & 15 & 14 & 14 & 3.712 & 3.709 & $L_{-2} \sigma'$ \\
 $V_{100}$ & 11 & 15 & 17 & 17 & 16 & 17 & 16 & 16 & 15 & 3.69 & 3.71 & $L_{-1} \bar{L}_{-1} \sigma'$ \\ \hline
 $V_{102}$ & 15 & 17 & 18 & 20 & 21 & 18 & 18 & 19 & 16 & 3.69 & 3.71 & $L_{-2} \sigma'$ \\
 $V_{102}$ & 10 & 13 & 15 & 18 & 18 & 19 & 19 & 20 & 19 & 4.070 & 4.082 & $L_{-3} \bar{L}_{-1} \sigma$ \\
 $V_{100}$ & 14 & 19 & 20 & 23 & 25 & 25 & 23 & 24 & 23 & 4.0841 & 4.0828 & $L_{-2} \bar{L}_{-2} \sigma$ \\
 $V_{102}$ & 16 & 20 & 21 & 24 & 26 & 26 & 24 & 25 & 24 & 4.075 & 4.083 & $L_{-3} \bar{L}_{-1} \sigma$ \\
 $V_{100}$ & 17 & 21 & 23 & 25 & 27 & 27 & 26 & 26 & 25 & 4.079 & 4.083 & $L_{-2} \bar{L}_{-2} \sigma$ \\
  \hline
\end{tabular}
\end{center}
\caption{Conformal spectrum in the sector $V_1$, for $Q=\frac12$.}
  \label{Tab_exp_V1}
\end{table}

Finally, Table~\ref{Tab_exp_V2} shows our results for the sector $V_2$. We see here the beginning of the conformal towers of the primaries
$\phi_{e,2} \times \phi_{e,-2}$ for $e = 0,\frac12,1,\frac32,2$. As for $V_1$, there are no singular vectors in this case neither.

\begin{table}
\begin{center}
\begin{tabular}{|c|cccccccc|ll|l|}
\cline{1-12}
  $V_{\ell,k,m}$
    & \multicolumn{8}{|c|}{$(i_5,i_6,\ldots,i_{12})$} & \multicolumn{2}{|c|}{$x_\ell$} & Identification \\  \cline{2-11}
   & 5 & 6 & 7 & 8 & 9 & 10 & 11 & 12 & Numerics & Exact & of scaling field \\
  \hline
 $V_{200}$ & 1 & 1 & 1 & 1 & 1 & 1 & 1 & 1 & 1.109570 & 1.109567 & $\phi_{0,2} \times \phi_{0,-2} \equiv \phi_0$ \\
 $V_{211}$ & 2 & 2 & 2 & 2 & 2 & 2 & 2 & 2 & 1.31286 & 1.31281 & $\phi_{1/2,2} \times \phi_{1/2,-2} \equiv \phi_{1/2}$ \\
 $V_{202}$ & 3 & 3 & 3 & 3 & 3 & 3 & 3 & 3 & 1.92257 & 1.92254 & $\phi_{1,2} \times \phi_{1,-2} \equiv \phi_1$ \\
 $V_{201}$ & 5 & 4 & 4 & 4 & 4 & 4 & 4 & 4 & 2.1099 & 2.1096 & $\bar{L}_{-1} \phi_{0}$ \\
 $V_{212}$ & 4 & 5 & 5 & 5 & 5 & 5 & 5 & 5 & 2.3117 & 2.3128 & $\bar{L}_{-1} \phi_{1/2}$ \\ \hline
 $V_{210}$ & 7 & 7 & 6 & 6 & 6 & 6 & 6 & 6 & 2.31299 & 2.31281 & $L_{-1} \phi_{1/2}$ \\
 $V_{203}$ & -- & 8 & 8 & 7 & 7 & 7 & 7 & 7 & 2.914 & 2.923 & $\bar{L}_{-1} \phi_{1}$ \\
 $V_{202}$ & 6 & 9 & 9 & 9 & 9 & 8 & 8 & 8 & 3.094 & 3.110 & $\bar{L}_{-2} \phi_{0}$ \\
 $V_{213}$ & -- &10 & 11 & 11 & 10 & 10 & 10 & 9 & 2.9365 & 2.9388 & $\phi_{3/2,2} \times \phi_{3/2,-2} \equiv \phi_{3/2}$ \\
 $V_{201}$ & 13 & 13 & 13 & 13 & 11 & 11 & 11 & 10 & 2.9228 & 2.9225 & $L_{-1} \phi_{1}$ \\ \hline
 $V_{213}$ & -- & 6 & 7 & 8 & 8 & 9 & 9 & 11 & 3.326 & 3.313 & $\bar{L}_{-2} \phi_{1/2}$ \\
 $V_{200}$ & 9 & 11 & 12 & 15 & 12 & 12 & 12 & 12 & 3.1088 & 3.1096 & $L_{-1} \bar{L}_{-1} \phi_{0}$ \\
 $V_{202}$ & 15 & 16 & 16 & 17 & 15 & 13 & 13 & 13 & 3.1098 & 3.1096 & $\bar{L}_{-2} \phi_{0}$ \\
 $V_{211}$ & 10 & 12 & 14 & 16 & 16 & 14 & 14 & 14 & 3.308 & 3.313 & $L_{-1} \bar{L}_{-1} \phi_{1/2}$ \\
 $V_{211}$ & 11 & 14 & 15 & 18 & 18 & 16 & 15 & 15 & 3.310 & 3.313 & $L_{-2} \phi_{1/2}$ \\ \hline
 $V_{213}$ & -- &15 & 17 & 19 & 19 & 18 & 16 & 16 & 3.311 & 3.313 & $\bar{L}_{-2} \phi_{1/2}$ \\
 $V_{211}$ & 17 & 21 & 20 & 20 & 21 & 21 & 18 & 17 & 3.3131 & 3.3128 & $L_{-2} \phi_{1/2}$ \\
 $V_{204}$ & -- & -- & -- & 12 & 14 & 17 & 17 & 18 & 3.907 & 3.922 & $\bar{L}_{-2} \phi_1$ \\
 $V_{204}$ & -- & -- & -- & 10 & 13 & 15 & 19 & 19 & 4.347 & 4.361 & $\phi_{2,2} \times \phi_{2,-2} \equiv \phi_{2}$ \\
 $V_{212}$ & 8 & 17 & 19 & 22 & 24 & 27 & 30 & 31 & 4.296 & 4.313 & $L_{-1} \bar{L}_{-2} \phi_{1/2}$ \\
  \hline
\end{tabular}
\end{center}
\caption{Conformal spectrum in the sector $V_2$, for $Q=\frac12$.}
  \label{Tab_exp_V2}
\end{table}

\subsubsection{The special case $Q=4$}
\label{sec:specQ4}

In Appendix~\ref{sec:appB} we show that the case $Q=4$ can be compared with an exact solution.
It is therefore a particularly important benchmark for the numerical method. Moreover, it is well known
that the $Q=4$ case is hampered by slow convergence, due to the logarithmic corrections produced by
a marginally irrelevant operator. To make sure that our generic analysis applies in this case as well,
we shall need the same kind of tables as above
with this value of $Q$. They are obtained using the same methods as before, and are shown in
Tables~\ref{Tab_exp_V0_Q4}--\ref{Tab_exp_V2_Q4}.

The precision of the extrapolated exponents $x_\ell$ suffers somewhat from the logarithmic corrections to scaling.
But assuming that the operator content of the generic case carries over, the assignment of sector labels
is nevertheless certain for the levels shown in the tables. To this end, it is particularly helpful that the
observed exponent difference between a descendent operator and its corresponding primary is determined
with considerably better precision than the exponents themselves.

This phenomenon is vividly illustrated in Table~\ref{Tab_exp_V0_Q4}. For example,
the primary $\varepsilon$ has the exact scaling dimension $\frac12$, but the numerically measured
exponent of $0.62$ is very imprecise. However, the corresponding measurements for the
descendents at level $1$, $2$ and $3$ come out as $1.64$, $2.64$ (or $2.67$) and $3.63$ (or $3.60$),
with the gaps being quite close to integers.

Similarly, the exponent of the primary $\varepsilon'$ is measured as $2.35$ instead of the exact value of $2$.
This is {\em per se} a catastrophic lack of precision---but it actually makes it easier to correctly identify the
descendent $L_{-1} \varepsilon'$, whose measured exponent comes out as $3.36$ with an almost perfect
integer gap! Without this phenomenon one could easily have mixed up the numerical value $3.36$ with the
other candidate exact value of $\frac72$.

In a similar fashion we have the Tables~\ref{Tab_exp_V1_Q4}--\ref{Tab_exp_V2_Q4} for the sectors $V_1$ and $V_2$,
respectively. 
Note that the rank of eigenvalues whole momentum labels are fixed by analytic continuation---using the PT symmetry (\ref{V_symmetry})---for
small values of $L$ are shown as {\tiny tiny} numbers in the tables. In the previous Tables~\ref{Tab_exp_V0}--\ref{Tab_exp_V2} (for $Q=\frac12$)
we have left blank such entries, although they can be determined in those cases as well.

\begin{table}
\begin{center}
\begin{tabular}{|c|ccccccccc|ll|l|}
\cline{1-13}
  $V_{\ell,k,m}$
    & \multicolumn{9}{|c|}{$(i_5,i_6,\ldots,i_{13})$} & \multicolumn{2}{|c|}{$x_\ell$} & Identification \\  \cline{2-12}
   & 5 & 6 & 7 & 8 & 9 & 10 & 11 & 12 & 13 & Numerics & Exact & of scaling field \\
  \hline
 $V_{000}$ & 1 & 1 & 1 & 1 & 1 & 1 & 1 & 1 & 1 & 0              & 0              & $\phi_{1,1} \times \phi_{1,1} \equiv I$ \\
 $V_{000}$ & 2 & 2 & 2 & 2 & 2 & 2 & 2 & 2 & 2 & 0.619 & 0.5 & $\phi_{2,1} \times \phi_{2,1} \equiv \varepsilon$ \\
 $V_{001}$ & 3 & 3 & 3 & 3 & 3 & 3 & 3 & 3 & 3 & 1.637 & 1.5 & $L_{-1} \varepsilon$ \\
 $V_{002}$ & 4 & 4 & 4 & 4 & 4 & 4 & 4 & 4 & 4 & 2.0003 & 2  & $L_{-2} I$ \\
 $V_{000}$ & 7 & 7 & 8 & 7 & 7 & 6 & 6 & 5 & 5 & 2.338 & 2    & $\phi_{3,1} \times \phi_{3,1} \equiv \varepsilon'$  \\ \hline
 $V_{002}$ & 5 & 5 & 5 & 5 & 5 & 5 & 5 & 6 & 6 & 2.646 & 2.5 & $L_{-2} \varepsilon$ \\
 $V_{003}$ & {\tiny 5} & 6 & 6 & 6 & 6 & 7 & 7 & 7 & 7 & 2.9994 & 3  & $L_{-3} I$ \\
 $V_{000}$ & 6 & 8 & 9 & 8 & 8 & 8 & 8 & 8 & 8 & 2.659 & 2.5 & $L_{-1} \bar{L}_{-1} \varepsilon$ \\
 $V_{003}$ & {\tiny 4} & 6 & 7 & 9 & 9 & 9 & 9 & 9 & 9 & 3.620 & 3.5 & $L_{-3} \varepsilon$ \\
 $V_{004}$ & {\tiny 3} & {\tiny 5} & {\tiny 7} & 10 & 10 & 10 & 10 & 10 & 10 & 3.994 & 4 & $L_{-4} I$ \\ \hline
 $V_{001}$ & 10 & 12 & 12 & 11 & 12 & 13 & 11 & 11 & 11 & 3.353 & 3 & $L_{-1} \varepsilon'$ \\
 $V_{003}$ & {\tiny 9} & 11 & 11 & 12 & 13 & 14 & 12 & 12 & 12 & 3.603 & 3.5 & $L_{-3} \varepsilon$ \\
 $V_{001}$ & 8 & 9 & 10 & 13 & 14 & 15 & 15 & 13 & 13 & 3.671 & 3.5 & $L_{-2} \bar{L}_{-1} \varepsilon$ \\
  \hline
\end{tabular}
\end{center}
\caption{Conformal spectrum in the sector $V_0$, for $Q=4$.}
  \label{Tab_exp_V0_Q4}
\end{table}

\begin{table}
\begin{center}
\begin{tabular}{|c|cccccccc|ll|l|}
\cline{1-12}
  $V_{\ell,k,m}$
    & \multicolumn{8}{|c|}{$(i_5,i_6,\ldots,i_{12})$} & \multicolumn{2}{|c|}{$x_\ell$} & Identification \\  \cline{2-11}
   & 5 & 6 & 7 & 8 & 9 & 10 & 11 & 12 & Numerics & Exact & of scaling field \\
  \hline
 $V_{100}$ & 1 & 1 & 1 & 1 & 1 & 1 & 1 & 1 & 0.133 & 0.125 & $\phi_{1/2,0} \times \phi_{1/2,0} \equiv \sigma$ \\
 $V_{101}$ & 2 & 2 & 2 & 2 & 2 & 2 & 2 & 2 & 1.138 & 1.125 & $L_{-1} \sigma$ \\
 $V_{100}$ & 3 & 3 & 3 & 3 & 3 & 3 & 3 & 3 & 1.212 & 1.125 & $\phi_{3/2,0} \times \phi_{3/2,0} \equiv \sigma'$ \\ 
 $V_{102}$ & 4 & 4 & 4 & 4 & 4 & 4 & 4 & 4 & 2.132 & 2.125 & $L_{-2} \sigma$ \\
 $V_{102}$ & 5 & 6 & 5 & 5 & 5 & 5 & 5 & 5 & 2.126 & 2.125 & $L_{-2} \sigma$ \\ \hline
 $V_{100}$ & 7 & 8 & 7 & 7 & 7 & 6 & 6 & 6 & 2.149 & 2.125 & $L_{-1} \bar{L}_{-1} \sigma$ \\
 $V_{101}$ & 6 & 7 & 6 & 6 & 6 & 7 & 7 & 7 & 2.218 & 2.125 & $L_{-1} \sigma'$ \\
 $V_{103}$ & {\tiny 4} & 5 & 8 & 8 & 8 & 8 & 8 & 8 & 3.107 & 3.125 & $L_{-3} \sigma$ \\
 $V_{103}$ & {\tiny 8} & 9 & 9 & 10 & 9 & 9 & 9 & 9 & 3.116 & 3.125 & $L_{-3} \sigma$ \\
 $V_{102}$ & 8 & 10 & 10 & 11 & 11 & 10 & 10 & 10 & 3.202 & 3.125 & $L_{-2} \sigma'$ \\ \hline
 $V_{101}$ & 12 & 13 & 14 & 12 & 12 & 12 &11 & 11 & 3.121 & 3.083 & $L_{-2} \bar{L}_{-1} \sigma$ \\
 $V_{103}$ & {\tiny 9} & 11 & 12 & 13 & 13 & 13 & 12 & 12 & 3.131 & 3.125 & $L_{-3} \sigma$ \\
 $V_{101}$ & 10 & 12 & 13 & 14 & 14 & 14 & 13 & 13 & 3.145 & 3.125 & $L_{-2} \bar{L}_{-1} \sigma$ \\
 $V_{100}$ & 11 & 14 & 15 & 17 & 15 & 16 & 15 & 14 & 3.225 & 3.125 & $L_{-1} \bar{L}_{-1} \sigma'$ \\
 $V_{102}$ & 13 & 16 & 16 & 18 & 16 & 17 & 16 & 15 & 3.236 & 3.125 & $L_{-2} \sigma'$ \\ \hline
 $V_{104}$ & {\tiny 2} & {\tiny 4} & {\tiny 8} & 9 & 10 & 11 & 14 & 16 & 4.114 & 4.125 & $L_{-4} \sigma$ \\
 $V_{100}$ & 15 & 17 & 17 & 19 & 19 & 18 & 17 & 17 & 3.37 & 3.125 & $\phi_{5/2,0} \times \phi_{5/2,0} \equiv \sigma''$ \\
 $V_{104}$ & {\tiny 6} & {\tiny 10} & {\tiny 11} & 15 & 17 & 19 & 19 & 18 & 4.07 & 4.125 & $L_{-4} \sigma$ \\
 $V_{105}$ & {\tiny 1} & {\tiny 2} & {\tiny 4} & {\tiny 8} & {\tiny 10} & 15 & 18 & 19 & 5.23 & 5.125 & $L_{-5} \sigma$ \\
 $V_{103}$ & {\tiny 5} & 9 & 11 & 16 & 18 & 20 & 20 & 20 & 4.16 & 4.125 & $L_{-3} \sigma'$ \\
  \hline
\end{tabular}
\end{center}
\caption{Conformal spectrum in the sector $V_1$, for $Q=4$.}
  \label{Tab_exp_V1_Q4}
\end{table}

\begin{table}
\begin{center}
\begin{tabular}{|c|ccccccc|ll|l|}
\cline{1-11}
  $V_{\ell,k,m}$
    & \multicolumn{7}{|c|}{$(i_5,i_6,\ldots,i_{11})$} & \multicolumn{2}{|c|}{$x_\ell$} & Identification \\  \cline{2-10}
   & 5 & 6 & 7 & 8 & 9 & 10 & 11 & Numerics & Exact & of scaling field \\
  \hline
 $V_{200}$ & 1 & 1 & 1 & 1 & 1 & 1 & 1 & 1.83 & 2 & $\phi_{0,2} \times \phi_{0,-2} \equiv \phi_0$ \\
 $V_{211}$ & 2 & 2 & 2 & 2 & 2 & 2 & 2 & 1.97 & 2.125 & $\phi_{1/2,2} \times \phi_{1/2,-2} \equiv \phi_{1/2}$ \\
 $V_{202}$ & 3 & 3 & 3 & 3 & 3 & 3 & 3 & 2.38 & 2.5 & $\phi_{1,2} \times \phi_{1,-2} \equiv \phi_1$ \\
 $V_{201}$ & 5 & 4 & 4 & 4 & 4 & 4 & 4 & 2.82 & 3 & $\bar{L}_{-1} \phi_{0}$ \\
 $V_{212}$ & 4 & 5 & 5 & 5 & 5 & 5 & 5 & 2.96 & 3.125 & $\bar{L}_{-1} \phi_{1/2}$ \\ \hline
 $V_{210}$ & 7 & 6 & 6 & 6 & 6 & 6 & 6 & 2.96 & 3.125 & $L_{-1} \phi_{1/2}$ \\
 $V_{213}$ & {\tiny 8} & 10 & 9 & 8 & 7 & 7 & 7 & 3.06 & 3.125 & $\phi_{3/2,2} \times \phi_{3/2,-2} \equiv \phi_{3/2}$ \\
 $V_{203}$ & {\tiny 6} & 8 & 7 & 7 & 8 & 8 & 8 & 3.36 & 3.5 & $\bar{L}_{-1} \phi_{1}$ \\
 $V_{201}$ & 10 &11 & 12 & 11 & 9 & 9 & 9 & 3.36 & 3.5 & $L_{-1} \phi_{1}$ \\
 $V_{202}$ & 6 & 9 & 10 & 10 & 10 & 10 & 10 & 3.84 & 4 & $\bar{L}_{-2} \phi_{0}$ \\ \hline
 $V_{213}$ & {\tiny 4} & 7 & 8 & 9 & 11 & 11 & 11 & 4.01 & 4.125 & $\bar{L}_{-2} \phi_{1/2}$ \\
 $V_{200}$ & 11 & 13 & 14 & 15 & 12 & 12 & 12 & 3.82 & 4 & $L_{-1} \bar{L}_{-1} \phi_{0}$ \\
 $V_{211}$ & 9 & 12 & 13 & 16 & 13 & 13 & 14 & 3.96 & 4.125 & $L_{-1} \bar{L}_{-1} \phi_{1/2}$ \\
 $V_{202}$ & 14 & 16 & 15 & 17 & 16 & 14 & 13 & 3.80 & 4 & $\bar{L}_{-2} \phi_{0}$ \\
  \hline
\end{tabular}
\end{center}
\caption{Conformal spectrum in the sector $V_2$, for $Q=4$.}
  \label{Tab_exp_V2_Q4}
\end{table}

\subsection{Ising model}
\label{sec:appA_Ising}

Finally we discuss the case of a unitary minimal model, namely the Ising model ($Q=2$). A special case of the first of the identities
(\ref{sumrules}) then reads \cite{DelfinoViti}
\begin{equation}
 G_{aaaa} =
 \langle \sigma_1 \sigma_2 \sigma_3 \sigma_4 \rangle = P_{aaaa} + P_{aabb} + P_{abba} + P_{abab} \,,
\end{equation}
%
where we have used the short-hand notation  $\sigma_i = Q \delta_{S_i,+} - 1$ for the usual FK spin operator ${\cal O}_a(\sigma_i)$ (\ref{spin-op})
and $S_i = \pm$ are the four Ising spins. Inserting the definition of $\sigma_i$, we can express
$G_{aaaa}$ in terms of the probabilities $P(S_1,S_2,S_3,S_4)$
of having fixed values of the $S_i$:
\begin{equation}
 \langle \sigma_1 \sigma_2 \sigma_3 \sigma_4 \rangle =
 \sum_{S_1,S_2,S_3,S_4 = \pm} (-1)^{\sum_{i=1}^4 \delta(S_i,+)} P(S_1,S_2,S_3,S_4) \,.
\end{equation}

It is straightforward to write a transfer matrix that computes the probabilities $P(S_1,S_2,S_3,S_4)$
in the Ising spin representation, by simply projecting on the required values of $S_i$ at the position
of each operator. Doing this we have verified the above identity to 4000 decimal places for various
systems. Since this relates very non-trivially the probabilities in the FK representation---whose computation
is intricate, as we have seen in section~\ref{sec:appA1}---to those in the Ising representation, this provides a strong test of the correctness
of our transfer matrix setup.

We can now compute $G_{aaaa}$ at larger sizes
and analyse it in terms of the eigenvalues of the Ising spin transfer matrix. Since two
spin operators are inserted simultaneously, the propagating states should only be those of the
$\mathbb{Z}_2$-even sector, and we have verified that this is indeed the case. 
\jesper{Added:} Note that the $\mathbb{Z}_2$-even sector is the simple module $\IrrJTL{0}{i} = \bAStTL{0}{\q^2=i} - \AStTL{3}{1} + \cdots$
which is the `top' corresponding to the leftmost diagram in Figure~\ref{figIsing1} (see \cite{GRSV1} for details).
In the conformal limit we expect the propagating
states to be the identity, $\phi_{1,1} \times \phi_{1,1} \equiv I$, and the energy operator,
$\phi_{2,1} \times \phi_{2,1} \equiv \varepsilon$.

We postpone the further discussion of these results to Appendix~\ref{sec:appB}. However, to refine
the analysis and parallel the discussion given above for generic $Q$ and the case $Q=4$, we
shall again need to establish the precise correspondence between the
finite-size and the conformal spectra. This is done in Table~\ref{Tab_exp_Q2}, where the rank of
eigenvalues now refer only to the $\mathbb{Z}_2$-even part of the spin representation (and not to a sector of the full FK transfer matrix spectrum).

Note that the spin transfer matrix does not contain sufficient information to attribute a momentum
label to the eigenvalues. However, since the Ising spectrum is included in the FK spectrum we can
still rely on the sizes $L=5,6,7$ to assign momentum labels to each level.

\begin{table}
\begin{scriptsize}
\begin{center}
\begin{tabular}{|c|cccccccccccccccc|ll|l|}
\cline{1-20}
  $V_{\ell,k,m}$
    & \multicolumn{16}{|c|}{$(i_5,i_6,\ldots,i_{20})$} & \multicolumn{2}{|c|}{$x_\ell$} & Identification \\  \cline{2-19}
   & 5 & 6 & 7 & 8 & 9 & 10 & 11 & 12 & 13 & 14 & 15 & 16 & 17 & 18 & 19 & 20 & Numerics & Exact & of scaling field \\
  \hline
 $V_{000}$ & 1 & 1 & 1 & 1 & 1 & 1 & 1 & 1 & 1 & 1 & 1 & 1 & 1 & 1 & 1 & 1 & 0              & 0              & $\phi_{1,1} \times \phi_{1,1} \equiv I$ \\
 $V_{000}$ & 2 & 2 & 2 & 2 & 2 & 2 & 2 & 2 & 2 & 2 & 2 & 2 & 2 & 2 & 2 & 2 & $1+10^{-12}$ & 1 & $\phi_{2,1} \times \phi_{2,1} \equiv \varepsilon$ \\
 $V_{001}, V_{002}$ & 3 & 3 & 3 & 3 & 3 & 3 & 3 & 3 & 3 & 3 & 3 & 3 & 3 & 3 & 3 & 3 & 2.0000002 & 2 & $L_{-1} \varepsilon, L_{-2} I$ \\
 $V_{002}, V_{003}$ & 4 & 4 & 4 & 4 & 4 & 4 & 4 & 4 & 4 & 4 & 4 & 4 & 4 & 4 & 4 & 4 & 3.000003 & 3  & $L_{-2} \varepsilon, L_{-3} I$ \\
 $V_{000}$ & 5 & 5 & 6 & 5 & 5 & 5 & 5 & 5 & 5 & 5 & 5 & 5 & 5 & 5 & 5 & 5 & 3.0000004 & 3    & $L_{-1} \bar{L}_{-1} \varepsilon$  \\ \hline
 $V_{003}, V_{004}$ & {\tiny 3} & {\tiny 4} & {\tiny 5} & 6 & 6 & 6 & 6 & 6 & 6 & 6 & 6 & 6 & 6 & 6 & 6 & 6 & 3.9995 & 4 & $L_{-3} \varepsilon, L_{-4} I$ \\
 $V_{001}$ & 6 & 6 & 7 & 7 & 8 & 8 & 8 & 7 & 7 & 7 & 7 & 7 & 7 & 7 & 7 & 7 & 4.000003 & 4  & $L_{-2} \bar{L}_{-1} \varepsilon$ \\
 $V_{000}$ & 7 & 8 & 9 & 9 & 9 & 9 & 10 & 10 & 9 & 8 & 8 & 8 & 8 & 8 & 8 & 8 & 4.0000002 & 4 & $L_{-2} \bar{L}_{-2} I$ \\
 $V_{005}$ & {\tiny 2} & {\tiny 3} & {\tiny 4} & {\tiny 6} & {\tiny 7} & 7 & 7 & 8 & 8 & 9 & 9 & 9 & 9 & 9 & 9 & 9 & 4.99994 & 5 & $L_{-5} I$ \\
 $V_{005}$ & {\tiny 5} & {\tiny 6} & {\tiny 8} & {\tiny 8} & {\tiny 10} & 10 & 11 & 11 & 12 & 11 & 11 & 10 & 10 & 10 & 10 & 10 & 4.9999 & 5 & $L_{-5} I$ \\ \hline
 $V_{000}$ & -- & 7 & 9 & 10 & 12 & 12 & 12 & 12 & 13 & 13 & 12 & 12 & 12 & 11 & 11 & 11 & 5.00002 & 5 & $L_{-2} \bar{L}_{-2} \varepsilon$ \\
 $V_{004}, V_{001}$ & {\tiny 8} & {\tiny 9} & {\tiny 11} & 12 & 13 & 13 & 15 & 14 & 14 & 14 & 15 & 13 & 13 & 13 & 12 & 12 & 5.000002 & 5 & $L_{-4} \varepsilon, L_{-3} \bar{L}_{-2} I$ \\
 $V_{006}$ & -- & {\tiny 2} & {\tiny 3} & {\tiny 4} & {\tiny 6} & {\tiny 7} & {\tiny 9} & 9 & 10 & 10 & 10 & 11 & 11 & 12 & 13 & 13 & 6.002 & 6 & $L_{-6} I$ \\
 $V_{006}$ & {\tiny 3} & {\tiny 5} & {\tiny 7} & {\tiny 8} & {\tiny 11} & {\tiny 11} & {\tiny 13} & 13 & 15 & 15 & 16 & 16 & 15 & 15 & 15 & 14 & 5.999 & 6 & $L_{-6} I$ \\
 $V_{007}$ & -- & {\tiny 1} & {\tiny 2} & {\tiny 3} & {\tiny 4} & {\tiny 6} & {\tiny 7} & {\tiny 9} & {\tiny 11} & 12 & 13 & 14 & 14 & 14 & 14 & 15 & 7.0004 & 7 & $L_{-7} I$ \\ \hline
 $V_{006}$ & {\tiny 6} & {\tiny 7} & {\tiny 10} & {\tiny 11} & {\tiny 14} & {\tiny 14} & {\tiny 16} & 16 & 18 & 17 & 17 & 17 & 18 & 17 & 16 & 16 & 5.9998 & 6 & $L_{-6} I$ \\
 $V_{005}$ & {\tiny 7} & {\tiny 9} & {\tiny 12} & {\tiny 14} & {\tiny 16} & 16 & 17 & 17 & 19 & 19 & 19 & 18 & 19 & 19 & 18 & 17 & 5.9999 & 6 & $L_{-5} \varepsilon$ \\
 $V_{000}$ & -- & 10 & 13 & 15 & 20 & 19 & 20 & 19 & 20 & 20 & 21 & 20 & 21 & 20 & 19 & 18 & 5.99997 & 6 & $L_{-3} \bar{L}_{-3} I$ \\
 $V_{007}$ & -- & -- & {\tiny 1} & {\tiny 2} & {\tiny 3} & {\tiny 4} & {\tiny 6} & {\tiny 8} & {\tiny 10} & 12 & 14 & 15 & 16 & 16 & 17 & 19 & 7.98 & 8 & $L_{-7} \varepsilon$ \\
 $V_{003}, V_{002}$ & 9 & 11 & 15 & 16 & 21 & 21 & 23 & 21 & 22 & 21 & 23 & 21 & 22 & 22 & 21 & 20 & 6.000003 & 6 & $L_{-4} \bar{L}_{-1} \varepsilon, L_{-4} \bar{L}_{-2} I$ \\ \hline
 $V_{007}$ & -- & {\tiny 3} & {\tiny 6} & {\tiny 7} & {\tiny 10} & {\tiny 11} & {\tiny 14} & {\tiny 15} & {\tiny 16} & 16 & 18 & 19 & 20 & 21 & 23 & 21 & 7.0002 & 7 & $L_{-7} I$ \\
 $V_{009}$ & -- & -- & -- & {\tiny 1} & {\tiny 2} & {\tiny 3} &{\tiny 4} & {\tiny 6} & {\tiny 8} & {\tiny 10} & {\tiny 13} & {\tiny 15} & 17 & 18 & 20 & 22 & 8.9 & 9 & $L_{-9} I$ \\
 $V_{009}$ & -- & -- & -- & -- & {\tiny 1} & {\tiny 2} & {\tiny 3} & {\tiny 4} & {\tiny 6 } & {\tiny 9} & {\tiny 10} & {\tiny 14} & {\tiny 16} & 18 & 22 & 23 & 9.85 & 10 & $L_{-9} \varepsilon$ \\
 $V_{007}$ & {\tiny 4} & {\tiny 6} & {\tiny 9} & {\tiny 11} & {\tiny 15} & {\tiny 15} & {\tiny 18} & {\tiny 18} & {\tiny 21} & {\tiny 22} & 24 & 24 & 24 & 23 &  24 & 24 & 6.999 & 7 & $L_{-7} I$ \\
 $V_{006}$ & {\tiny 6} & {\tiny 8} & {\tiny 11} & {\tiny 14} & {\tiny 18} & {\tiny 18} & {\tiny 22} & 23 & 24 & 23 & 25 & 25 & 25 & 25 & 26 & 25 & 6.99995 & 7 & $L_{-6} \varepsilon$ \\ \hline
 $V_{???}$ & -- & -- & -- & 13 & 17 & 17 & 21 & 22 & 23 & 24 & 26 & 26 & 27 & 26 & 27 & 26 & 6.9993 & 7 & \\
 $V_{008}$ & -- & -- & {\tiny 3} & {\tiny 6} & {\tiny 8} & 10 & 13 & 15 & 17 & 18 & 20 & 22 & 23 & 24 & 25 & 27 & 7.9992 & 8 & $L_{-8} I$  \\
 $V_{006}$ & {\tiny 8} & {\tiny 10} & {\tiny 14} & {\tiny 17} & {\tiny 22} & {\tiny 23} & {\tiny 27} & 26 & 29 & 28 & 28 & 27 & 29 & 28 & 28 & 28 & 6.9997 & 7 & $L_{-6} \varepsilon$ \\
 $V_{???}$ & 9 & 11 & 16 & 18 & 24 & 24 & 29 & 28 & 31 & 29 & 30 & 29 & 31 & 30 & 30 & 29 & 6.9996 & 7 & \\
  \hline
\end{tabular}
\end{center}
\end{scriptsize}
\caption{Conformal spectrum in the $\mathbb{Z}_2$-even sector for the Ising model ($Q=2$).
The rank $i_k$ of the eigenvalues refer to this sector only, and not to the full spectrum of the
corresponding FK model.}
  \label{Tab_exp_Q2}
\end{table}

Doing this we encounter an interesting phenomenon: In some cases a given eigenvalue corresponds
to two different momentum labels. For instance, the 3rd eigenvalue belongs simultaneously to $V_{001}$
and $V_{002}$ for $L=5,6,7$; the 4th eigenvalue belongs to $V_{002}$ and $V_{003}$ for $L=6,7$; and
the 7th eigenvalue belongs to $V_{001}$ and $V_{003}$ for $L=7$. We recall that we count here only
distinct eigenvalues, so statements of this type mean that the eigenvalue is degenerate (in addition to the
usual degeneracy coming from the sign of the momentum), with different
momentum identifications for each of the degenerate states.
The momentum labels $m$ assigned to the lowest-rank eigenvalues for $L=5,6,7$ are shown
in Table~\ref{Tab_momentum_Q2}.

\begin{table}
\begin{center}
\begin{tabular}{|c|ccc|}
\cline{1-4}
  Rank & \multicolumn{3}{|c|}{$L$} \\ \cline{2-4}
  $i_k$ & 5 & 6 & 7 \\  \hline
  1 & 0 & 0 & 0 \\
  2 & 0 & 0 & 0 \\
  3 & $1,2$ & $1,2$ & $1,2$ \\
  4 & 2 & $2,3$ & $2,3$ \\
  5 & 0 & 0 & 3 \\ \hline
  6 & 1 & 1 & 0 \\
  7 & 0 & 0 & $1,3$ \\
  8 & 1 & 0 & 2 \\
  9 & 2 & $1,2$ & 0 \\
  10 & & 0 & 1 \\ \hline
  11 & & $2,3$ & $1,3$ \\
  12 & & $1,2$ & 2 \\
  13 & & 0 & 0 \\
  14 & & 0 & 1 \\
  15 & & & $2,3$ \\ \hline
\end{tabular}
\end{center}
\caption{Momentum labels $m$ (modulo $L$) for the first 15 eigenvalues in the $\mathbb{Z}_2$-even sector
for the Ising model ($Q=2$). We have not been able to identify the labels $i_L$ for line 17 of the table.}
\label{Tab_momentum_Q2}
\end{table}

With all these ingredients, a close inspection of Table~\ref{Tab_exp_Q2} reveals that we observe precisely the minimal
characters $\chi_{r,s}(q)$ corresponding to the identity and energy operators, namely $\widetilde{\chi}_{r,s}(q) \equiv q^{-h_{r,s}+c/24} \chi_{r,s}(q)$
with the expansions
\begin{subequations}
\begin{eqnarray}
 (r,s) = (1,1) & : & 1+q^2+q^3+2q^4+2q^5+3q^6+\cdots \,, \\
 (r,s) = (2,1) & : & 1+q+q^2+q^3+2q^4+2q^5+\cdots \,.
\end{eqnarray}
\end{subequations}

\subsection{Practical remarks}
\label{sec:practical-remarks}

\subsubsection{Diagonalisation of $T$}
\label{sec:arnoldi}

Our diagonalisation of the transfer matrix $T$ is based on its decomposition (\ref{sparse_matrix_factorisation}) as a product of sparse matrices.
Indeed, the elementary operators ${\sf J}_i$ and ${\sf D}_i$ have at most one non-zero entry per column. It is therefore feasible to compute
$w = T v$---i.e., the action of $T$ on a vector of weights $v$---without ever storing $T$ and working only on its non-zero entries.
It follows that it is highly efficient to diagonalise $T$ by iterative methods that require only the operator $w = T v$ and not $T$ itself.
Among such methods, we have found that the Arnoldi method is well suited for our situation, where $T$ is a non-symmetric real matrix.

We have used the {\sc C++} interface for the {\sc Fortran} library {\sc Arpack} \cite{Arpack} for producing both eigenvalues and eigenvectors, in cases
where standard numerical precision (16 digits) is sufficient. This applies in particular to the scalar product method of section~\ref{sec:appA2}.
For the more direct method of section~\ref{sec:appA1}, a considerably higher numerical precision is needed. To this end we have used the
{\sc CLN} package \cite{CLN} (again written in {\sc C++}) that performs floating point operators to any desired numerical precision. We are greatly
indebted to Christian R.\ Scullard who has provided a pure {\sc C++} version of {\sc Arpack} with templates that are compatible with {\sc CLN};
this is unpublished work, but it has been described in recent publications on a different subject \cite{Scullard1,Scullard2,Scullard3}.

\subsubsection{Boundary conditions}
\label{sec:M_bdry_cond}

In the first method (see section~\ref{sec:1st_method}) we compute the probabilities $P_{\cal P}$ corresponding to FK cluster
correlation functions on the cylinder (see Figure~\ref{fig:cylinder}). As described in section~\ref{sec:appA1} this is done in practice by
expressing the probability as the the ratio of partition functions $W_{\cal P}$ on finitely long cylinders, where free boundary conditions
have been imposed at a distance $M$ from the insertion points of the cluster operators.

To ensure that the result does not depend on $M$, the latter much be chosen large enough. Let $d$ denote the number of decimal digits in
the desired numerical precision. Let $\Lambda_I$ and $\Lambda_\varepsilon$ be the largest and next-largest eigenvalues of $T$ within the sector
$V_{0}$; it is easy to compute their values beforehand. We now claim that we must chose $M$ so that $(\Lambda_\varepsilon / \Lambda_I)^M \le 10^{-d}$.

As the notation shows, these two eigenvalues are associated, throughout the critical regime $0 \le Q \le 4$, with the scaling levels that determine
respectively the identity $I$ and the energy operator $\varepsilon$. It should be noticed that $\Lambda_\varepsilon$ thus defined is most definitely
{\em not} the next-largest eigenvalue of $T$, if we take into account all sectors. This latter eigenvalue would rather be $\Lambda_\sigma$, the largest
eigenvalue within the sector $V_1$, associated with the scaling level of the leading order parameter operator $\sigma$, which has the smallest
non-zero scaling dimension in the CFT. In other words, we have $\Lambda_I > \Lambda_\sigma > \Lambda_\varepsilon$.
The point of the claim made above is then that outside the region $t \in [t_1,t_2]$ of the cylinder (see Section~\ref{sec:appA1} for notations)
that contains the operator insertions,
there are no constraints enforcing the propagation of clusters. This implies that the dominant decaying mode is given by the ratio
$\Lambda_\varepsilon / \Lambda_I$, as can easily be checked by direct inspection of the numerically computed probabilities.

\subsubsection{Attainable sizes}

The largest sizes $L$ attainable in the numerical work are essentially limited by the exponentially growing dimensions of the corresponding transfer
matrices. For the first method (see section~\ref{sec:1st_method}) there is however an additional complication, since we do not obtain the amplitudes
$A_i$ directly but acquire them by solving the linear system (\ref{PAeval}).

Suppose first that there are $K$ eigenvalues coupling to $P_{\cal P}$, and call $\Lambda_{\rm max}$ and $\Lambda_{\rm min}$ the largest
and smallest of those. If all computations are done to $d$ decimal digits, and we wish to obtain the final values of the amplitudes to $d_0$ digits,
then we must chose $d$ large enough that $( \Lambda_{\rm min} / \Lambda_{\rm max})^K > 10^{d_0 - d}$. Indeed, we will need the separation
between operators to take the values $l = 1,2,\ldots,K$ in order to have enough equations to solve the linear system for $A_i$. And assuming all
$A_i$ to be of the order unity, the smallest term on the right-hand side of (\ref{PAeval}) should not be below the level of numerical precision.

Similar considerations can be made for the case where $K$ is so large that we are unable (or do not want) to determine all the amplitudes. In
that case, we truncate the sum in (\ref{PAeval}) at some $i_{\rm max}$, and provide $i_{\rm max}$ different values of $l$. Since the formula
is no longer exact, we must make sure to use it in the proper asymptotic regime. We therefore take $l=i_{\rm min},i_{\rm min}+1,\ldots,i_{\rm max}$
for some suitable large $i_{\rm min}$. In practice we have chosen $i_{\rm min} = 100$ and repeated the computation for a slightly higher value
(e.g., $i_{\rm min} = 120$), in order to check that the $A_i$ were unchanged to the desired numerical precision.

In our largest computations, for $L=7$, we determined more than 500 amplitudes in this way. This computation needed as much as
$d=4000$ digits of numerical precision and required the work of $\sim 100$ computers simultaneously for a period of $\sim 3$ months.
See section~\ref{sec:big-computation} for further details on this computation.

For the second scalar product method (see section~\ref{sec:2nd_method}) it was enough to work with standard double precision ($d=16$ digits).
This method is altogether more efficient and could be done up to $L=11$ in a much more reasonable time (at most a few days on a single computer).

\subsubsection{Jordan blocks}

With the first method it is possible to determine the Jordan block structure of correlation functions, by using a fit of the form (\ref{generalised_amplitudes}).
To do this in practice requires some reasonable prior knowledge about where to look for the Jordan blocks, and about their expected order $r_i$.
The analytical understanding of indecomposability in the Temperley-Lieb algebra has grown steadily over recent years, including in the affine TL
case \cite{GrahamLehrer}, so the search for Jordan blocks is certainly not devoid of any guidance. However, one can also take a more naive numerical
approach to the problem.

Imagine that we are interested in some specific value $Q = Q_0$, where indecomposability is known to occur. Consider a correlation function
$P_{\cal P}$ for which we have previously established, for {\em generic} $Q$, which transfer matrix sectors $V_{\ell,k,m}$ occur with a non-zero amplitude.
At the non-generic $Q_0$, some of the eigenvalues from different sectors will collide, as can be seen by explicit diagonalisation using the methods
of section~\ref{sec:momentum_sectors} (or predicted analytically). The number of eigenvalues that collide at a given value $\Lambda_i$ is a natural
guess for the largest possible rank $r_i$ of the Jordan block corresponding to the generalised amplitude (\ref{generalised_amplitudes}). Trying the corresponding
Ansatz for different values of $i_{\rm min}$ (see above), we can examine whether the amplitudes are consistently determined (and non-zero) and arrive at a
quantitative understanding of the Jordan block structure.

\subsubsection{Extrapolation of amplitudes}

Whichever method is used for obtaining the eigenvalue amplitudes $A_i$ on the cylinder, the results need to be extrapolated to infinite size
($L \to \infty$) in order to be compared with predictions of conformal field theory. We shall see in Appendix~\ref{sec:appB} that in many cases
the finite-size results are quite far away from their CFT limits, more often than not by a factor in the range 2--5. Obtaining reliable extrapolations
is important, in particular because we only possess rather small sizes ($L \le 11$). Moreover, most
amplitudes exhibit mod 2 parity effects, so that even and odd $L$ must be treated separately.

Some guidance can be taken from our results for the Ising model  where larger sizes can be obtained ($L \le 16$).
We find the following method to provide quite accurate results. First we produce finite-size results with $\frac{2a}{L}$ equal to or close to a constant,
typically $\frac12$. Since it turns out that some amplitudes vanish exactly, by symmetry reasons, for $\frac{2a}{L}=\frac12$ precisely, we often take
$2a = (L-1)/2$ for odd $L$, and $2a = (L-2)/2$ for even $L$.
These results are then corrected by a conformal factor, namely the powers of $\sin \frac{2 \pi a}{L}$ appearing in (\ref{mainexp}).
Finally, we extrapolate the corrected results, separately for odd and even $L$, by fitting all available data against a polynomial in $1/L$ (or in some
cases leaving out the first or first few data points). The reliability of the extrapolation can then be estimated by comparing the two parities, which are expected
to give identical results.

\subsection{The split-attach algebra}
\label{sec:split-attach}

The scalar product method described in section~\ref{sec:2nd_method} and appendix~\ref{sec:appA2} requires us to find the left eigenvectors of $T$.
One way to attain this is to take the transpose of the right eigenvectors of the transposed matrix $T^{\rm t}$.
 
The scalar product method described in section~\ref{sec:2nd_method} and appendix~\ref{sec:appA2} requires us to find the left eigenvectors of $T$.
One way to attain this is to take the transpose of the right eigenvectors of the transposed matrix $T^{\rm t}$.

In the join-detach algebra for the FK cluster model, $T$ acts on basis states which are set partitions of $L$ points. 
The basic building block of the sparse matrix factorisation scheme (\ref{sparse_matrix_factorisation}) is the
action of the elementary join and detach operators, ${\sf J}_i$ and ${\sf D}_i$, on these states.
If this was presented in explicit matrix form, it would of course be trivial to take the transpose of the corresponding matrices.
The point is, however, that when we use an iterative diagonalisation scheme (such as the Arnoldi method; see section~\ref{sec:arnoldi}), it is inefficient
to represent ${\sf J}_i$ and ${\sf D}_i$ as explicit matrices. Rather, we just need to provide their action on any state:
for any in-state $v_{\rm in}$, the definition of ${\sf J}_i$ (or ${\sf D}_i$) in terms of set partitions provides the corresponding
 out-state $v_{\rm out} = {\sf J}_i v_{\rm in}$ (or ${\sf D}_i v_{\rm in}$).
To treat the transpose in a similar setting means that we should instead answer the question: given $v_{\rm out}$, what are the possible
$v_{\rm in}$ that could lead to it (and with which transition weights)?

We therefore formally define the split operator ${\sf S}_i = {\sf J}_i^{\rm t}$ and the attach operator ${\sf A}_i = {\sf D}_i^{\rm t}$ as the transpose
of the join and detach operator, respectively, and provide now their transition elements in terms of the basis states.

\subsubsection{Without marked clusters}

We first consider the analogue of the usual join-detach algebra, that is, without any marked clusters.

The operator ${\sf S}_i$ performs a split between sites $i$ and $i+1$. In particular, ${\sf S}_i$ is zero on states in which sites $i$ and $i+1$ are not
connected (i.e., in the same block of the set partition). If the two sites are connected, ${\sf S}_i$ acts by breaking up the block containing $i$ and $i+1$
in all possible ways. To be precise, assume the block consist of $k \ge 2$ points, ordered cyclically: $i_1 < i_2 < \cdots < i_k < i_1$ (the inequalities being
considered modulo $L$), with $i_1 = i$ and $i_2 = i+1$. Place one ``separator'' between $i_1$ and $i_2$. Place a second separator at the position of
any of the above inequalities, i.e., between $i_\ell$ and $i_{\ell+1}$ for any $\ell=1,2\ldots,k$, with the subscripts considered mod $k$. Note that
this includes the case where the two separators are at identical positions. Then break up the block by cutting it at the
positions of the two separators. This produces the $k$ possible out-states of ${\sf S}_i$, which each occur with transition weight $1$. Note that the
block is being broken into precisely two pieces in the $k-1$ cases where the two separators are different,%
\footnote{To be precise, if one separator is placed between $i_{s_1-1}$ and $i_{s_1}$ and the other between $i_{s_2-1}$ and $i_{s_2}$
(with $s_2 \neq s_1$), then the resulting blocks
are $\{i_{s_1},i_{s_1+1},\ldots,i_{s_2-1}\}$ and $\{i_{s_2},i_{s_2+1},\ldots,i_{s_1-1}\}$, with all subscripts considered mod $k$.}
whereas nothing happens (i.e., ${\sf S}_i$ acts as the identity) in the remaining case where the separators coincide.

The operator ${\sf A}_i$ attaches a singleton at site $i$ to certain other blocks of the set partition, as we now describe. In particular, ${\sf A}_i$ is zero
on states in which site $i$ is not a singleton. If $i$ is an singleton, ${\sf A}_i$ leaves the state unchanged with weight $Q$, and connects $i$ to any other
``visible'' block in the set partition with weight $1$. To determine whether a block is visible by $i$, draw the set partition as a hypergraph with vertex
set $1,2,\ldots,L$ embedded in the half-infinite cylinder (concretely, in Figure~\ref{fig:cylinder} this would be the part of the cylinder situated to the left
of the current time slice, with the vertices on the slice). If a block is adjacent to the face where $i$ resides, the block is said to be visible. Another, less
formal way to state the same, is that we connect $i$ to any block that is not nested within another block, seen from the position of $i$.

It is straightforward (albeit somewhat laborious) to verify that the split-detach algebra thus defined satisfies the algebraic relations which are the transpose
of those defining the join-detach algebra:
\begin{subequations}  \label{split-attach}
\begin{eqnarray}
 {\sf S}_i^2 &=& {\sf S}_i \,, \\
 {\sf A}_i^2 &=& A {\sf D}_i \,, \\
 {\sf S}_i {\sf A}_j {\sf S}_i &=& {\sf S}_i \mbox{ for } j=i,i+1 \,, \\
 {\sf A}_i {\sf S}_j {\sf A}_i &=& {\sf A}_i \mbox{ for } j=i-1,i \,.
\end{eqnarray}
\end{subequations}

\subsubsection{With marked clusters}

We now provide the definition of the split-attach algebra in the representation with marked clusters having a fixed spin label. This is analogous
to section~\ref{sec:appA2}.

The split operator ${\sf S}_i$ acts as zero, unless sites $i$ and $i+1$ belong to the same block (which can be unmarked or marked). Otherwise:
\begin{itemize}
 \item If $i$ and $i+1$ belong to the same unmarked block, ${\sf S}_i$ leaves the block unchanged or splits it into two unmarked blocks,
  as before, with a total of $k$ possibilities for a block of size $k$.
 \item If $i$ and $i+1$ belong to the same marked block, ${\sf S}_i$ leaves the block unchanged or splits it into two blocks, of which one
 is unmarked and the other keeps the same mark as the original block. Both choices for which of the two blocks should carry the mark are realised.
\end{itemize}

The attach operator ${\sf A}_i$ acts as zero, unless site $i$ is an unmarked singleton. Otherwise:
\begin{itemize}
 \item With weight $Q$, the operator ${\sf A}_i$ leaves the singleton block at $i$ unchanged and unmarked.
 \item With weight $1$, it marks site $i$ by each possible colour of the mark which is used in none of the other blocks in the partition.
 \item With weight $1$, it attaches site $i$ to each of its visible blocks (which can be marked or unmarked).
\end{itemize}
The two first rules cover the cases where the corresponding detach operation is trivial, in the sence that ${\sf D}_i$ would turn
a (marked or unmarked) singleton into an unmarked singleton. The third rule covers the non-trivial case, where the corresponding
detach operation acts on a (marked or unmarked) block of size $\ge 2$.

With these modifications one can again check that the defining relations (\ref{split-attach}) are satisfied, but within the larger representation that
allows clusters to be marked.

\section{Basic checks}
\label{sec:appB}

In this appendix we check our general method for computing the amplitudes $A_i$ against a series of exact results.
First we discuss the easy case of two-point functions and quantify the conformal content of the lattice spin operator.
Next we compare our results for the $s$-channel spectrum of four-point functions against exact solutions for $Q=2$, $Q=4$,
and the limit $Q \to 0$.

\subsection{Lattice observables and scaling fields}
\label{sec:appB1}

As discussed in the main text, one of the basic difficulties in our problem is that lattice observables correspond, in the scaling limit, to combinations of scaling fields weighed with appropriate powers of the cut-off. The question we want to investigate briefly here is how this might affect the measurement of amplitudes. 

According to (\ref{G_2-point}), the two-point function  of the lattice spin operator becomes, in the geometrical formulation, proportional to the probability that two points belong to the same cluster. The leading dimension controlling the large distance behavior of this probability  is $h_{1/2,0}$ and thus we expect, in the scaling limit,  to have on the cylinder 
\begin{equation}
P_{aa}\;\xrightarrow{\rm \, scaling}\; \left({\xi\over (1-\xi)^2}{\bar{\xi}\over (1-\bar{\xi})^2}\right)^{h_{1/2,0}} \,. \label{twoptfctgene}
\end{equation}
We have here used the same notation (\ref{eq:w1w2w3w4}) as for the four-point function, so $w_1=ia$, $w_3=i(a+x)+l$, and $w_{13}=-ix-l$. As before,
we set $\xi=e^{-2\pi(l+ix)/L}$, and the above dependence on $\xi, \bar{\xi}$ follows from a reasoning similar to the one leading to (\ref{CylExpan1}). We find numerically that only the sector $F_{0,-1}$ contributes, with
\begin{equation}
\displaystyle F_{0,-1} =\frac{q^{-c/24}\bar{q}^{-c/24}}{P(q) P(\bar{q})} \sum_{e \in \mathbb{Z}} q^{h_{e+1/2,0}} \bar{q}^{h_{e+1/2,0}} \,. \label{spec2}
\end{equation}
As is usually the case in lattice models, the order operator on the lattice is not purely represented by a conformal field of weight $(h_{1/2,0},h_{1/2,0})$. In general, one expects instead that this field is only the first in a sum of the type
\begin{equation}
 {\cal O}_a(\sigma_i)\sim \sum_{e\in\mathbb{Z}} A_e\epsilon^{2h_{e+1/2,0}} \Phi_{e+1/2,0}(z)\Phi_{e+1/2,0}(\bar{z})+\hbox{descendants} \,, \label{junk}
\end{equation}
where $z,\bar{z}$ are the complex coordinates corresponding to $\sigma_i$, we have introduced the notation $\Phi_{r,s}$ for conformal chiral fields with weight $h_{rs}$, and $\epsilon$ stands for the lattice cutoff ($\epsilon$ was set equal to unity so far). In the limit where $L,l\to\infty$ (that is, $L/\epsilon,l/\epsilon\to\infty$ if $L,l$ are measured in units of length) the contributions of the excited (higher) spin fields will scale away. For finite values of the parameters, they are unavoidably there. Note that 
\begin{equation}
h_{3/2,0}-h_{1/2,0}={m+1\over 2m} \,.
\end{equation}
This is larger for small values of $m$, i.e., smaller values of $Q$: therefore, the closer we will get to $Q=0$, the faster these contributions will decrease with $\epsilon$ ($\epsilon/L$ on the cylinder). 

We do not know how the $Q$-dependent amplitudes $A_e$ in (\ref{junk}) behave a priori, but we have studied numerically the two-point function in order to understand the amount of ``mixing'' of the order operator with the leading correction at weight $h_{3/2,0}$. We have checked that this mixing is small and decreases significantly with increasing $L$. 

\smallskip 
As an example, we give here results for  $Q=3/2$---a case for which $m$ is irrational, and all operators in the spectrum can be uniquely identified. 
 The dimension of the order parameter is  $h_{1/2,0}=0.0583892$, and by (\ref{twoptfctgene}) the two-point function in the conformal limit expands as 
\begin{equation}
P_{aa}^{\rm scaling}\propto \xi^{h_{1/2,0}}\bar{\xi}^{h_{1/2,0}}\left(1+0.116778 (\xi+\bar{\xi})+0.0136372 \xi\bar{\xi}+0.0652078(\xi^2+\bar{\xi}^2)+\ldots\right) \,, \label{twoptfct32}
\end{equation}
where all numerical constants have been given to six-digit precision.
The leading terms at momentum $h-\bar{h}=1$ and $h-\bar{h}=2$ are easily identified as the ground states of the corresponding momentum sectors. With no knowledge of the possible mixing with the term $\xi^{h_{3/2,0}}\bar{\xi}^{h_{3/2,0}}$ one would look for  the $\xi\bar{\xi}$ term in (\ref{twoptfct32}) in the lattice data as the contribution of the  first excited state in the sector at vanishing momentum, but a careful analysis of the amplitudes as well as the scaling dimensions shows this is not what happens. On the lattice, the two point function contains, in addition to the terms in (\ref{twoptfct32}), a term in the bracket proportional to 
$(\xi\bar{\xi})^{h_{3/2,0}-h_{1/2,0}}$.  This follows from (\ref{junk}), which leads to 
\begin{equation}
P_{aa}^{\rm lattice}\propto {1\over |z|^{4h_{1/2,0}}}\left[1+\left({A_1\over A_0}\right)^2\left({\epsilon\over |z|}\right)^{4(h_{3/2,0}-h_{1/2,0})}+\ldots\right] \,,
\end{equation}
and, after mapping on the cylinder, to 
\begin{eqnarray}
P_{aa}^{\rm lattice} &\propto& \left({2\pi\over L}\right)^{4h_{1/2,0}} 
 \left[\left({\xi\over (1-\xi)^2}\right)^{h_{1/2,0}} \left({\bar{\xi}\over (1-\bar{\xi})^2}\right)^{h_{1/2,0}} \right. \nonumber \\
 &+& \left. \left({A_1\over A_0}\right)^2\left({2\pi \epsilon\over L}\right)^{4(h_{3/2,0}-h_{1/2,0})}\left({\xi\over (1-\xi)^2}\right)^{h_{3/2,0}} \left({\bar{\xi}\over (1-\bar{\xi})^2}\right)^{h_{3/2,0}}+\ldots\right] \,.
 \label{2ptfun_mapped_cyl}
\end{eqnarray}
The amplitude ratio $A_1/A_0$ is non-universal, but, for a given lattice, it is a fixed quantity. The term $L^{-4(h_{3/2,0}-h_{1/2,0})}$ goes to zero as $L$ goes to infinity, guaranteeing the disappearance of the unwanted terms in this limit. The question is how much this might affect the results for $L$ finite. 

Measurement of the amplitudes---or rather the ratio with respect to the leading term---gives the following results:


\begin{align}
\setlength{\arraycolsep}{4mm}
\renewcommand{\arraystretch}{1.2}
 \begin{array}{c|llll}
   L & \xi+\bar{\xi} & \xi\bar{\xi} & \xi^2+\bar{\xi}^2 &(h_{3/2,0},h_{3/2,0})
   \\ \hline
 5 & 0.1009697881 & 0.004693387836 & 0.05633421252 & -0.0007594762012 \\
 6 & 0.1044947727 & 0.006252504965 & 0.05534757764 & -0.0006074838333 \\
 7 & 0.1070768500 & 0.007528362715 & 0.05578623767 & -0.0004817486169 \\  
 8 & 0.1089775943 & 0.008549110109 & 0.05665273685 & -0.0003834115301 \\
 9 & 0.1103983091 & 0.009361639062 & 0.05759817991 & -0.0003078006915 \\
10 & 0.1114793562 & 0.010010585922 & 0.05849501339 & -0.0002497183860 \\
11 & 0.1123166454 & 0.010532656068 & 0.05930089355 & -0.0002048250690 \\
12 & 0.1129759889 & 0.010956414072 & 0.06000781281 & -0.0001698028289 \\
\infty & 0.11679 & 0.13644 & 0.06519 & < 10^{-6} \\
{\rm CFT} & 0.116778 & 0.0136372 & 0.0652078 & 0 \\
   \hline
 \end{array} \nonumber
\end{align}

The extrapolation to $L \to \infty$ was made via a polynomial of order 7 in $1/L$, using all data points, since there are no discernable parity effects in this case.
We observe that on one hand the expected ratios converge to their conformal values to a very good precision (4 or 5 digits).
On the other, the one that is not expected---corresponding to the coupling to $(h_{3/2,0},h_{3/2,0})$---decreases fast, and converges to a value close to zero.%
\footnote{To make this last extrapolation, we have {\em not} used the knowledge of (\ref{2ptfun_mapped_cyl}) concerning the exact power-law dependence on $L$.
Rather we performed just the usual polynomial fit, in order to establish the same methodology for other less trivial cases.}
This example shows that one can obtain fine extrapolations, even though for the sizes considered in this paper (and the value $Q=3/2$ taken here) the mixing of the
field $(h_{3/2,0},h_{3/2,0})$ with the order parameter has a rather large relative coefficient of the order of $10^{-2}$.

\subsection{The case $Q=2$}
\label{sec:appB-Q2}

We now consider the case $Q=2$ (and later $Q=4$) to check the consistency of our approach, and assess the quality of convergence of the amplitudes
in the four-point functions. While in general, the geometrical probabilities cannot be expressed in terms of simple four-point functions in rational CFTs,
the situation is better for these two values of $Q$. This has to do with the relationship \cite{DelfinoViti} between
correlation functions of spins in the Potts model and the geometrical objects, see eq.~(\ref{sumrules}).
We stress that for $Q$ arbitrary, the left-hand sides of these equations are only formally defined: it is in fact the right-hand sides that give them a meaning.
In the $Q=2$ case the first relation (\ref{sumrules}a) reads simply
\begin{equation}
G_{aaaa}=P_{aaaa}+P_{aabb}+P_{abba}+P_{abab} \,, \quad \mbox{for } Q=2 \,. \label{combiIsing}
\end{equation}
On the other hand, recall from (\ref{order-param-corr}) that 
\begin{equation}
G_{aaaa} = \langle \prod_{i=1}^4 (Q\delta_{\sigma_i,a}-1)\rangle
\end{equation}
As discussed in \ref{sec:appA_Ising}, for $Q=2$, the order parameter $Q\delta_{\sigma_i,a}-1$ coincides with the Ising spins $S_i = \pm 1$, so this four-point function is nothing
but the four-point function of the spin operator in the Ising model \cite{MCW}
%
\begin{eqnarray}
\langle \sigma(z_1) \sigma(z_2) \sigma(z_3) \sigma(z_4) \rangle &=& {1\over 2}\left|
     {z_{13}z_{24}\over 
	  z_{12}z_{23}z_{34}z_{41}}\right|^{1/4}\left(\left|1+\sqrt{1-z}\right|+\left|1-\sqrt{1-z}\right|\right)\nonumber\\
	  &=& {1\over 2}|z_{13}|^{-1/4}|z_{24}|^{-1/4} {1\over |z(1-z)|^{1/4}}\left(\left|1+\sqrt{1-z}\right|+\left|1-\sqrt{1-z}\right|\right) \,. \label{Ising_ssss}
\end{eqnarray}
We also recall the structure constant $C_{\sigma\sigma\epsilon}={1\over 2}$, where $\epsilon$ denotes the energy operator; the latter appears as
the non-trivial part of the scaling limit of $S_i S_{i+1}$. The four-point function (\ref{Ising_ssss}) involves two conformal blocks, corresponding to the
fusion channels $\sigma\sigma\sim 1$ and $\sigma\sigma\sim \epsilon$.  Expanding the function $\G$ as in (\ref{leadingexpan}) gives then
 \begin{equation}
 |z|^{1/4} \G(z,\bar{z})=1+{1\over4} z^{1/2}\bar{z}^{1/2}+{1\over 16}(z^{1/2}\bar{z}^{3/2}+z^{3/2}\bar{z}^{1/2})+{1\over 64}z^{3/2}\bar{z}^{3/2}+{1\over 64} (z^2+\bar{z}^2)+\ldots \,.
 \end{equation}
 The normalised correlation function on the lattice should then be 
 \begin{equation}
 1+s^2 (\xi \bar{\xi})^{1/2} +(s^2-s^4) (\xi \bar{\xi})^{1/2} (\xi + \bar{\xi})+{s^4\over 4}(\xi^2+\bar{\xi}^2)+(s^2-2s^4+s^6) (\xi \bar{\xi})^{3/2} + \frac{s^8}{16} (\xi \bar{\xi})^2 + \ldots \,,\label{isinglat}
 \end{equation}
where $s=\sin {2\pi a\over L}$, and the extra terms arise from the conformal mapping as in (\ref{CylExpan1}).
 
\subsubsection{Reduction from the generic case}
 
\begin{figure}[ht]
\begin{center}
    \includegraphics[width=10cm]{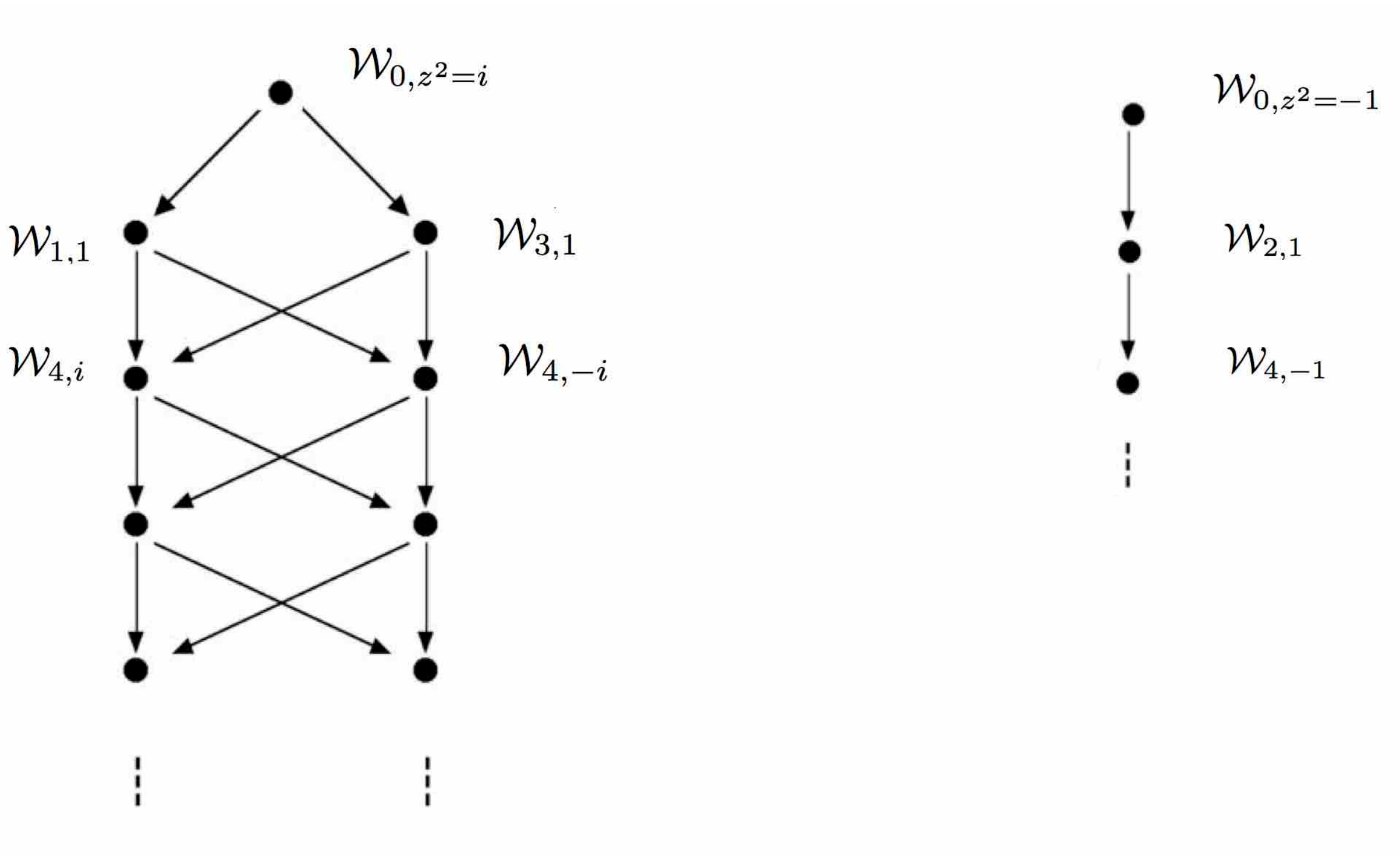}
     \caption{The simple modules $\IrrJTL{0}{i}$ and $\IrrJTL{0}{-1}$ are the tops of these two diagrams representing the structure of the standard modules for $\q=e^{i\pi/4}$. }\label{figIsing1}
\end{center}
\end{figure}

We first  checked---by computing the four-spin correlation function numerically for the ordinary (spin representation) Ising model on the cylinder---that the identity 
(\ref{combiIsing}) holds exactly in finite size; details are given in section~\ref{sec:appA_Ising}. This computation is certainly a rather stringent test of the program that
determines the $P_{a_1,a_2,a_3,a_4}$ numerically.  Note  that the (large) set of eigenvalues that contribute  to the geometrical correlations for generic values of $Q$ 
reduces drastically---as expected---when we consider the combination (\ref{combiIsing}).  In algebraic terms, generically irreducible representations of the Jones
algebra  become reducible, and a complex structure of submodules of  the relevant $ \bAStTL{j}{z^2}$ develops. It turns out that only two   {\sl simple} quotients contribute to the Ising model: $\IrrJTL{0}{i}$ and $\IrrJTL{0}{-1}$, which are obtained as the the `tops' of chains of modules according to the  diagrams in Figure \ref{figIsing1} (for a detailed discussion of the emergence of simple modules of the Jones algebra describing minimal models when $\q$ is a root of unity, see \cite{GRSV1}). 

The continuum limit  of the simple modules of the Jones algebra is fully expressed in terms of irreducible modules $\IrrV{r,s}$ of the Virasoro algebra
\begin{subequations}
\begin{eqnarray}
\IrrJTL{0}{i} &\mapsto& \IrrV{1,1}\otimes\IrrVb{1,1}\bigoplus 
\IrrV{21}\otimes\IrrVb{2,1}=\IrrV{h=0}\otimes\IrrVb{h=0}\bigoplus
\IrrV{h=1/2}\otimes\IrrVb{h=1/2} \,, \\
\IrrJTL{0}{-1} &\mapsto& 2 \, \IrrV{12}\otimes\IrrVb{12}=2 \, \IrrV{h=1/16}\otimes\IrrVb{h=1/16} \,,
\end{eqnarray}
\end{subequations}
where the factor 2 indicates that, in fact, the representation splits into two isomorphic direct summands for the subalgebra generated by $e_i$ and $u^2$, and  $\IrrV{r,s}$ denotes the irreducible Virasoro module with conformal weight $h_{rs}$ (and, e.g., character given by the well-known Rocha-Caridi formula \cite{RochaCaridi}).

We checked that the combination (\ref{combiIsing}) receives contributions only from the tops in the diagrams in Figure \ref{figIsing1} , which corresponds, in the continuum limit, to the required channels of the minimal Ising model. Note however  that  individual  probabilities in  (\ref{combiIsing}), unlike their sum,   do involve contributions outside the simple representations.

\subsubsection{Numerical checks of the amplitudes}

We now turn to the question of the convergence of the amplitudes measured on the lattice. Surprisingly, it seems this question has not been investigated much in the past (see \cite{ReinickeVescan} for early work).
Since the Ising model is much easier to handle numerically than the general FK cluster model, we can study  in this case much larger sizes (up to $L_{\rm max} = 16$), and explore in particular the nature of the convergence of the coefficients in the expansion (\ref{isinglat}). 

We present in the following Figures~\ref{figIsing2}--\ref{figIsing2dist2} two different ways of handling the data. First, we consider the case where $a$ is changed as $L$ increases.
For definiteness we consider $2a={L\over 2}$ when $L$ is even and $2a={L-1\over 2}$ when $L$ is odd.
In both cases, $\sin{2\pi a\over L}\rightarrow 1$ as $L\rightarrow\infty$, but we should of course expect even/odd parity effects in $L$
because of the different choices of $2a$.

A priori the aim would be to compute the amplitude ratios, with respect to the ground state amplitude, corresponding
to the five last terms in (\ref{isinglat}), namely the contributions to $G_{aaaa}$ of $\epsilon$, $(L_{-1}+\bar{L}_{-1})\epsilon$, $T+\bar{T}$, 
$\bar{L}_{-1} L_{-1} \epsilon$, and $T \bar{T}$, respectively. However it turns out that $(L_{-1}+\bar{L}_{-1})\epsilon$ and $T+\bar{T}$
correspond to eigenvalues that are exactly degenerate in finite size, so the corresponding contributions cannot be disentangeled.
We thus list the amplitude ratios in (\ref{isinglat}) that we can access numerically along with their corresponding analytical predictions:
\begin{subequations}
\label{Ising4ptCFTpredictions}
\begin{eqnarray}
 \epsilon & : & s^2 \\
 (L_{-1}+\bar{L}_{-1})\epsilon + (T+\bar{T}) & : & 2 \left( s^2 - \frac34 s^4 \right) \\
 \bar{L}_{-1} L_{-1} \epsilon & : & s^2 - 2 s^4 + s^6 \\
 T \bar{T} & : & \frac{s^8}{16}
\end{eqnarray}
\end{subequations}
 
The four panels of Figure~\ref{figIsing2} show the corresponding amplitudes ratios, where in each case we have divided the ratio by the expected analytical result
just given. It is seen that in each case the result tends to $1$ after
extrapolation, thus confirming the whole analysis.

\begin{figure}[ht]
\begin{center}
    \includegraphics[width=7cm]{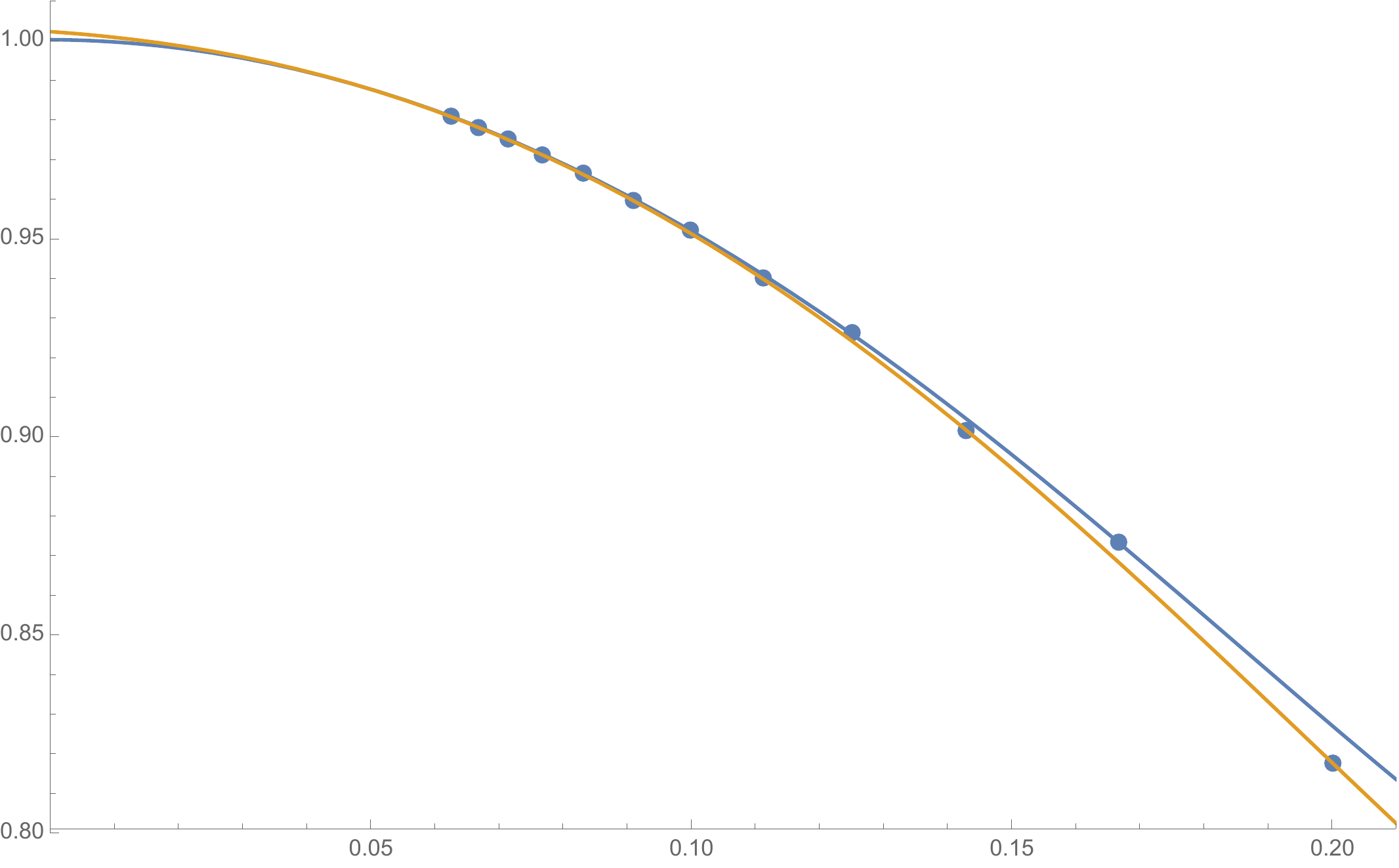} \qquad \qquad
    \includegraphics[width=7cm]{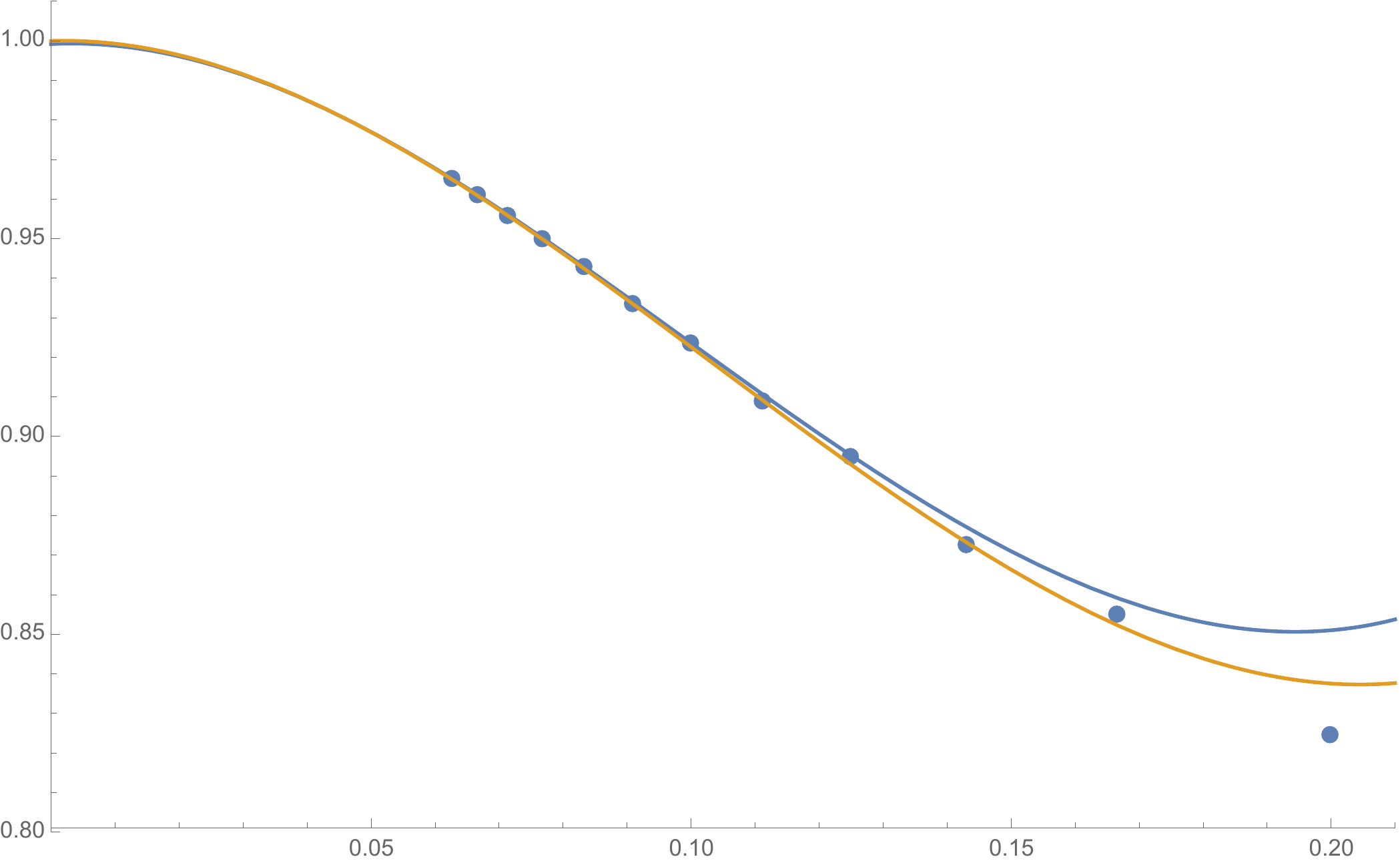}

    \vspace{-0.0cm}\hspace{-0.1cm}(a)\hspace{8.1cm}(b)\vspace{0.5cm}

    \includegraphics[width=7cm]{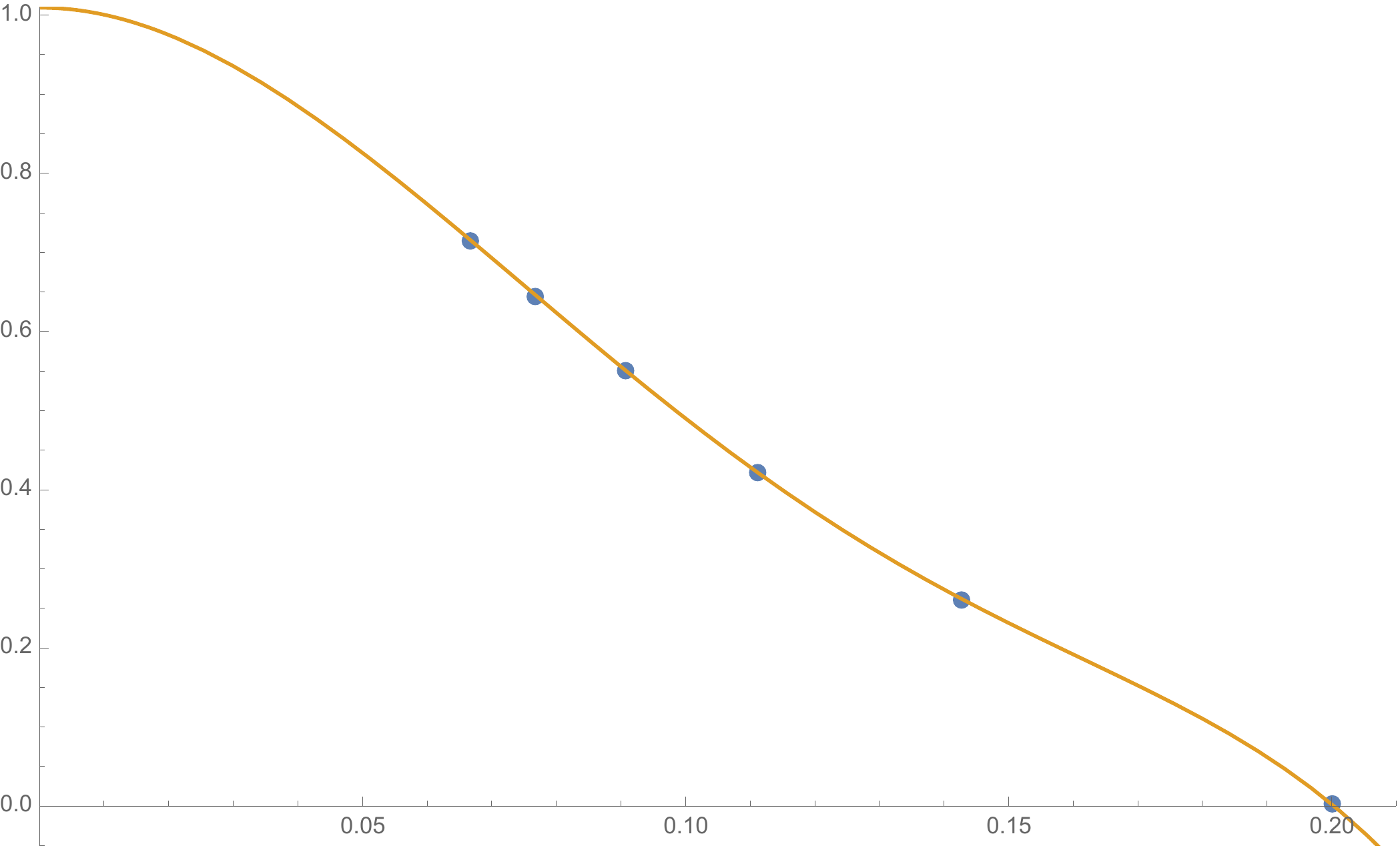} \qquad \qquad
    \includegraphics[width=7cm]{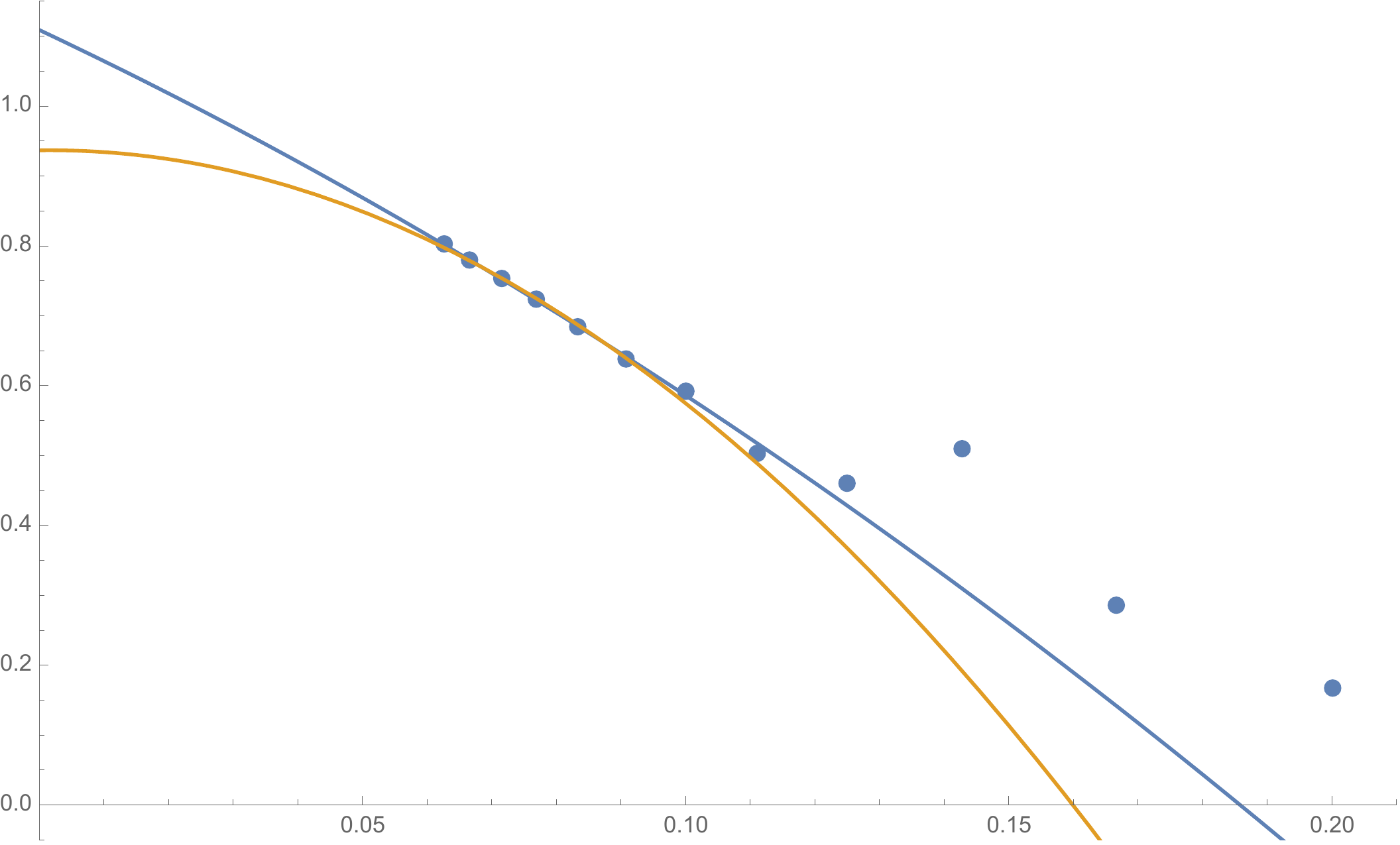}

    \vspace{-0.0cm}\hspace{-0.1cm}(c)\hspace{8.1cm}(d)\vspace{0.0cm}

  \caption{Measures of the first four amplitude ratios in $G_{aaaa}$ for the Ising model, divided by the expected analytical results. The 
  respective panels show the contributions from (a) the energy operator $\epsilon$, (b) its first descendent $(L_{-1}+\bar{L}_{-1})\epsilon$
  along with the stress tensor $T+\bar{T}$, (c) its second descendent $\bar{L}_{-1} L_{-1} \epsilon$, and (d) the quantity $T \bar{T}$.
  All data is shown against $1/L$. The extrapolations for even and odd $L$ are shown respectively as blue and orange curves.}
  \label{figIsing2}
\end{center}
\end{figure}

A number of remarks can be made about Figure~\ref{figIsing2}. First, the extrapolations have to be carried out using some amount of common sense. In
some cases the obvious non-monotonicity of the finite-size results (panel d) makes it clear that one should leave out the first few points from
the fits. In other cases, (panel c) including all points in a high-order polynomial fit leads to the best results. Second, the comparison between fits
through even and odd sizes appears to be a good measure for the ``error bar'' on the extrapolated value. Third, there is a tendency for the finite-size
effects to grow as one goes to higher descendents. Thus, if we were limited to smaller sizes---such as $L_{\rm max}=11$, as would be the case for
the FK cluster model at generic values of $Q$---it should be expected that the last data point might very well be off the true extrapolated value by
a factor of 2 or more.


\begin{figure}[ht]
\begin{center}
    \includegraphics[width=7cm]{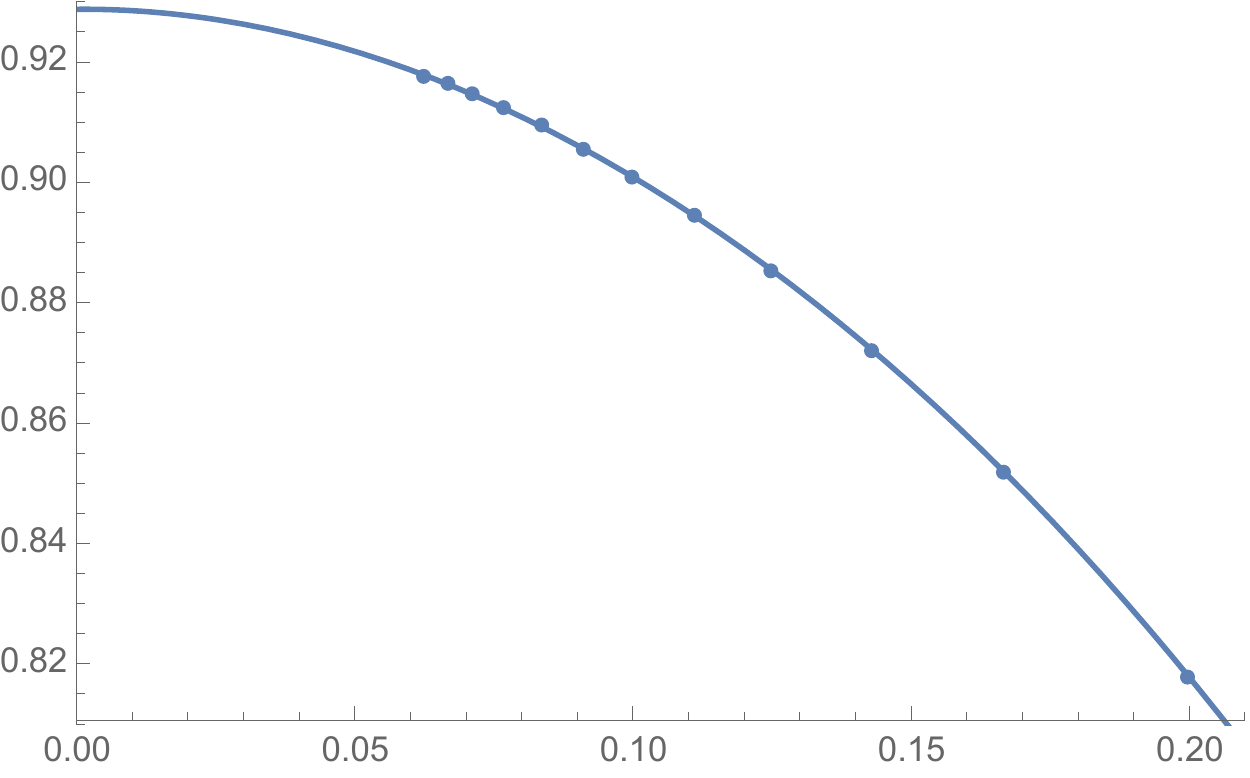} \qquad \qquad
    \includegraphics[width=7cm]{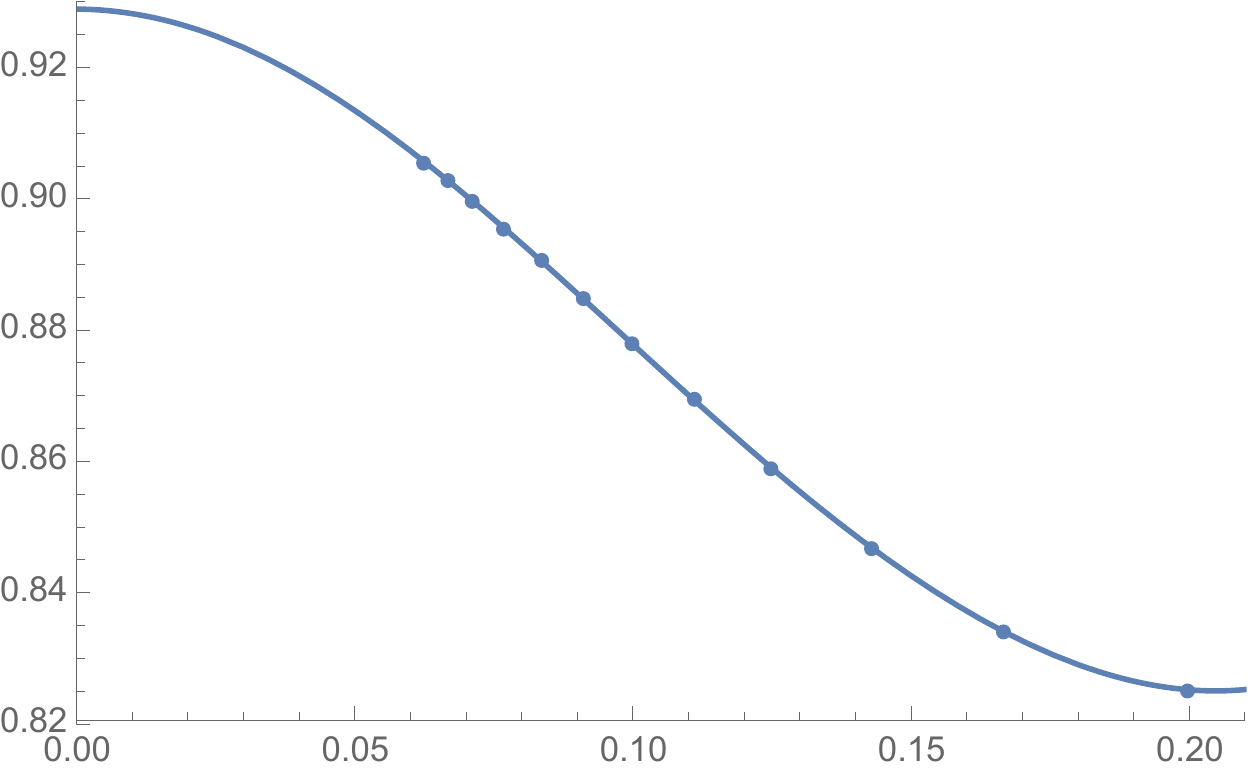}

    \vspace{-0.0cm}\hspace{-0.1cm}(a)\hspace{8.1cm}(b)\vspace{0.5cm}

    \includegraphics[width=7cm]{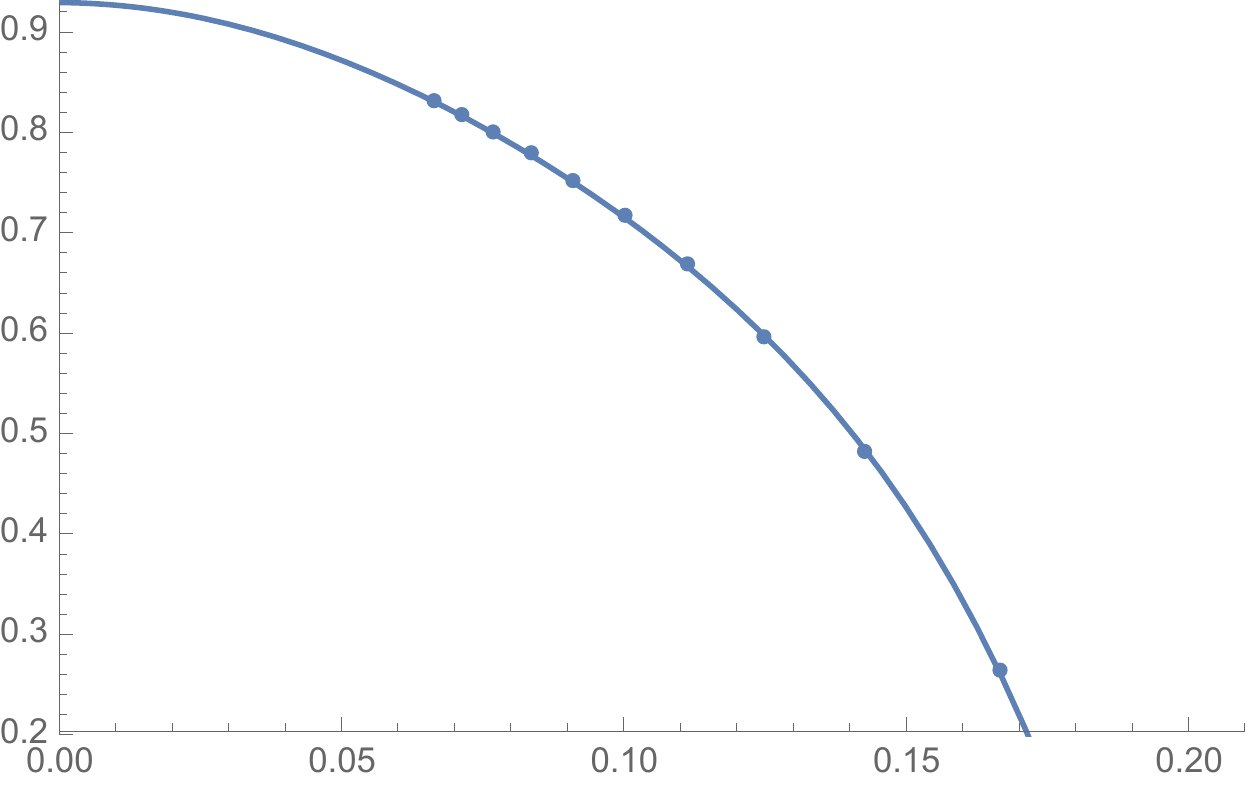} \qquad \qquad
    \includegraphics[width=7cm]{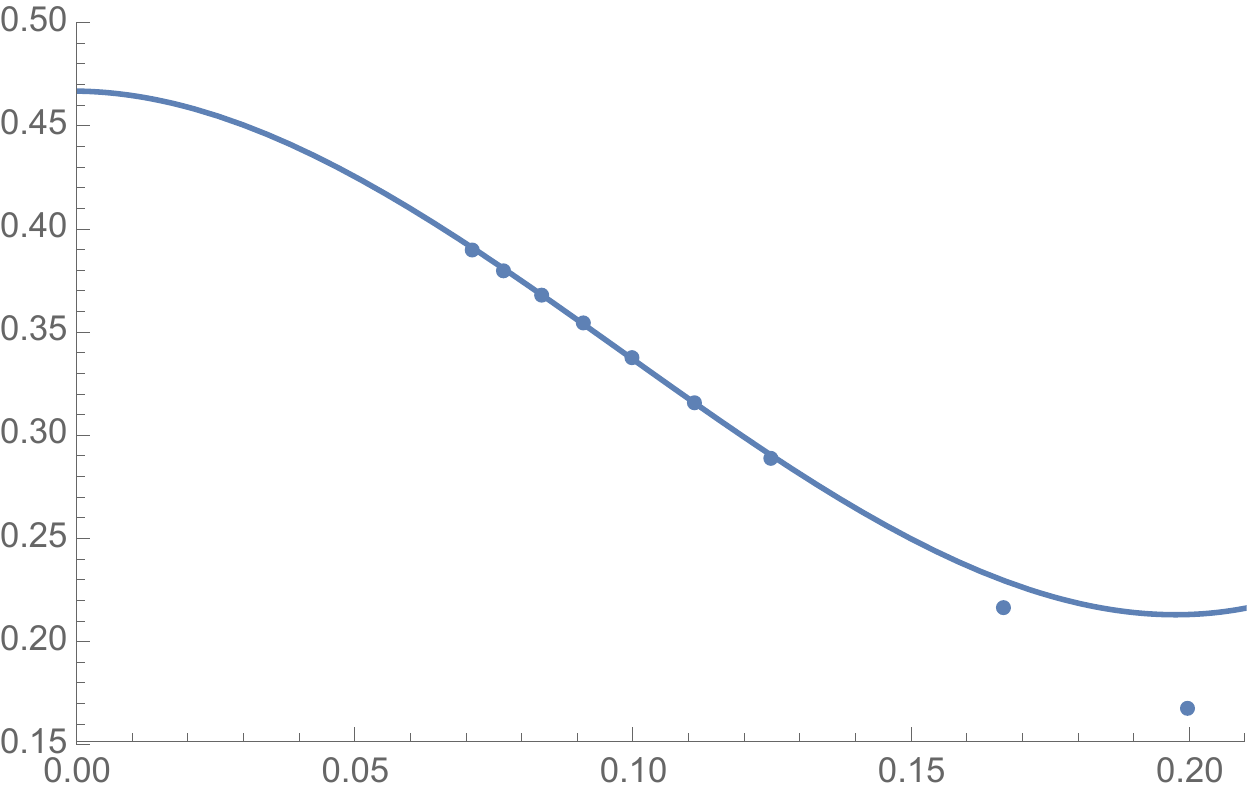}

    \vspace{-0.0cm}\hspace{-0.1cm}(c)\hspace{8.1cm}(d)\vspace{0.0cm}

  \caption{Same as Figure~\ref{figIsing2}, but for a fixed distance $2a=2$ between each pair of operators.
  The panels are as in Figure~\ref{figIsing2}. The extrapolations, shown as blue curves, now uses sizes $L$ of both parities.}
  \label{figIsing2dist2}
\end{center}
\end{figure}

It is also interesting to study what happens when $a$ is fixed while $L$ is increased.
In this case, one does not expect the amplitude ratios to converge to the four-point CFT values reported in (\ref{Ising4ptCFTpredictions}), since
the latter require $a,L \gg 1$ (in units of the lattice spacing) with fixed value of the ratio $\frac{a}{L}$ (and hence $s$ fixed).
To illustrate this, we consider for instance the case of $2a=2$ fixed, with $L$ increasing---note that we do not expect parity effects using this
protocol. To make the analysis comparable to the one above,
we again divide the ratio by the CFT predictions (\ref{Ising4ptCFTpredictions}). The results are shown in Figure~\ref{figIsing2dist2}, to be compared
with the previous Figure~\ref{figIsing2}. Clearly the convergence is no longer towards one. It is however remarkable that the numerical data still
appear to converge to a finite value. For the energy operator and its first two descendents (panels a, b, c) we get fine extrapolations to the values
$0.9292$, $0.9289$ and $0.9292$ respectively, which are all compatible among themselves. For $T \bar{T}$ the extrapolations are more problematic,
but still appear to converge to a finite value $\simeq 0.53$, which definitely appears to be different from the previous one.

We are not sure whether these values can be interpreted within CFT---neither whether the observed convergence is real or only apparent.
Repeating the computations for $2a=3$, a similar extrapolation gives the consistent values $0.9686$ and $0.9680$ for the energy field and its first descendent,
whereas $T \bar{T}$ gives another value $\simeq 0.71$.


\subsection{The case $Q=4$}
\label{sec:appB-Q4}

As pointed out in \cite{Ribault}, the Potts model at $Q=4$  is particularly interesting, since in this case all geometrical correlation functions can be expressed in terms of spin correlation functions. This provides analytical results for all the $P_{abcd}$, and thus provides a simple benchmark against which to test our approach. There is however a drawback to this idea: numerics  for the Potts model at $Q=4$ are known to be affected by stronger than usual corrections to scaling due to the presence of a marginal operator. As a result, convergence to the expected amplitudes appears to be slower than for other values of $Q$, especially small ones. 

In any case, we start by observing that the  Potts model partition function, can be expressed in the continuum limit as the sum \cite{DFSZ,SKY},
\begin{equation}
Z_{\rm Potts~ Q=4}=\sum_{n\in \mathbb{Z}} F_{2n,1}-{1\over 2} F_{2n+1,1}+{3\over 2}F_{2n,-1} \,, \label{Pottsfour}
\end{equation}
where the generating functions are obtained by taking the limit $m\to\infty$ of our general result (\ref{bspec1})
\begin{subequations}
\begin{eqnarray}
F_{j,1} &=& {1\over \eta\bar{\eta}}\sum_{k\in\mathbb{Z}} q^{(j+k)^2/4}\bar{q}^{(j-k)^2/4} \,, \\
F_{j,-1} &=& {1\over \eta\bar{\eta}}\sum_{k\in\mathbb{Z}} q^{(j+k+1/2)^2/4}\bar{q}^{(j-k-1/2)^2/4} \,.
\end{eqnarray}
\end{subequations}
In contrast with the case of generic $Q$ \cite{DFSZ}, we see that only  a very small subset of the $F_{j,e^{2i\pi p/M}}$ generating functions contribute. This corresponds to the fact---which we checked explicitly in finite size, and which has a simple representation theoretic interpretation; see below---that of all the usual Jones algebra  modules, only those  corresponding to $z^2=\pm 1$ play a role at $Q=4$.   

\subsubsection{Exact results via the Ashkin-Teller model}

After these preliminaries, we now consider four-point functions per se. In \cite{Ribault}, some of the probabilities at $Q=4$ are given based on results in \cite{DelfinoViti}, combined with the careful reading of a paper by Al.B.\ Zamolodchikov \cite{AlyoshaZ}. Calling $\tau_1,\tau_2$ the two spins of the Ashkin-Teller model, we introduce, following \cite{AlyoshaZ}, the four quantities
\begin{subequations}
\begin{eqnarray}
R_1 &=& \langle \tau_1\tau_2\tau_2\tau_1\rangle \,, \\
R_2 &=& \langle \tau_1\tau_1\tau_2\tau_2\rangle \,, \\
R_3 &=& \langle \tau_1\tau_2\tau_1\tau_2\rangle \,, \\
G &=& \langle \tau_1\tau_1\tau_1\tau_1\rangle \,;
\end{eqnarray}
\end{subequations}
note that $G$ here must not be confused with $\G$ in (\ref{GenFptf}).
It is then expected that%
\footnote{Note there is a slight change of notations between \cite{Ribault} and \cite{AlyoshaZ}: $R_1$ and $R_2$ are interchanged.}
\begin{subequations}
\begin{eqnarray}
P_{aabb} &\propto& G+R_2-R_1-R_3 \,, \\
P_{aaaa} &\propto& -G+R_1+R_2+R_3 \,, \\
P_{abab}-P_{abba} &\propto& R_1-R_3 \,.
\end{eqnarray}
\end{subequations}
The quantities on the right-hand side are then calculated in \cite{AlyoshaZ}. We introduce first
\begin{equation}
F_0(z)\equiv [z(1-z)]^{-1/8}\theta_3^{-1}(\qq) \,,
\end{equation}
where $z=\frac{z_{12}z_{34}}{z_{13}z_{24}}$ as usual.
We shall also need the Jacobi theta functions
\begin{subequations}
\begin{eqnarray}
\theta_3(\qq)&=&\sum_{-\infty}^\infty \qq^{n^2} \,, \\
\theta_2(\qq)&=&\sum_{-\infty}^\infty \qq^{(n+1/2)^2} \,.
\end{eqnarray}
\end{subequations}
These two functions are used to define $\qq(z)$ implicitly via
\begin{equation}
z\equiv {\theta_2^4(\qq)\over \theta^4_3(\qq)}
\end{equation}
Note that the function $\qq(z)$ is analytic in $z$:%
\footnote{Thus $\qq$ must not be confused with either the nome $q$ in generating functions of conformal weights, not with the quantum group deformation parameter $\q$.}  
\begin{equation}
\qq(z)=\frac{z}{16}+\frac{z^2}{32}+\frac{21 z^3}{1024}+\frac{31 z^4}{2048}+\frac{6257
   z^5}{524288}+\frac{10293 z^6}{1048576}+\frac{279025 z^7}{33554432}+\frac{483127
   z^8}{67108864}+\ldots \,.
\end{equation}
We then set
\begin{subequations} \label{combi}
\begin{eqnarray}
G&=&F_0\bar{F}_0\sum_{m,n=-\infty}^\infty \qq^{(\beta_+m+\beta_-n)^2}\bar{\qq}^{(\beta_+m-\beta_-n)^2} \,, \\
R_1&=&F_0\bar{F}_0\sum_{m,n=-\infty}^\infty \qq^{(\beta_+m+\beta_-(n+1/2))^2}\bar{\qq}^{(\beta_+m-\beta_-(n+1/2))^2} \,, \\
R_2&=&F_0\bar{F}_0\sum_{m,n=-\infty}^\infty (-1)^m\qq^{(\beta_+m+\beta_-n)^2}\bar{\qq}^{(\beta_+m-\beta_-n)^2} \,, \\
R_3&=&F_0\bar{F}_0\sum_{m,n=-\infty}^\infty (-1)^m\qq^{(\beta_+m+\beta_-(n+1/2))^2}\bar{\qq}^{(\beta_+m-\beta_-(n+1/2))^2} \,. 
\end{eqnarray}
\end{subequations}
Finally we choose
\begin{subequations}
\begin{eqnarray}
\beta_+=1 \,, \\
\beta_-={1\over 2} \,.
\end{eqnarray}
\end{subequations}
The claim is then that the $z,\bar{z}$ dependent part of the four-point function (the function $\G$ of section~\ref{sec:generalities}) is given by corresponding combinations of the functions $G$ and $R_i$. So for instance 
\begin{equation}
P_{aabb}\propto |z_{12}z_{34}|^{-1/4}\G_{aabb}(z,\bar{z}) \,,
\end{equation}
with
\begin{eqnarray}
\G_{aabb}(z,\bar{z}) &=& G+R_2-R_1-R_3 \nonumber \\
&=& 2F_0\bar{F}_0\left(\sum_{m,n=-\infty}^\infty
\qq^{(2m+n/2)^2}\bar{\qq}^{(2m-n/2)^2}-\qq^{(2m+n/2+1/4)^2}\bar{\qq}^{(2m-n/2-1/4)^2}\right) \,.
\end{eqnarray}
What we must do then is to expand the combinations (\ref{combi}) in powers of $z$. This gives expressions of the form 
\begin{equation}
(z\bar{z})^{-1/8} \sum_{\Delta\bar{\Delta}} z^\Delta \bar{z}^{\bar{\Delta}} \,. \label{sumexpi}
\end{equation}
The point is that the set of weights $(\Delta,\bar{\Delta})$ is {\sl the set of conformal weights appearing in the s-channel expansion of the four point function}. (It is not necessarily the set of primary fields, as the exact result does not distinguish between primaries and descendents. Hence our use of the notation $\Delta$ is slightly abusive when compared with the general case.) This must be the same set as the set of {\sl conformal weights contributing to the lattice observable}. Hence, we must compare the sets occurring in (\ref{combi}) with the sets determined directly. These sets can be determined numerically, as for $Q$ generic, and are found to be:
%
%
\begin{subequations} \label{modcont}
\begin{eqnarray}
\hbox{Spec~} P_{aabb} &\subset& \hbox{Spec~}\AStTL{0}{1}/\AStTL{1}{1}+\hbox{Spec~}\AStTL{0}{-1}+\hbox{Spec~}\AStTL{2}{1}/\AStTL{3}{1}+\ldots \,, \\
\hbox{Spec~} P_{aaaa} &\subset& \hbox{Spec~} \AStTL{0}{-1}+\hbox{Spec~} \AStTL{2}{1}/\AStTL{3}{1}+\ldots \,, \\
\hbox{Spec~}(P_{abab}-P_{abba}) &\subset& \hbox{Spec~} \AStTL{2}{-1}+\ldots \,, \\
\hbox{Spec~}(P_{abab}+P_{abba}) &\subset& \hbox{Spec~}\AStTL{2}{1}/\AStTL{3}{1}+\ldots \,.
\end{eqnarray}
\end{subequations}
Here the quotients of modules have a direct meaning in terms of subtracting eigenvalues common to various sectors in finite size.
Like for other non-generic values of $\q$, the modules contributing to the probabilities are no longer irreducible, and the
$Q=4$ results correspond to taking the simple irreducible tops.  Indeed we find in general
that $\hbox{Spec~}\AStTL{2k+1}{1} \subset \hbox{Spec~}\AStTL{2k}{1}$ for $k=1,2,\ldots$.

We have moreover established numerically that 
$\hbox{Spec~}(P_{abab}-P_{abba})$ does not contain $\hbox{Spec~} \AStTL{4}{-1}$, which would otherwise have been a viable candidate in view of
symmetries and the results for generic $Q$. For these reasons we conjecture that the complete result generalising (\ref{modcont}) should in fact read:
\begin{subequations} \label{modcont_conj}
\begin{eqnarray}
\hbox{Spec~} P_{aabb} &=& \hbox{Spec~}\AStTL{0}{1}/\AStTL{1}{1}+\hbox{Spec~}\AStTL{0}{-1}+ \sum_{k \ge 1} \hbox{Spec~}\AStTL{2k}{1}/\AStTL{2k+1}{1} \,, \\
\hbox{Spec~} P_{aaaa} &=& \hbox{Spec~} \AStTL{0}{-1}+\sum_{k \ge 1} \hbox{Spec~} \AStTL{2k}{1}/\AStTL{2k+1}{1} \,, \\
\hbox{Spec~}(P_{abab}-P_{abba}) &=& \hbox{Spec~} \AStTL{2}{-1} \,, \\
\hbox{Spec~}(P_{abab}+P_{abba}) &=& \sum_{k \ge 1} \hbox{Spec~}\AStTL{2k}{1}/\AStTL{2k+1}{1}+ \,.
\end{eqnarray}
\end{subequations}

\subsubsection{Probability $P_{aaaa}$}

First we consider $P_{aaaa}$. The spectrum predicted from the decomposition in (\ref{modcont}) is 
\begin{subequations}
\begin{eqnarray}
(q\bq)^{1/24} F_{0,-1}&=&2 q^{1/16}\bq^{1/16}\left(1+q+\bq+(q\bq)^{1/2}+q\bq+\ldots\right) \,, \\
(q\bq)^{1/24} \left(F_{2,1}-F_{3,1}+\ldots\right)&=&q\bq+q^{1/4}\bq^{9/4}+q^{9/4}\bq^{1/4}+\ldots \,.
\end{eqnarray}
\end{subequations}
Corresponding to this we find
\begin{eqnarray}
-G+R_1+R_2+R_3 &=& (z\bar{z})^{-1/8}\left(\sqrt{2} z^{1/16}\bar{z}^{1/16}+{1\over 16\sqrt{2}}(z^{1/16}\bar{z}^{17/16}+z^{17/16}\bar{z}^{1/16})+{1\over 8\sqrt{2}}(z\bar{z})^{9/16}+\ldots\right.\nonumber\\
 & & \left.- {1\over 128}z\bar{z}-{1\over 512} (z^{1/4}\bar{z}^{9/4}+z^{9/4}\bar{z}^{1/4})+\ldots\right) \,,
\end{eqnarray}
and thus, after mapping to the cylinder 
\begin{eqnarray} \label{Paaaa_cyl_Q4}
P_{aaaa} &\propto& (\xi \bar{\xi})^{1/16} + {\cos^2{2\pi a\over L}\over 4\sqrt{2}} (\xi \bar{\xi})^{1/16}(\xi + \bar{\xi})+{\sin^2{2\pi a\over L}\over 2\sqrt{2}}(\xi\bar{\xi})^{9/16}+\ldots\nonumber\\
& & -{(4\sin^2{2\pi a\over L})^{15/8}\over 128\sqrt{2}}\xi\bar{\xi} -{(4\sin^2{2\pi a\over L})^{19/8}\over 512\sqrt{2}} (\xi \bar{\xi})^{1/4} (\xi^2 + \bar{\xi}^2) + \ldots \,,
\end{eqnarray}
where we recall that
\begin{equation}
\xi=e^{-2\pi (l+ix)/L}\,, \quad \bar{\xi}=e^{-2\pi(l-ix)/L} \,.
\end{equation}
%


The first and third  terms in this expression are associated with the largest eigenvalue in $F_{0,-1}$ at zero momentum (this is denoted $V_{100}$
in section~\ref{sec:momentum_sectors}). The second term also belongs to $F_{0,-1}$, but at non-zero momentum (namely $V_{101}$). Finally,
the fourth and fifth  terms are the leading ones in $F_{2,1}$ (in respective momentum sectors $V_{200}$ and $V_{202}$). To associate these terms
to definite eigenvalues, for each size $L$, requires going through the detailed analysis of section~\ref{sec:specT_CFTlimit}.
We moreover note that one has to be careful in the comparison, since on  the lattice we do not individually observe the amplitudes of conjugate terms,
such as the two terms in $(\xi \bar{\xi})^{1/16}(\xi + \bar{\xi})$, since they correspond to the same eigenvalue. Therefore, to make the comparison
we must divide the numerically observed amplitude by two in such cases (cf.\ remark \ref{rem:degeneracy})---this has been done tacitly below
and in subsequent similar cases.
This leads to the following results for the second to fifth terms in (\ref{Paaaa_cyl_Q4}):


\begin{align}
\setlength{\arraycolsep}{4mm}
\renewcommand{\arraystretch}{1.2}
 \begin{array}{ll|lllll}
   L & & (\xi\bar{\xi})^{1/16}(\xi+\bar{\xi} )&( \xi\bar{\xi})^{9/16} & \xi\bar{\xi} & (\xi \bar{\xi})^{1/4} (\xi^2 + \bar{\xi}^2)
   \\
   \hline
 5 & {\rm lattice} & 0.01025716807 & 0.2499038005 & -0.1702401995 & -0.05899812501 \\
    & {\rm CFT}   & 0.01688067229 & 0.3197920460 & -0.0615759664 & -0.02928111207 \\
    & {\rm ratio}   & 0.6076 & 0.7815 & 2.7647 & 2.0149 \\ \hline
 7 & {\rm lattice} & 0.005692994702 & 0.2821337922 & -0.1820689962 & -0.06496561673 \\
    & {\rm CFT}   & 0.008753198131 & 0.3360469943 & -0.0675747438 & -0.03294025194 \\
    & {\rm ratio}   & 0.6504 & 0.8396 & 2.6943 & 1.9722 \\ \hline
 9 & {\rm lattice} & 0.003595596367 & 0.2949643343 & -0.1820598325 & -0.06747242144 \\
    & {\rm CFT}   & 0.005330469599 & 0.3428924513 & -0.0701787272 & -0.03455627732 \\
    & {\rm ratio}   & 0.6745 & 0.8602 & 2.5942 & 1.9525 \\ \hline
 \end{array} \nonumber
\end{align}
We see that the first two columns come out satisfactorily, and the ratios can be rather convincingly extrapolated to a number close to one,
despite of the small number of sizes. For the higher-order terms (the last two columns of the table) the convergence is definitely slower,
but still goes in the right direction. We have already seen (witness Figure~\ref{figIsing2} in the Ising case) that ratios of the order of $2$ at
small sizes ($L \le 9$ here) are not uncommon, and it is quite plausible that also these columns could be extrapolated to one, provided one
could obtain a few more sizes. In conclusion, the test of $P_{aaaa}$ is fully compatible with the exact results.

\subsubsection{Probability $P_{aabb}$}

We next consider the probability $P_{aabb}$ of a pair of ``short'' clusters.  The spectrum predicted from the decomposition in (\ref{modcont}) is 
\begin{eqnarray}
(q\bq)^{1/24} F_{0,-1}&=&2 q^{1/16}\bq^{1/16}\left(1+q+\bq+(q\bq)^{1/2}+q\bq+\ldots\right) \,, \\
(q\bq)^{1/24} \left(F_{0,1}-F_{1,1}+F_{2,1}-F_{3,1}+\ldots\right)&=&1+(q \bq)^{1/4}+ (q \bq)^{1/4}(q+\bq)+q\bq+(q \bq)^{1/4}(q^2+\bq^2)+\ldots \,. \nonumber
\end{eqnarray}
Correspondingly, we find
\begin{eqnarray}
G-R_1+R_2-R_3 &=& (z\bar{z})^{-1/8}\left( -\sqrt{2} (z \bar{z})^{1/16} - {1\over 16\sqrt{2}} (z \bar{z})^{1/16} (z + \bar{z}) + \ldots\right.\nonumber\\
&+& \left. 1+{1\over 2} (z \bar{z})^{1/4} +{1\over 16} (z \bar{z})^{1/4} (z + \bar{z})+\ldots\right) \,,
\end{eqnarray}
and thus
\begin{eqnarray}
P_{aabb} &\propto& -\sqrt{2} (\xi \bar{\xi})^{1/16} +\ldots\nonumber\\
&+& 1+{(4\sin^2{2\pi a\over L})^{3/8}\over 2} (\xi \bar{\xi})^{1/4} + \ldots \,.
\end{eqnarray}
The first term is the leading eigenvalue in $F_{0,-1}$, corresponding in finite size to the largest eigenvalue in $V_{100}$.
The second and third terms are the two leading eigenvalues in  $\bar{F}_{0,1}$, corresponding to the two largest eigenvalues in $V_{000}$.
The comparison with numerics---presented in the by now familiar format---here comes out as:

\begin{align}
\setlength{\arraycolsep}{4mm}
\renewcommand{\arraystretch}{1.2}
 \begin{array}{ll|ll}
   L & & (\xi\bar{\xi})^{1/16}&(\xi\bar{\xi})^{1/4}   \\
   \hline
 5 & {\rm lattice} & -1.342154491 & 0.6006793776 \\
    & {\rm CFT}   & -1.414213562 & 0.8098363105    \\
    & {\rm ratio}   & 0.9490 & 0.7417 \\ \hline
 7 & {\rm lattice} & -1.400211084 & 0.7065845314  \\
    & {\rm CFT}   & -1.414213562 & 0.8250340627 \\
    & {\rm ratio}   & 0.9901 & 0.8564 \\ \hline
 9 & {\rm lattice} & -1.431374714 & 0.7598690826 \\
    & {\rm CFT}   & -1.414213562 & 0.8312967743 \\
    & {\rm ratio}   & 1.0121 & 0.9141 \\ \hline
 \end{array} \nonumber
\end{align}
It again appears convincing that the ratios will converge to one, i.e., that the numerical results confirm the analytical ones.

\subsubsection{The combination $P_{abab}-P_{abba}$}

As a last example we discuss antisymmetric combination $P_{abab}-P_{abba}$ of the probabilities of having two ``long'' propagating clusters,
as shown in Figure~\ref{fig:2clusters}.

It is easy to expand the spectrum of exponents  arising from (\ref{modcont}) on the one hand: the first few terms are 
\begin{eqnarray*}
(q \bq)^{1/24} F_{2,-1} = (q \bq)^{9/16} (q + \bq + q^2 + \bq^2 + 2 q \bq + \ldots ) + (q \bq)^{1/16} (q^3 + \bq^3 + \ldots) \,.
\end{eqnarray*}
The powers of $q,\bq$ 
on the right-hand side are expected to be exactly the exponents appearing in the sum (\ref{sumexpi}). Using the foregoing discussion and the expressions of $R_1,R_3$, we can on the other hand  identify these exponents by performing a direct expansion whose first few terms are:
\begin{eqnarray*}
R_1-R_3={(z\bar{z})^{-1/8} \over 128\sqrt{2}}\left\lbrace (z \bar{z})^{9/16} \left(z + \bar{z} + {25\over 32}(z^2 + \bar{z}^2) + {9\over 16} z \bar{z} + \ldots \right) +
(z \bar{z})^{1/16} \left(  {1\over 16} (z^3 + \bar{z}^3) + \ldots \right) \right\rbrace \,.
\end{eqnarray*}
This plays the role of the function $\G$ 
in our general discussion (\ref{CylExpan1}). It is easy to check that the two sets agree. Moreover, we have also checked that there is no gap in the spectrum, that is, all exponents predicted from the generating function (\ref{modcont}) are indeed present with non-zero coupling constant. 

To check the numerical values of the amplitudes, we need as usual to map the four-point function on the cylinder. We find the first few terms
\begin{eqnarray}
P_{abab}-P_{abba} &\propto& (\xi \bar{\xi})^{9/16} (\xi + \bar{\xi}) + \left[{9\over 4}-{9\sin^2 {2\pi a\over L}\over 4}\right] (\xi \bar{\xi})^{25/16} \nonumber\\
&+& \left[{25\over 8}-{25\over 8} \sin^2{2\pi a\over L}\right] (\xi \bar{\xi})^{9/16} (\xi^2+\bar{\xi}^2) + {\sin^2 {2\pi a \over L} \over 4} (\xi \bar{\xi})^{1/16} (\xi^3 + \bar{\xi}^3) +\ldots) \,.
\end{eqnarray}

The first term corresponds to the lowest eigenvalue in the $F_{2,-1}$ sector with conformal spin $h-\bar{h}=1$, originating from $V_{211}$ in finite size.
The second and third term correspond to the leading eigenvalue with $h-\bar{h}=0$ (resp.\ $h-\bar{h}=2$), originating from $V_{210}$ (resp.\ $V_{212}$)---note
that since the scaling dimension $h+\bar{h} = \frac{25}{8}$ is the same for these two terms their ordering in the finite-size data cannot be guessed
straightaway, and one has to make use of the lattice momentum to reveal the correct conformal spin.
Finally, the fourth term corresponds to the lowest eigenvalue in the sector with $h-\bar{h}=3$---we
notice that this state is not present for $L=5$, since this size is yet too small to accommodate the high value of the momentum.
The numerical results for the last three terms being discussed (taking ratios with respect to the first term) now run as follows:

\begin{align}
\setlength{\arraycolsep}{4mm}
\renewcommand{\arraystretch}{1.2}
 \begin{array}{ll|lll}
  L & & (\xi\bar{\xi})^{25/16}& (\xi \bar{\xi})^{9/16} (\xi^2 + \bar{\xi}^2) &(\xi \bar{\xi})^{1/16} (\xi^3 + \bar{\xi}^3)  \\
   \hline
5 & {\rm lattice} & 0.05719383440 & 0.2474873290 & 0    \\
   & {\rm CFT}   & 0.10742794066 & 0.2984109462 & 0    \\
   & {\rm ratio}   & 0.5324 & 0.8294 & \\ \hline
7 & {\rm lattice} & 0.03267744359 & 0.1238111030 & 0.1246090804  \\
   & {\rm CFT}   & 0.05570501180 & 0.1547361439 & 0.2376211084  \\
   & {\rm ratio}   & 0.5866 & 0.8001 & 0.5244 \\ \hline
9 & {\rm lattice} & 0.02135602073 & 0.07680979880 & 0.1355927174 \\
   & {\rm CFT}   & 0.03392290080 & 0.09423028002 & 0.2424615776 \\
   & {\rm ratio}   & 0.6295 & 0.8151 & 0.5592 \\ \hline
 \end{array} \nonumber
\end{align}

Finally, observe that the  Kac parametrisation
\begin{equation}
h_{rs}={[(m+1)r-ms]^2-1\over 4m(m+1)}
\end{equation}
becomes, in the limit $m\to\infty$, $h_{rs}={(r-s)^2\over 4}$. The exponents appearing in the various Ashkin-Teller correlators are thus in agreement with  the spectra conjectured in \cite{Ribault}
\begin{eqnarray}
G&:&~~ \mathcal{S}_{\mathbb{Z},2\mathbb{Z},}\nonumber\\
R_1&:& ~~\mathcal{S}_{\mathbb{Z}+{1\over 2},2\mathbb{Z},}\nonumber\\
R_2&:&~~ \mathcal{S}_{\mathbb{Z},2\mathbb{Z}}\nonumber\\
R_3&:& ~~\mathcal{S}_{\mathbb{Z}+{1\over 2},2\mathbb{Z}}
\end{eqnarray}
after the switch of $r,s$ labels mentioned earlier in remark~\ref{rem:switch}.

\subsection{The case $Q=0$}
\label{sec:appB-Q0}


The case $Q=0$ (or rather the limit $Q \to 0$) is connected to the combinatorics of spanning trees and forests (see, e.g., \cite{JSS05,CJSSS04,JS05}). It is interesting in the present context for two reasons.  On the one hand, it gives rise to Jordan cells in the transfer matrix, and thus provides an opportunity to study their effect on the determination of amplitudes. On the other hand, it also turns out that the combination $P_{abab}-P_{abba}$ can be explicitly calculated in this case, providing another non-trivial check  of our method, this time in the region of  small values of $Q$.

While for $Q$ generic,  the two-point function of the Potts spin operator is proportional to the probability that the two points belong to the same cluster
[see (\ref{G_2-point})], the limit $Q\to 0$ needs to be handled with  care, since the partition function itself vanishes. At leading order, the only terms left in the partition
function are spanning trees: a  naive definition---requiring that the two points still belong to the same cluster---would then lead to a trivially constant two-point function.
To obtain non-trivial results it is better to change the normalisation by one factor of $Q$, that is, to redefine the partition function as the number of spanning trees.
A natural redefinition of the two-point function will be given below; it follows by using the equivalence of the $Q\to 0$ limit with symplectic fermions
\cite{SaleurN=2,Ivashkevich} and the theory of spanning trees.

The simplest, in fact, is to start by discussing the combination $P_{abab}-P_{abba}$. 
Indeed,  consider four points labelled $1,2,3,4$ in the plane. As usual, we consider the square lattice in concrete calculations.
We denote by ${\cal N}_{13,24}$ (resp.\ ${\cal N}_{14,23}$) the numbers of configurations of spanning trees where one tree connects points $1,3$
and {\sl a different} tree connects points $2,4$ (resp.\ one tree connects points $1,4$ and a different ones points $2,3$).
It is  possible to show, by generalising Kirchoff's original discussion \cite{Kirchhoff} along the lines of \cite{Priezzhev}, that
%
\begin{equation}
{\cal N}_{13,24}-{\cal N}_{14,23}=\hbox{Det}~ \Delta_{(12)(34)} \,, \label{corrfct}
\end{equation}
where the right-hand side is the determinant of the lattice Laplacian after having removed the  lines corresponding to points $1,2$ and columns corresponding to points $3,4$. Here, the Laplacian is defined as follows. It is a matrix denoted $\Delta$ whose linear size is the number of vertices of the graph, here a planar lattice, that we suppose loopless (i.e., no vertex is connected to itself) for simplicity. The  diagonal elements $\Delta_{ii}$ are the number of edges incident on vertex $i$. The off-diagonal elements $\Delta_{ij}$ are equal to minus the number of edges connecting the vertices $i$ and $j$. We also denote by $\Delta^{(kl)}$ the minor of $\Delta$ obtained by erasing row $k$ and column $l$. We recall that $\hbox{det} \, \Delta=0$, since by definition the sum of all rows (or the sum of all columns) is zero. Moreover, by the Kirchhoff matrix-tree theorem \cite{Kirchhoff}, $\hbox{det} \, \Delta^{(kl)}$ is equal to the number of spanning trees on the graph. If we now go back to the $Q\to 0$ limit of the Potts model and the definition (\ref{P_corr_def}) of the probabilities $P_{a_1,a_2,a_3,a_4}$ in terms of the cluster expansion, we see that (\ref{corrfct}) is the leading contribution to $P_{abab}-P_{abba}$ as $Q\to 0$: in this limit indeed, any non-connected additional cluster gets cancelled by a power of $Q$, and so do cycles 
within the clusters (so each cluster is a tree indeed).

Meanwhile, (\ref{corrfct}) can be calculated in the continuum limit, which is simply described by a pair of symplectic fermions $(\theta^+,\theta^-)$ with Euclidean action
\begin{equation}
S=\int d^2x~ d_{\alpha\beta}\partial_\mu\theta^\alpha\partial_\mu\theta^\beta \,,
\end{equation}
subject to summation over repeated indices, and with $d_{+-}=1,d_{-+}=-1$. The quantity in (\ref{corrfct}) is then nothing but $\langle\theta^+(1)\theta^+(2)\theta^-(3)\theta^-(4)\rangle$ and gives%
\footnote{Note that Wick's theorem has to be handled with care, since the partition function vanishes.}
\cite{Kausch1,Kausch2}
\begin{equation}
\langle\theta^+(1)\theta^+(2)\theta^-(3)\theta^-(4)\rangle=\ln \left|{z_{14}z_{23}\over z_{13}z_{24}}\right|^2 \,. \label{psi4}
\end{equation}

\subsubsection{Two-point function}

Before studying $P_{abab}-P_{abba}$, we can extract from this a useful definition of the two-point function as well. Indeed imagine sending point $1$ to $3$, and point $2$ to $4$. In this case, the configurations in ${\cal N}_{14,23}$ become negligible, while those in ${\cal N}_{13,24}$ describe a situation where $1=3$ belongs to a tree and $2=4$  to a different tree.  
We use this to define a two-point function:
\begin{equation}
g_2(1,2)\equiv  {{\cal N}_{1,2}\over {\cal N}_1} \,,
\end{equation}
where ${\cal N}_{1,2}$ is the number of configurations of spanning forests with two trees only, with points 1 and 2 
 belonging to different trees, while ${\cal N}_1$ denotes the number of spanning trees. In the continuum limit we therefore have from (\ref{psi4})
\begin{equation}
g_2(z,\bar{z})= 2\ln\left| z/\epsilon\right|^2 \,, \label{logtwopt}
\end{equation}
where $\epsilon$ is a short-distance cutoff. Note that $\epsilon$ cannot be eliminated by multiplicative renormalisation, as is usually done in CFT.
We could, alternatively, set $\epsilon=1$, and say that the two-point function takes the form (\ref{logtwopt}) up to a non-universal additive constant. 
 
On the cylinder, we expect the following behavior in the conformal limit \cite{JeanMarc}:
\begin{equation}
g_2(w,\bar{w})=2\ln\left( {L\over\pi\epsilon}\sinh{\pi w\over L}\right)+2\ln\left( {L\over\pi\epsilon}\sinh{\pi \overline{w}\over L}\right) \,.
\end{equation}
We shall measure $w,L$ in units of the lattice spacing. This means that $\epsilon$ will be a (non-universal) numerical constant of order unity. We now expand $g_2$ to see how it connects with the results from the transfer matrix:
\begin{equation}
g_2(\xi,\bar{\xi})=4\ln {L\over 2\pi\epsilon} + {4\pi l\over L}  -2\sum_{n=1}^\infty \left({\xi^n\over n}+{\bar{\xi}^n\over n}\right) \,. \label{expandtwoptf}
\end{equation}

It is seen that $g_2$ consists of a linear term and a sum of exponentials.  All the amplitudes can  be compared with lattice calculations: the sum of exponentials arises from a sum over eigenvalues of the transfer matrix as usual, while the linear term arises because of the presence of a Jordan cell of rank two.  Write (\ref{expandtwoptf}) as 
\begin{equation}
g_2 \propto A l + \sum_{i \ge 0} B_i (\Lambda_i/\lambda_0)^l \,, \label{g13cylexp}
\end{equation}
where $\Lambda_0$ denotes the ground-state eigenvalue. 
We get from (\ref{expandtwoptf}) that $A={4\pi\over L}$ while $B_0=4 \ln {L\over 2\pi\epsilon}$, $B_1=-4$, and $B_2=-2$ (where the last term comes
from $n=2$ in the sum). However, if we change the normalisation of (\ref{g13cylexp}) so that $A={1 \over L}$, we get instead:
\begin{equation}
  B_0 = \frac{1}{\pi} \ln \frac{L}{2 \pi \epsilon} \,, \qquad
  B_1 = -\frac{1}{\pi} \,, \qquad
  B_2 = -\frac{1}{2\pi} \,, \qquad
  B_3 = -\frac{1}{3\pi} \,, \cdots
\end{equation}


As usual we can confront this with the numerical computations. We find that $A={1 \over L}$ {\em exactly} in finite size, which motivates the above choice
of normalisation. Moreover, $g(1,3)$ is found to couple only to a very small set of eigenvalues: apart from the rank-two Jordan cell parameterised by
$(B_0,A)$, there are only two (resp.\ three) simple eigenvalues for $L=5$ and $2a=2$ (resp.\ $L=7$ and $2a=3$). Presented in the usual table form we find:
\begin{align}
\setlength{\arraycolsep}{4mm}
\renewcommand{\arraystretch}{1.2}
 \begin{array}{ll|llll}
  L & & B_0 & B_1 & B_2 & B_3 \\
   \hline
5 & {\rm lattice} & 0.4457211803 & -0.2933396913 & -0.1523814889 & - \\
   & {\rm CFT}   & 0.6560432383 & -0.3183098862 & -0.1591549431 & \\
   & {\rm ratio}   & 0.6794 & 0.9215 & 0.9574 & \\ \hline
7 & {\rm lattice} & 0.5509145352 & -0.3020464547 & -0.1439480976 & -0.1049199829 \\
   & {\rm CFT}   & 0.7631456776 & -0.3183098862 & -0.1591549431 & -0.1061032954 \\
   & {\rm ratio}   & 0.7219 & 0.9489 & 0.9045 & 0.9888 \\ \hline
 \end{array} \nonumber
\end{align}
Note that we have here set $\epsilon = 1$ in the CFT result for $B_0$ in order to get an order of magnitude estimate. Since this term is not expected to be
universal, the agreement for $B_0$ is here seen to be reasonable. The agreement for the other terms ($B_1$, $B_2$, $B_3$) is seen to be very good,
even at these small sizes.


We now comment more generally on the structure of  the two-point function in terms of the transfer matrix.
The linear term in (\ref{expandtwoptf}) indicates the presence of a Jordan cell of rank two (since we get a term linear in $l$), and that  the two corresponding (pseudo) eigenvalues must be in the ground-state sector since there is no exponential decay. 
The sum of exponentials involves only integer conformal weights, and corresponds simply to coupling to descendants of the identity. Remarkably, note that  we  couple only to purely chiral or purely antichiral fields. This explains the scarcity of non-zero amplitudes observed numerically.

\begin{figure}[ht]
\begin{center}
    \includegraphics[width=10cm]{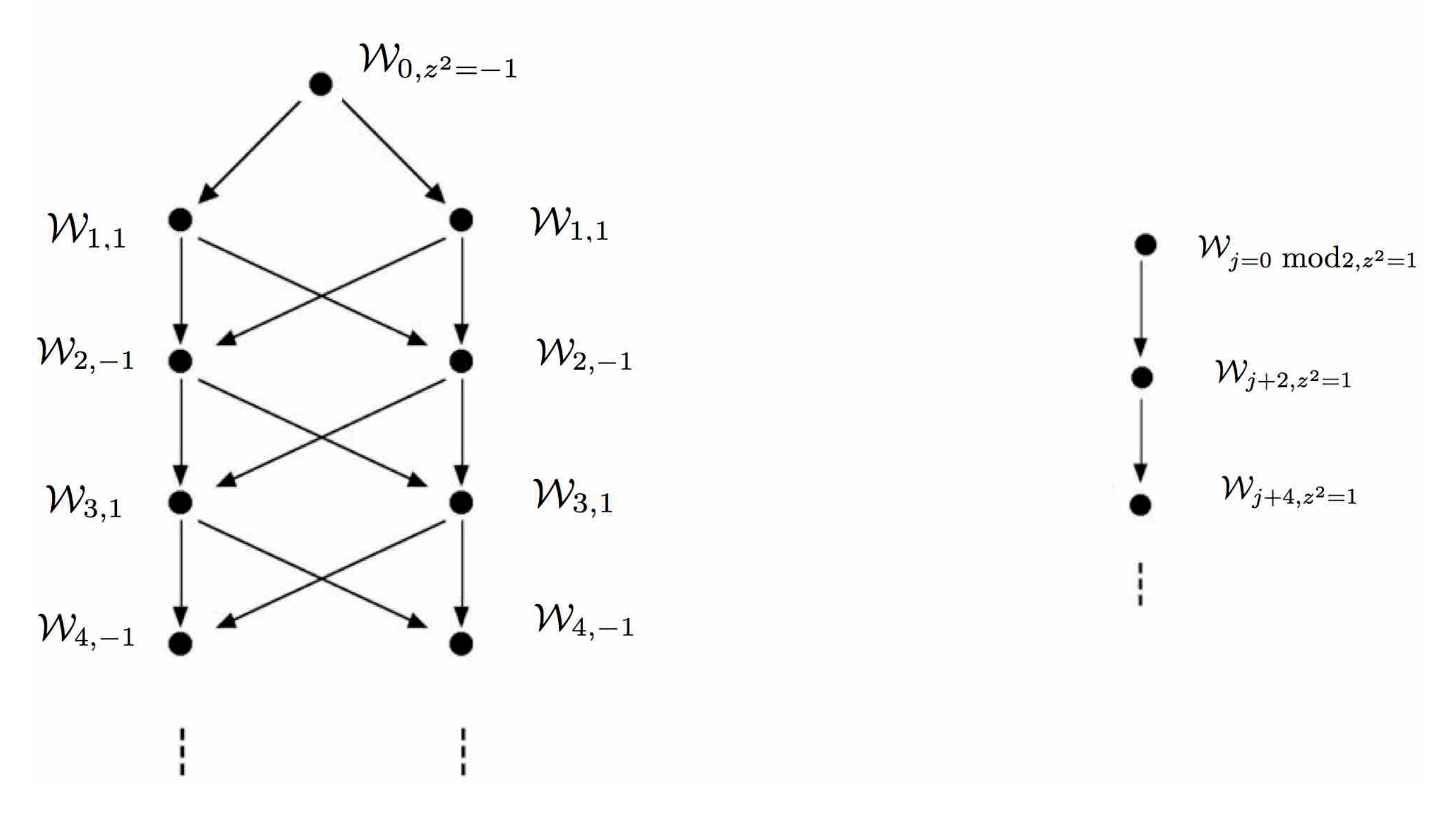}
     \caption{
     Structure of standard modules for $\q=e^{i\pi/2}$. }\label{fig1}
\end{center}
\end{figure}

It is also useful here to  recall the structure of some of the modules for $Q=0$. Since $\q$ is a root of unity, the $\AStTL{j}{z^2=e^{2iK}}$ are reducible, and have a structure of submodules that depends on the values of $j,z$. For instance 
we get the structure shown in Figure~\ref{fig1}.

Meanwhile, we can read the dimension content from the general formulae:
\begin{equation}
F_{j,1}={1\over (q\bar{q})^{-2/24}P(q)P(\bar{q})}\sum_{e\in \mathbb{Z}} q^{{(j-2e)^2-1\over 8}}\bar{q}^{{(j+2e)^2-1\over 8}} \,.
\end{equation}
Let us consider in more detail $F_{1,1}$. This contains the exponents $(h,\bar{h})$, here ordered by the total value of the scaling dimension $h+\bar{h}$:
\begin{equation}
(0,0);(0,1)\times 2;(1,0)\times 2;(1,1)\times 3;(2,0)\times 3;(0,2)\times 3; \ldots \,.
\end{equation}
We see that the first eigenvalue with $h-\bar{h}=1$ is $(h,\bar{h}) = (1,0)$, the first eigenvalue with $h-\bar{h}=2$  is $(h,\bar{h})=(2,0)$, etc. This agrees with the expansion (\ref{expandtwoptf}). Since only purely chiral contributions appear in this expansion, this means that, for a given value of $n=h-\bar{h}$, we have only the smallest realisation, $(h,\bar{h})=(n,0)$. The same applies to all the other sectors. 

\subsubsection{The four-point function $P_{abab}-P_{abba}$}

We now go back to the combination $P_{abab}-P_{abba}$, which we write  in terms of the variable $z\equiv {z_{12}z_{34}\over z_{13}z_{24}}$:
\begin{equation}
P_{abab}-P_{abba}\propto -[\ln(1-z)+\ln(1-\bar{z})] \,, \label{simple1}
\end{equation}
and expanding this we get
\begin{equation}
P_{abab}-P_{abba}\propto \sum_{n=1}^\infty {z^n+\bar{z}^n\over n} \,.
\end{equation}
Note that, if we exchange points 1 and 2, we have $z={z_{12}z_{34}\over z_{13}z_{24}}\to {z_{21}z_{34}\over z_{23}z_{14}}={z\over z-1}$. This corresponds to sending $1-z\to {1\over 1-z}$, and thus to changing the sign of $P_{abab}-P_{abba}$, as required from the geometrical interpretation. This does {\sl not mean} that only odd spins appear in the expression (\ref{leadingexpan}),  because in this case there are large degeneracies (in fact, only a few of the terms correspond to primary fields). 

%
Going to the cylinder as usual we find now 
\begin{equation}
P_{abab}-P_{abba}\propto\sum_{n=1}^\infty {(-1)^n\over n} \left(4\sin^2 {2\pi a\over L}\right)^n {\xi^n\over (1-\xi)^{2n}}+\hbox{ h.c.}\label{4ptftcq0}
\end{equation}
%


To compare with the numerical work, we rewrite (\ref{4ptftcq0}) as
\begin{equation}
 P_{abab}-P_{abba}\propto\sum_{n=1}^\infty C_n \xi^n + \hbox{ h.c.} \label{4ptftcq0rewritten}
\end{equation}
and choose the normalisation $C_1 = 1$, so that the amplitude ratios can be read off directly from the following coefficients. The first few then read
explicitly
\begin{subequations}
\begin{eqnarray}
 C_2 &=& 2 \left( \cos \frac{2 \pi a}{L} \right)^2 \,, \\
 C_3 &=& \frac13 \left( 1 + 2 \cos \frac{4 \pi a}{L} \right)^2 \,, \\
 C_4 &=& \left( \cos \frac{2 \pi a}{L} + \cos \frac{6 \pi a}{L} \right)^2 \,.
\end{eqnarray}
\end{subequations}

As above, the numerical work was done for $L=5$ and $2a=2$ (resp.\ $L=7$ and $2a=3$). The contributing eigenvalues are then found to be exactly
the same as for the two-point function $g_2(1,2)$, 
corresponding to the terms $n=1,2$ (resp.\ $n=1,2,3$) in (\ref{4ptftcq0rewritten}). There is no longer
any Jordan cell (the term $n=0$). Note also that in both cases, we expect at all orders to have only chiral or antichiral contributions.
The numerical results compared with the conformal predictions run as follows:
\begin{align}
\setlength{\arraycolsep}{4mm}
\renewcommand{\arraystretch}{1.2}
 \begin{array}{ll|lll}
  L & & C_2/C_1 & C_3/C_1 \\
   \hline
5 & {\rm lattice} & 0.1984202998 & - \\
   & {\rm CFT}   & 0.1909830056 & \\
   & {\rm ratio}   & 1.0389 &  \\ \hline
7 & {\rm lattice} & 0.09439172580 & 0.2233910496 \\
   & {\rm CFT}   & 0.09903113209 & 0.2143680440 \\
   & {\rm ratio}   & 0.9532 & 1.0421 \\ \hline
 \end{array} \nonumber
\end{align}
Once again the agreement is very reasonable, given the rather small sizes.

%

A peculiarity of this four-point function---as well as of the two-point function $g_2$ considered above---is that only fields with $h=0$ or $\bar{h}=0$ appear. Since the spin $h-\bar{h}$ is limited to values $0,1,\ldots, L-1$ for a finite size $L$, this means only a very small number of eigenvalues contribute to the correlation function at finite $L$. This is quite different from the usual case, where having $h-\bar{h}$ fixed does not preclude eigenvalues with larger $h+\bar{h}$ from contributing. Of course, it could be that some eigenvalues would contribute in finite $L$, although their amplitude would tend to zero when $L$ increases, because they would not be present in the continuum limit. But this does not seem to be the case here---a fact that  can probably be proven exactly in finite size using the Laplacian on the cylinder. 

\subsubsection{The combination $P_{aabb}-P_{abba}$}

It is also interesting to observe that, under crossing $z_2\leftrightarrow z_3$, ie $z\to {1\over z}$, the combination $P_{abab}-P_{abba}$ that
we have just discussed becomes $P_{aabb}-P_{abba}$. We have studied the latter combination independently.
From our analytical result (\ref{simple1}) we now find 
\begin{equation}
P_{aabb}-P_{abba}\propto \ln z-\ln(1-z)+\ln \bar{z}-\ln(1-\bar{z}) \,.
\end{equation}
Hence on the cylinder we expect 
%
\begin{equation}
P_{aabb}-P_{abba}\propto \ln(4\sin^2(2\pi a/L))+\ln\xi+\sum_{n=1}^\infty \left(2{\xi^n\over n}+{(-4\sin^2(2\pi a/L))^n\over n}
{\xi^n\over (1-\xi)^{2n}}\right)+\hbox{h.c.} \,.
\end{equation}

It is easy to ascertain that the correct sectors are observed in the numerical study.
Indeed we observe now again a Jordan cell on the ground state, plus the same 2 (resp.\ 3) simple eigenvalues as before for the case $L=5$ and $2a=2$
(resp.\ $L=7$ and $2a=3$). The indecomposability parameter, i.e., the coefficient multiplying $l$ in the correlation function, is found to be $A = \frac{1}{L}$ 
exactly in finite size, like it was for the two-point function. Let us write the remaining terms, concerning the simple eigenvalues, as
$\sum_{n=1}^\infty D_n \xi^n + \hbox{ h.c.}$ The comparison with numerics can then be summarised as follows:
\begin{align}
\setlength{\arraycolsep}{4mm}
\renewcommand{\arraystretch}{1.2}
 \begin{array}{ll|lll}
  L & & D_2/D_1 & D_3/D_1 \\
   \hline
5 & {\rm lattice} & -0.1984202998 & - \\
   & {\rm CFT}   & -0.1909830056 & \\
   & {\rm ratio}   & 1.0389 &  \\ \hline
7 & {\rm lattice} & -0.3298008395 & 0.08579175620 \\
   & {\rm CFT}   & -0.3460107358 & 0.08232653457 \\
   & {\rm ratio}   & 0.9532 & 1.0421 \\ \hline
 \end{array} \nonumber
\end{align}
It will not have escaped the reader's attention that the ratios are exactly the {\em same} as for the previously considered case $P_{abab}-P_{abba}$.
This presumably extends to arbitrary $L$ and implies that the symmetry $z \to \frac{1}{z}$ is respected exactly on the lattice---something that we
again suspect can be proven by a careful study of the Laplacian on the cylinder. Per se, this close relationship between the foregoing two
four-point functions is pretty remarkable, since after all, in the lattice study of $P_{abab}$ we have two long clusters,
while in $P_{aabb}$ we have two short ones.
 
%
%
%
%
%
%
%

\end{document}